\documentclass[aasmacros]{aa}

\usepackage{amsmath}
\usepackage{wasysym}
\usepackage{marvosym}
\usepackage{txfonts}
\usepackage{mathtools}
\usepackage{placeins}
\usepackage{siunitx}
\usepackage{enumitem}
\DeclareSIUnit[]\astronomicalunit{\text{au}}

\usepackage{natbib}
\bibpunct{(}{)}{;}{a}{}{,} 

\usepackage{epstopdf}
\usepackage{graphicx}
\pdfoutput=1
\usepackage[pdftex]{color,xcolor}
\usepackage{subcaption}

\usepackage{scalerel}
\usepackage{tikz}
\usetikzlibrary{svg.path}

\definecolor{orcidlogocol}{HTML}{A6CE39}
\tikzset{
  orcidlogo/.pic={
    \fill[orcidlogocol] svg{M256,128c0,70.7-57.3,128-128,128C57.3,256,0,198.7,0,128C0,57.3,57.3,0,128,0C198.7,0,256,57.3,256,128z};
    \fill[white] svg{M86.3,186.2H70.9V79.1h15.4v48.4V186.2z}
                 svg{M108.9,79.1h41.6c39.6,0,57,28.3,57,53.6c0,27.5-21.5,53.6-56.8,53.6h-41.8V79.1z M124.3,172.4h24.5c34.9,0,42.9-26.5,42.9-39.7c0-21.5-13.7-39.7-43.7-39.7h-23.7V172.4z}
                 svg{M88.7,56.8c0,5.5-4.5,10.1-10.1,10.1c-5.6,0-10.1-4.6-10.1-10.1c0-5.6,4.5-10.1,10.1-10.1C84.2,46.7,88.7,51.3,88.7,56.8z};
  }
}

\newcommand\orcidicon[1]{\href{https://orcid.org/#1}{\mbox{\scalerel*{
\begin{tikzpicture}[yscale=-1,transform shape]
\pic{orcidlogo};
\end{tikzpicture}
}{|}}}}

\usepackage{hyperref}
\hypersetup{draft=False, colorlinks=true, citecolor=blue, urlcolor=blue, linkcolor=red, linktoc=all}

\usepackage{pifont}
\def\Haek{\ding{51}}

\DeclareMathOperator\erf{erf}

\def\mearth{M_\oplus}

\def\msun{M_\odot}

\def\f1{f_{\rm I}}
\def\mstar{M_*}
\def\rstar{R_*}

\def\beq{\begin{equation}}
\def\eeq{\end{equation}}

\def\t2{\tau_{\rm II}}

\def\sigmas0{\Sigma_{\rm s,0}}
\def\mj{M_{\textrm{\tiny \jupiter }}}

\def\s0{S_0}

\def\apjl{ApJ}                
\def\apjs{ApJS}               
\def\ao{Appl.Optics}          
\def\aap{A\&A}                

\def\({\left(}
\def\){\right)}
\def\<{\left<}
\def\>{\right>}

\defcitealias{2018MNRAS.475.5618B}{B18}
\defcitealias{Kimura2016}{K16}

\begin{document}

\title{The influence of infall on the  properties of protoplanetary discs}
\subtitle{Statistics of masses, sizes, lifetimes, and fragmentation}

   \author{O. Schib\inst{\ref{bern},\ref{uzh}}
          \and
          C.~Mordasini\inst{\ref{bern}}
          \and
          N.~Wenger\inst{\ref{bern}}
          \and
          G.-D.~Marleau\inst{\ref{tueb},\ref{bern},\ref{mpia}}
          \and
          R.~Helled\inst{\ref{uzh}}
          }

   \institute{
Physikalisches Institut,
Universit\"at Bern, Gesellschaftsstrasse 6, 3012 Bern, Switzerland\\
\email{oliver.schib@space.unibe.ch}
\label{bern}
\and
Center for Theoretical Astrophysics and Cosmology,
Institute for Computational Science,
Universit\"at Z\"urich, Winterthurerstrasse~190, 8057~Z\"urich, Switzerland
\label{uzh}
\and Institut f\"ur Astronomie und Astrophysik, Universit\"at T\"{u}bingen, Auf der Morgenstelle 10, 72076 T\"{u}bingen, Germany
\label{tueb}
\and Max-Planck-Institut f\"ur Astronomie, K\"onigstuhl~17, 69117~Heidelberg, Germany
\label{mpia}
             }

\date{Received August 11, 2020 / Accepted November 8, 2020}

\abstract
{The properties of protoplanetary discs determine the conditions for planet formation. In addition, planets can already form during the early stages of infall.}
{We constrain physical quantities such as the mass, radius, lifetime, and gravitational stability of protoplanetary discs by studying their evolution from formation to dispersal.}
{We perform a population synthesis of protoplanetary discs with a total of $\num{50000}$ simulations using a 1D vertically integrated viscous evolution code, studying a parameter space of final stellar mass from $\num{0.05}$ to $\SI{5}{\msun}$. 
Each star-and-disc system is set up shortly after the formation of the protostar and fed by infalling material from the parent molecular cloud core. Initial conditions and infall locations are chosen based on the results from a radiation-hydrodynamic population synthesis of circumstellar discs. We also consider a different infall prescription based on a magnetohydrodynamic (MHD) collapse simulation in order to assess the influence of magnetic fields  on disc formation. The duration of the infall phase is chosen to produce a stellar mass distribution in agreement with the observationally determined stellar initial mass function.}
{We find that protoplanetary discs are very massive early in their lives. When averaged over the entire stellar population, the discs have masses of $\sim \num{0.3}$ and $\SI{0.1}{\msun}$ for systems based on hydrodynamic or MHD initial conditions, respectively. In systems characterised by a final stellar mass $\sim \SI{1}{\msun}$, we find disc masses of $\sim \SI{0.7}{\msun}$ for the `hydro' case and $\sim \SI{0.2}{\msun}$ for the `MHD' case at the end of the infall phase. Furthermore, the inferred total disc lifetimes are long, $\approx \num{5}$--$\SI{7}{Myr}$ on average. This is despite our choice of a  high value of $\num{e-2}$ for the background viscosity $\alpha$-parameter.
In addition, we find that fragmentation is common in systems that are simulated using hydrodynamic cloud collapse, with more fragments of larger mass formed in more massive systems. In contrast, if disc formation is limited by magnetic fields, fragmentation may be suppressed entirely.}
{Our work draws a picture quite different from the one often assumed in planet formation studies: Protoplanetary discs are more massive and live longer. This means that more mass is available for  planet formation.
Additionally, when fragmentation occurs, it can affect the disc's evolution by transporting large amounts of mass radially. We suggest that the early phases in the lives of protoplanetary discs should be included in studies of planet formation. Furthermore, the evolution of the central star, including its accretion history, should be taken into account when comparing theoretical predictions of disc lifetimes with observations.}


\maketitle

\section{Introduction}\label{sect:Introduction}
Planet formation takes place in protoplanetary discs around young stars. Two prevalent formation mechanisms have been proposed. The first is core accretion (CA) \citep{safronov1972,Pollack1996}, where planetesimals, formed from dust, coagulate to form rocky cores and continue to become terrestrial planets or, by accretion of gas from the disc, gas giant planets. The second is gravitational instability (GI) \citep{1951PNAS...37....1K,Cameron1978,1997Sci...276.1836B,1998ApJ...503..923B}, where entire regions of protoplanetary discs collapse under their own gravity (fragmentation) to form bound clumps consisting mainly of gas (fragments).
Most if not all protoplanetary discs are self-gravitating early in their lives. In this phase, angular momentum transport is dominated by global instabilities (Sect.~\ref{subsubsect:gobaltrans}) or spiral arms and the discs may fragment (\citealt{Harsono2011}, \citealt{2005ApJ...633L.137V}, \citealt{2018MNRAS.477.3273N}, especially the introduction and references therein).
This has a profound effect on properties like the temperature in the disc during this time --- and maybe also much later. In particular, if protoplanets form during this early phase, their properties will be strongly influenced by the conditions in the disc. In some cases the disc may fragment and produce bound clumps that could survive to become gaseous planets.
\citet{Manara2018} compared masses of observed exoplanets to measured disc masses and found that exoplanetary systems masses are comparable to or higher than the most massive discs. They conclude that either dust disc masses are underestimated, planets form very rapidly or discs are being continuously replenished from the environment. Including the formation phase of the discs, when discs are most massive, could explain this apparent lack of planet formation material.

The formation of protoplanetary discs is coupled to the formation of stars: The collapse of molecular cloud cores (MCC hereafter) leads to the formation of a protostar at the centre, surrounded by a disc of gas (and dust) around it, or possibly multiple protostars. Gravitational instability typically begins at early times in the disc's evolution (few $\num{e3}$ to $\SI{e4}{yr}$) when the MCC collapse is still ongoing \citep{2008NewAR..52...21L}. This means that if a fragment forms, the material removed from the disc can be replenished by infalling matter from the cloud, which may lead to further fragmentation.

Many planet formation studies begin from a given disc profile and do not consider any infall of envelope material. 
Initial disc masses are often chosen around a value obtained from the minimum mass solar nebula (MMSN) \citep{weidenschilling1977,1981PThPS..70...35H} of ($\sim \SI{0.01}{\msun}$) or a few times this value \citep{2018haex.bookE.143M,Mutter2017,Coleman2017,Hopkins2016,Lines2015,Simon2015a}. 
Some studies do consider infall, but use very simple  models like the classic Shu collapse model \citep{shu1977b} or the Bonnor--Ebert sphere \citep{1955ZA.....37..217E,1956MNRAS.116..351B}: \citet{2005A&A...442..703H,Jin2014,Xiao2015,Kimura2016}.
\citet{2018MNRAS.477.3273N} make a strong case against using the MMSN prescription and instead introduce the ``maximum mass solar nebula''. 
They argue that the most realistic time to start planet formation simulations is when the disc reaches the maximum mass. 
This typically corresponds to the end of the infall phase. However, as we discuss below, many discs become self-gravitating long before this stage. This typically leads to transfer of significant amounts of mass and angular momentum in the disc either by global instabilities (Sect.~\ref{subsubsect:gobaltrans}, spiral arms (Sect.~\ref{subsubsect:spiral}), or gravitationally bound clumps, which  significantly changes the conditions in the disc. 
Furthermore, the properties of the host star are also influenced by accretion of disc material and/or clumps onto  the star.

There are several other aspects in the formation process of protoplanetary discs that are often neglected in planet formation studies:
\begin{enumerate}
\item Star formation is a highly turbulent process and it is not self-evident that protoplanetary discs can be adequately described by simple power-law density profiles \citepalias{2018MNRAS.475.5618B}.
\item The infalling material reaches the disc at sub-Keplerian velocity and therefore changes the angular momentum budget of the disc \citep{2005A&A...442..703H}.
\item The infalling material collides with the disc at some point along its trajectory. Adding the disc material at a distance from the star that corresponds to the streamlines intersecting the disc's midplane, as done in several studies, is incorrect \citep{Visser2010}.
\item Collapsing MCCs are characterised by strong outflows from the inner region of the forming disc \citep{1996ARA&A..34..207H}. This can change the infall streamlines and prevent matter from falling near or directly onto the star as predicted for example by the Shu collapse model.
\end{enumerate}

In this work we investigate the influence of the infall prescription on the statistical properties of the protoplanetary disc at the end of the infall phase, on their lifetimes, and on the number and properties of fragments formed. We perform a population synthesis of discs. Our model includes the formation of the discs and their evolution until their dispersal. We choose the infall times (duration of the infall phase) such that the resulting population of stars agrees with the initial mass function of \citet{Chabrier2007}. Our inferred distributions of disc properties can be compared to observations.

The paper is organised as follows: in Sect.~\ref{sect:Model} we describe the model used in this study. Section~\ref{sec:param} details the parameter space investigated and the initial conditions used. In Sect.~\ref{sec:example} we describe the formation and evolution of a system as an example. In Sect.~\ref{sect:Results} we present our results, which we compare to other studies and observations in Sect.~\ref{Sect:comp}. Our conclusions are summarised in Sect.~\ref{sect:Conclusions}.

\section{Model description}\label{sect:Model}

Here we present the numerical model used in this work. After a brief overview we give some details about our treatment of the disc's evolution, temperature model, viscosity, inner edge, stellar evolution, disc dispersal, and fragmentation.

The core of our simulation consists of a rotationally symmetric disc of gas described by the vertically integrated surface density $\Sigma$. The evolution of the surface density is calculated by solving the diffusion equation with an effective viscosity (see \S~\ref{subsect:Evol}). The disc is truncated at the inner edge. 
Matter evolving across this truncation radius is assumed to accrete onto the central star of mass $\mstar \equiv \mstar(t)$, with a fraction considered as outflow (see \S~\ref{subsect:Inneredge}).

For simplicity, in the following we refer to the young stellar object at the centre of the disc as ``star'', even though not all such objects qualify for this term in a strict sense. In the early stages of the simulations the central objects are not on the pre-main sequence, and some objects never reach hydrogen fusion.

We start our simulation at a time of $\SI{1}{kyr}$ after the formation of the central object with an initial disc profile obtained from the data in \citet[][hereafter \citetalias{2018MNRAS.475.5618B}]{2018MNRAS.475.5618B}, as detailed in \S~\ref{subsect:Initial}. The disc evolves viscously and is fed by infalling material from the MCC as described in \S~\ref{subs:Infall}.

As the disc evolves, the conditions for fragmentation (\S~\ref{subsect:Fragmentation}) are continuously monitored. When they are fulfilled, we expect the formation of a bound clump of disc material by gravitational collapse. Therefore, we remove the corresponding clump's mass from the disc. At this stage of our project, we do not model the clump as an evolving protoplanet. Instead, we follow a similar approach as in \citet{Kratter2008} and add the clump's mass to the star. The modelling of the clump would entail a number of complex physical processes related to the clump's evolution such as  mass accretion, mass loss, migration and gap formation. These processes are of great importance to the question whether some of these clumps can survive to become giant planets and we hope to address this topic in future research. 
In this study we focus on the conditions in the disc during and after clump formation, and we therefore assume that the clumps  migrate inwards on a short timescale and accrete on the central star. Based on the results of hydrodynamic simulations we expect this to be a likely outcome of fragmentation \citep{2011MNRAS.416.1971B,Malik2015}. Therefore, the material removed from the disc by fragmentation is simply added to the star after a type-I migration time scale $\tau_0$ \citep{Baruteau2016}:
\begin{equation}
\begin{aligned}
	\tau_0 &= \SI{1.3}{Myr} \, \times \, \\
	& \left( \frac{h}{0.05} \right) ^2 \left( \frac{\Sigma}{\SI{200}{g.cm^{-2}}} \right) ^{-1} \left(\frac{\mstar}{\msun}\right) ^{3/2} \left( \frac{r_\mathrm{p}}{\SI{5}{au}} \right) ^{-1/2} \left( \frac{M_\mathrm{p}}{\mearth} \right) ^{-1},
\end{aligned}
\end{equation}
with $h$ the disc's aspect ratio, $r_\mathrm{p}$ and $M_\mathrm{p}$ the clump's semi-major axis and mass, respectively. We consider $10 \%$ of the clump's mass as outflow (see also \S\ref{subsect:Inneredge}).
The central star is evolved according to the evolution tracks of \citet{2008ASPC..387..189Y} as described in \S~\ref{subsect:Stellarmodel}. We calculate the disc's vertical temperature structure  by taking into account a number of physical effects (see \S~\ref{subsect:Temperaturemodel}).

\subsection{Evolution of the protoplanetary disc}\label{subsect:Evol}

The evolution of the disc's surface density $\Sigma$ is described by the diffusion equation \citep{1952ZNatA...7...87L,1974MNRAS.168..603L}:
\begin{equation}
\label{evogeneral}
	\frac{\partial \Sigma}{\partial t} = \frac{3}{r}\frac{\partial}{\partial r}\left[ r^{1/2} \frac{\partial}{\partial r}\left(\nu \Sigma r^{1/2}\right)\right] + S.
\end{equation}
Here, $r$ denotes the distance from the central star. $S \equiv S(r,t)$ is a source/sink term:
\begin{equation}
    S \equiv S(r,t) = S_\mathrm{inf}(r,t) - S_\mathrm{int}(r,t) - S_\mathrm{ext}(r,t) - S_\mathrm{frag}(r,t).
\end{equation}
$S_\mathrm{inf}$ describes the infall from the MCC (see \S~\ref{subs:Infall}), $S_\mathrm{int}$ and $S_\mathrm{ext}$ the rates of internal and external photoevaporation, respectively (\S~\ref{subs:evap}) and $S_\mathrm{frag}$ the removal of mass due to fragmentation (\S~\ref{subsect:Fragmentation}).
We solve Eq.~\ref{evogeneral} using the implicit donor cell advection-diffusion scheme from \citet{Birnstiel2010}.

\subsubsection{Autogravitation}\label{subsubsect:AG1}

Since at early stages the discs are massive, we cannot neglect their self-gravity.
We follow \citet{2005A&A...442..703H} in assuming a vertical disc structure that is isothermal ($p=\rho c_\mathrm{s}^2$) and in hydrostatic equilibrium, and consider the vertical component of the star's gravity as well as the local gravitation of the disc:
\begin{equation}
\frac{1}{\rho}\frac{\mathrm{d}p}{\mathrm{d}z} = -\frac{G \mstar}{R^3} z - 4\pi G\Sigma,
\label{hydrostat}
\end{equation}
where $\rho \equiv \rho(r,z)$ and $p \equiv p(r,z)$  are the density and pressure in the disc, respectively, $G$ is the gravitational constant.
The angular frequency of the disc becomes:
\begin{equation}\label{omegah}
\Omega(r,t)=\bigg[\frac{G \mstar}{r^3}+\frac{1}{r}\frac{\mathrm{d}V_\mathrm{d}}{\mathrm{d}r}\bigg]^{1/2}
\end{equation}
The Keplerian angular frequency therefore receives a modification through the disc's gravitational potential $V_\mathrm{d}$ \citep{2005A&A...442..703H}:
\begin{equation}
	V_\mathrm{d}(r)=\int_{\rstar}^{\infty}\!-G\frac{4\mathrm{K}\,\big[-\frac{4r/r_1}{(r/r_1-1)^2}\big]}{\left|(r/r_1-1)\right|}\Sigma(r_1)\,\mathrm{d}r_1.
\end{equation}
$\mstar$ and $\rstar$ denote the stellar mass and radius, respectively. $\mathrm{K}$ is the elliptic integral of the first kind.

The solution to Eq.~\ref{hydrostat} is:
\begin{equation}
\label{vertstruct}
\rho(r,z)=\rho_0(r) \exp \bigg(-\bigg(\frac{\left|z\right|}{H_0}+\bigg(\frac{z}{H_1}\bigg)^2\bigg)\bigg)
\end{equation}
where $\rho_0(r) \equiv \rho(r,0)$ is the density in the midplane, $H_0$ represents the influence of the autogravitation of the disc and $H_1$ the gravitation of the star
\footnote{Note the special notation in \cite{2005A&A...442..703H}. The denominator of their Eq.~(19) should be $\sqrt{\mathrm{G}\mstar/r^3}$, and so should their $\Omega_k$ in their Eq.~(16). Otherwise their Eq.~(17) is not a solution to their Eq.~(16) and/or their Eq.~(16) is not consistent with Eq.~(5) in \citet{2000A&A...358..378H} and Eq.~(5) in \citet{Paczynski1978}.}:
\begin{equation}
H_0=\frac{c_\mathrm{s}^2}{4\pi G\Sigma}
\end{equation}
\begin{equation}
H_1=\frac{\sqrt{2}c_\mathrm{s}}{\sqrt{\frac{\mathrm{G} \mstar}{r^3}}}
\end{equation}
with $c_\mathrm{s}$ the isothermal sound speed:
\begin{equation}
c_\mathrm{s}=\sqrt{\frac{k_\mathrm{B}T}{\mu \mathrm{u}}},
\end{equation}
$k_\mathrm{B}$ is  the Boltzmann constant, $T$ the midplane temperature, $\mu$ the mean molecular weight which we set to $\num{2.3}$ in this study, and $\mathrm{u}$ the atomic mass unit.

For the relationship between $\rho(r,z)$ and $\Sigma(r)$ the following needs to hold:
\begin{equation}
\Sigma(r)=\int_{-\infty}^{\infty} \! \rho(r,z) \, \mathrm{d}z
\end{equation}
Requiring
\begin{equation}
\rho_0(r)=\frac{\Sigma(r)}{\sqrt{2\pi}H},
\label{sigma_h}
\end{equation}
we get for the vertical scale height $H$:
\begin{equation}
H=\frac{H_1}{\sqrt{2}}\exp\bigg(\frac{H_1^2}{4H_0^2}\bigg)\bigg(1-\erf\bigg(\frac{H_1}{2H_0}\bigg)\bigg).
\end{equation}
Equation~\ref{evogeneral} was derived assuming a Keplerian orbit. However, with the definition of $\Omega$ in Eq.~\ref{omegah} we introduce an angular frequency that is not Keplerian. Therefore our treatment of the disc's autogravitation on its evolution is not completely self-consistent  
and should be regarded as an approximation (as in \citealt{2005A&A...442..703H}).

\subsection{Temperature model}\label{subsect:Temperaturemodel}

Based on the vertical structure of the disc (Eq.~\ref{vertstruct}) we consider the following physical processes when determining the disc's midplane temperature:
\begin{itemize}
\item viscous heating
\item irradiation from the remaining envelope of the MCC
\item shock heating from gas infalling on the disc's surface
\item irradiation due to the star's intrinsic luminosity
\item irradiation due to shock heating from accretion of material onto the star.
\end{itemize}

Following \citet{nakamotonakagawa1994a} in considering an optically-thick as well as an optically-thin regime we have an energy balance at the disc's surface:
\begin{equation}
	\sigma T_\mathrm{S}^4 = \frac{1}{2}\left( 1 + \frac{1}{2 \Sigma \kappa_\mathrm{p}} \right) (\dot{E}_\nu + \dot{E}_\mathrm{s}) + \sigma T_\mathrm{env}^4 + \sigma T_\mathrm{irr}^4
\end{equation}
with $T_\mathrm{S}$ the surface temperature and we obtain the following expression for the midplane temperature $T_\mathrm{mid}$:
\begin{equation}
\label{temp}
\begin{aligned}
	\sigma T_\mathrm{mid}^4 &=
	\frac{1}{2} \left[ \left( \frac{3 \Sigma \kappa_\mathrm{R}}{8} + \frac{1}{2 \Sigma \kappa_\mathrm{P}} \right) \dot{E}_\nu + \left( 1 + \frac{1}{2 \Sigma \kappa_\mathrm{P}} \right) \dot{E}_\mathrm{s} \right] \\
	&+ \sigma T_\mathrm{env}^4 \\
	&+ \sigma T_\mathrm{irr}^4.
\end{aligned}
\end{equation}
Here, $\kappa_\mathrm{R} = \kappa_\mathrm{R}(\rho_0,T_\mathrm{mid})$ and $\kappa_\mathrm{P} = \kappa_\mathrm{P}(\rho_0,T_\mathrm{mid})$ denote the Rosseland and Planck opacity evaluated at the midplane temperature, respectively, $\dot{E}_\nu(T_\mathrm{mid}) = \Sigma \nu(T_\mathrm{mid}) \left( r \frac{\mathrm{d}\Omega}{\mathrm{d}r}\right)$ the energy dissipation rate due to viscous transport (we do not assume  Keplerian frequencies), $\dot{E}_\mathrm{s} = S_\mathrm{inf} (r \Omega)^2 / 2$ the shock heating due to infalling material \citep{Kimura2016} (K16 from now on). The expression for the infall source term $S_\mathrm{inf}$ is given in \S~\ref{subs:Infall}. $T_\mathrm{env}$ (set to $\SI{10}{K}$ in this study) is the background temperature in the envelope, and $T_\mathrm{irr}$ the irradiation temperature from the central star. $T_\mathrm{irr}$ contains the contribution of the star's intrinsic luminosity as well as the luminosity from accreted material (see \S~\ref{subsect:Stellarmodel}). We use the irradiation model from \citet{2005A&A...442..703H}.
Eq.~\ref{temp} is solved numerically using Brent's method to obtain the midplane temperature $T_\mathrm{mid}$.
We use gas opacities from \citet{Malygin2014} and dust opacities from \citet{Semenov2003} to calculate $\kappa_\mathrm{R}$ and $\kappa_\mathrm{P}$, as in \citet{2017ApJ...836..221M,2019ApJ...881..144M}\footnote{The iron mass fraction given in the caption of Fig.~1 in \citet{2017ApJ...836..221M} should be \num{0.3} and not \num{0.4}.}. For the dust opacities we assume dust grains made of ``normal silicates'' (NRM model in \citealt{Semenov2003}) as homogeneous spheres.

\subsubsection{Photoevaporation}\label{subs:evap}

The discs are subject to both external and internal photoevaporation. We use an adaptation of the model for external FUV photoevaporation by massive stars from \citet{2003ApJ...582..893M}. Our model for internal photoevaporation includes EUV irradiation by the host star. We closely follow \cite{Clarke2001}. The sink terms $S_\mathrm{ext}(r,t)$ and $S_\mathrm{int}(r,t)$ are given in Appendix~\ref{app:evap}.

\subsection{Viscosity}\label{subsect:viscosity}

We use the $\alpha$ prescription of viscosity \citep{1973A&A....24..337S}:
\begin{equation}
    \nu = \alpha \frac{c_\mathrm{s}^2}{\Omega}.
\end{equation}
In the calculation of the $\alpha$-parameter we consider gravitational torques ($\alpha_\mathrm{G}$) as well as torques due to other (``background'', $\alpha_\mathrm{bg}$) processes (see \S~\ref{backgroundalpha}) and calculate:
\begin{equation}
    \alpha = \max (\alpha_\mathrm{G},\alpha_\mathrm{bg}),
\end{equation}
where $\alpha_\mathrm{G} = \alpha_\mathrm{d}$ during infall phase and $\alpha_\mathrm{G} = \alpha_\mathrm{GI}$ afterwards. The contributions are explained in the following.

\subsubsection{Global transport of angular momentum}\label{subsubsect:gobaltrans}

At the beginning of our simulations, the disc is fed by infalling material and its mass is comparable to that of the host star and therefore the centre of mass can be shifted away from the central star.
In this regime the disc can become globally unstable \citep{Harsono2011}. The resulting indirect gravitational potential leads to very efficient transport of mass and angular momentum. \citet{Kratter2010a} developed a parametrisation that can be used to simulate this regime. They performed a numerical parameter study of rapidly accreting, gravitationally unstable discs. They propose a characterisation of such systems by two dimensionless parameters $\xi$ and $\Gamma$. The thermal parameter $\xi$ is defined as
\begin{equation}
    \xi = \frac{\dot{M}_\mathrm{in} G}{c_\mathrm{s}^3}
\end{equation}
and relates the infall rate of material from the MCC, $\dot{M}_\mathrm{in}$, to the disc's sound speed $c_\mathrm{s}$ (evaluated at the infall location). The rotational parameter $\Gamma$,
\begin{equation}
    \Gamma = \frac{\dot{M}_\mathrm{in}}{M_{*\mathrm{d}} \Omega_\mathrm{in}},
\end{equation}
where $M_{*\mathrm{d}}$ is the combined mass of disc and star, and $\Omega_\mathrm{in}$ the angular frequency in the disc at the infall location, compares the system's infall time scale to its orbital time scale.
\citet{Kratter2010a} use this parametrisation to derive an effective Shakura--Sunyaev $\alpha$ to describe the transport of angular momentum in the globally unstable phase:
\begin{equation}
    \alpha_\mathrm{d} = \frac{1}{18(2-k_\Sigma)^2 (1 + l_j)^2} \frac{\xi^{2/3}}{\Gamma^{1/3}}.
\end{equation}
Here, $k_\Sigma$ is the power-law index of the disc's surface density, $l_j$ a parameter related to the infalling material's angular momentum. We follow \citet{Kratter2010a} in setting $k_\Sigma = 3/2$ and $l_j = 1$. We thus set $\alpha_\mathrm{G} = \alpha_\mathrm{d}$ during the infall phase.

\subsubsection{Transport by spiral arms}\label{subsubsect:spiral}
At the end of the infall phase, when the disc mass $M_d$ and disc-to-star mass ratio $q$ are at their maximum and then start to decrease, the above description  is no longer applicable ($\xi = 0$). We therefore follow \citetalias{Kimura2016} in using the approach of \citet{Zhu2010} to describe the effect of spiral arms on the angular momentum transport and set
\begin{equation}
    \alpha_\mathrm{GI} = \exp{\left(- Q_\mathrm{Toomre}^4\right)}
\end{equation}
globally. We use the minimum value of $Q_\mathrm{Toomre}$ (defined in Sect.~\ref{ssub:regimes}) across the entire disc. There are also local prescriptions of this transport mechanism, such as in \citet{Kratter2008}, so we may be overestimating $\alpha_\mathrm{GI}$. However, using a local prescription has an negligible influence on the evolution of the disc mass, and we want to avoid an underestimate of the angular momentum transport given the large extent of spiral arms seen in hydrodynamic simulations. As a result, we set $\alpha_\mathrm{G} = \alpha_\mathrm{GI}$ after the infall phase.

\subsubsection[Background alpha]{Background $\alpha$}\label{backgroundalpha}

Like \citetalias{Kimura2016} we use a high value of $\alpha_\mathrm{bg} = \num{e-2}$ in most of the simulations (see Sect.~\ref{sec:param}). $\alpha_\mathrm{bg}$ is used to describe the torque due to any mechanism other than the global instability of the disc or spiral arms described above. For an overview of possible mechanisms, see \citet{2014prpl.conf..411T}, see also \citet{2020ApJ...891..154D}. The effect of MHD wind-driven accretion (magnetohydrodynamic disc winds; \citealp{2009ApJ...691L..49S,2013ApJ...769...76B,2015ApJ...801...84G}) will be addressed in future work.

\subsection{Inner disc edge}\label{subsect:Inneredge}

The disc is truncated at the inner edge. This inner truncation radius is fixed at $\SI{0.05}{au}$. Mass flowing across this radius is accreted on the central star, with \SI{10}{\percent} considered lost as outflow \citep{Vorobyov2010a}.
This is a crude treatment of the inner region of the disc. However, since we are mostly interested in the regions at of tens of $\si{au}$ in the disc, a detailed description of the innermost region is not necessary. We note, though, that the behaviour of the disc at its inner edge may have an influence on the disc's observed lifetime.

\subsection{Stellar model}\label{subsect:Stellarmodel}

During the simulation the star accretes a considerable amount of mass and its properties change significantly. We compute the stellar radius and its photospheric temperature at each time step, given its age and mass, according to pre-calculated tables from \citet{2008ASPC..387..189Y}.
Furthermore, we consider heating of the disc by accretion of disc material onto the star, where the total stellar luminosity is given by
\begin{equation}
	L_\mathrm{tot} = L_\mathrm{int} + \frac{G \mstar \dot{\mstar}}{2 \rstar},
\end{equation}
where $L_\mathrm{int}$ is the intrinsic luminosity from the stellar photosphere, $\mstar$ the mass, $\rstar$ the radius, and $\dot{\mstar}$ the accretion rate onto the star (\citetalias{Kimura2016};\citealt{2018A&A...614A..98V}). This assumes that one half of the potential energy of the gas relative to the star is dissipated in the accretion disc and only the remaining half at the surface of the star.

\subsection{Disc dispersal condition and disc lifetime}\label{subs:disp}

We follow \citetalias{Kimura2016} in defining the condition for the dispersal of the disc based on the optical depth in the near-infrared (NIR) emitting region. This region is taken to be the location where the mid-plane temperature is above $\SI{300}{K}$. The time of dispersal $t_\mathrm{NIR}$ is then defined as the moment when the vertical optical depth drops below unity in the NIR region.

\citetalias{Kimura2016} then go on to define the disc lifetime as:
\begin{equation}
	t_\mathrm{life} = t_\mathrm{NIR} - t_\mathrm{pms},
\end{equation}
 where $t_\mathrm{pms}$ is the starting time of the pre-main sequence phase given by the condition that $t_\mathrm{grow} \geq 3 t_\mathrm{evap}$, where $t_\mathrm{evap} = M_\mathrm{disc}/\dot{M}_\mathrm{X}$ and $t_\mathrm{grow} = \mstar / \dot{\mstar}$, $\dot{M}_\mathrm{X} = \num{1.62e-8} (\mstar/\msun)^{\num{-1.57}} \msun / \SI{}{yr}$ are the disc's evaporation and growth time scale, respectively. 
 Thus, their lifetimes are reduce with respect to $t_\mathrm{NIR}$. They argue that the evolutionary tracks of pre-main sequence stars are used to determine the age of young clusters observationally, and the age is therefore defined as  the time after the pre-main sequence phase starts, not as the time after the collapse of the MCC. This is indeed how disc lifetimes are typically determined \citep{HaischJr.2001}.
 However, their definition of $t_\mathrm{pms}$ delays the start of the pre-main sequence to very late stages of the disc's evolution, $> \SI{5}{Myr}$ in some cases, which is too long. 
 Therefore we use $t_\mathrm{NIR}$ as the disc's lifetime in Sect.~\ref{sect:Results}. We discuss the influence of this reduction in Sect.~\ref{Sect:comp}.
 
\subsection{Fragmentation}
\label{subsect:Fragmentation}

\subsubsection{Two regimes of fragmentation}
\label{ssub:regimes}

The main condition for fragmentation is given by the Toomre $Q$ parameter \citep{Toomre1964}:
\begin{equation}
\label{qtoomre}
	Q_\mathrm{Toomre} = \frac{c_s \kappa}{\pi G \Sigma},
\end{equation}
where $\kappa$ denotes the epicyclic frequency. Our discs are not Keplerian and therefore $\kappa$ is not equal to $\Omega$. An axisymmetric razor-thin disc has exponentially growing modes when $Q_\mathrm{Toomre}<1$.

The conditions for disc fragmentation have been studied extensively (see \citealt{2016ARA&A..54..271K} for a review).
Our discs evolve in time while being fed by infalling material continuously.
If matter reaches the disc at a rate faster than it can be transported away viscously, the disc will fragment independently of cooling \citep{Boley2009}. The $\xi$ and $\Gamma$ parameters introduced in Sect.~\ref{subsubsect:gobaltrans} can be used to judge if this is the case. The condition is given in \citet{2016ARA&A..54..271K}:
\begin{equation}\label{regime}
	\Gamma < \frac{\xi^{2.5}}{850}.
\end{equation}
This condition is typically satisfied at the beginning of the disc's evolution, when infall rates are high. We refer to this phase as the ``infall-dominated regime''. Since we assume a short infall phase at constant infall rate, the condition from Eq.~\ref{regime} is satisfied during most of the infall phase in almost all systems in our study. We assume that the disc fragments when $Q_\mathrm{Toomre} < 1$ in the infall-dominated regime, (i.e., when Eq. \ref{regime} is satisfied). We note that the exact $Q$ value for fragmentation is uncertain, but is often taken as unity.

If the condition from Eq.~\ref{regime} is no longer satisfied, we transition into the ``cooling-dominated regime''. There, the disc can only fragment if it cools efficiently. This is stated in the Gammie criterion \citep{gammie2001}:
\begin{equation}
\label{gammie}
	\beta \equiv t_\mathrm{cool} \Omega \lesssim \beta_c.
\end{equation}
In the cooling-dominated regime we therefore assume that the disc fragments when $Q_\mathrm{Toomre}<1$ and the condition from Eq.~\ref{gammie} is satisfied \citep{Kimura2012,2007astro.ph..1485A,2017ApJ...848...40B}.
In Eq.~\ref{gammie} we have \citep{2012A&A...547A.112M}:
\begin{equation}
t_\mathrm{cool} = \frac{3 \gamma \Sigma c_\mathrm{s}^2}{32 (\gamma - 1) \sigma T^4}\tau_\mathrm{eff}
\end{equation}
with an effective optical depth $\tau_\mathrm{eff} = \kappa_\mathrm{R} \Sigma /2 + 2/(\kappa_\mathrm{R} \Sigma) $, and $\beta_c \approx 3$ the critical cooling parameter \citep{2017ApJ...847...43D,2017ApJ...848...40B}.
The adiabatic index $\gamma$ is fixed at $\num{1.45}$. In reality it depends on temperature and on the ortho-to-para ratio of molecular hydrogen. The latter can vary substantially \citep{2013ApJ...778...77D} and its precise value in protoplanetary discs is currently unknown \citep{2007ApJ...656L..89B}.
This may have an important effect on fragmentation and planet migration and should be studied in more detail in the future.

\subsubsection{Initial fragment mass}
\label{ssub:mfrag}

When the conditions for fragmentation are satisfied, a region of the disc can collapse under its own gravity and form one or several bound clumps with a given initial mass. The number of clumps formed corresponds to the dominant azimuthal wave number $m$. \citet{2015ApJ...812L..32D} find $m \sim \mstar / M_\mathrm{disc}$. We therefore remove $m$ times the initial fragment mass on a free-fall time scale $\tau_\mathrm{ff} = \sqrt{3 \pi /(32 G \rho_\mathrm{c})}^{-1}$, where we substitute $\rho_\mathrm{c}$ with the density at the midplane.

There is currently no agreement about the value of the initial fragment mass. A rough estimate is the Toomre mass \citep{Nelson2006}
\begin{equation}
	M_T = \frac{\pi c_\mathrm{s}^4}{G^2 \Sigma}.
\end{equation}
\citet{Forgan2011} estimate the local Jeans mass inside the spiral structure of a self-gravitating disc (using our convention for $H$, $H=c_\mathrm{s}/\Omega$ in absence of autogravitation) by\footnote{Note that \citet{Forgan2011} are missing a square root over $\left(1~+~\frac{\Delta \Sigma}{\Sigma}\right)$ starting from their Eq.~(9).}
\begin{equation}
    M_\mathrm{J,FR} = \frac{4}{3} \frac{2^{1/4} \pi^{11/4} c_s^3 H^{1/2}}{G^{3/2} \Sigma^{1/2} \sqrt{1 + \beta^{-1}}}.
\end{equation}
\citet{Boley2009} calculated a different initial fragment mass of
\begin{equation}
    M_\mathrm{F} = \frac{1.6 \, c_s^3}{G \Omega}.
\end{equation}
In a Keplerian disc with $Q_\mathrm{Toomre} = 1$ it holds that
\begin{equation*}
    M_\mathrm{F} \approx 0.16 M_\mathrm{T} \approx 0.04 M_\mathrm{J,FR}.
\end{equation*}
The result from \citet{Boley2009} is confirmed by their SPH simulations as well as in \citet{Tamburello2015} (albeit in a different context) for a large parameter space. Therefore we use $M_\mathrm{F}$ as the default value, but also investigate the other extreme $M_\mathrm{J,FR}$.

\subsection{Infall from the molecular cloud core}\label{subs:Infall}

The simulations of \citetalias{2018MNRAS.475.5618B} infer  disc-to-star mass ratios of order unity shortly after the disc formation. Since these conditions are favourable for fragmentation, our simulations start as early as possible.

\subsubsection{Disc formation}\label{subsubs:discform}

The results from \citetalias{2018MNRAS.475.5618B} show one feature in particular: the process of disc formation is chaotic. Accretion from the environment may be interrupted and restart later. This may lead to inner and outer discs that are misaligned. 
Interaction between stars may lead to multiple systems or let multiple systems become unbound. Discs may be truncated or stripped by such processes. Due to the nature of our model, which includes a rotationally symmetric disc around a single star, we cannot reproduce this diversity. However, while \citetalias{2018MNRAS.475.5618B}'s simulation is a population synthesis of \emph{protostellar} discs, our focus lies on \emph{protoplanetary} discs. 
After all, for planet formation to proceed, favourable conditions will need to be in place for some time. 
We therefore concentrate on discs that survive the chaotic early phase described in \citetalias{2018MNRAS.475.5618B} and evolve into `well-behaved' discs that can be reasonably well described by rotationally symmetric structures (although they could develop spiral arms).  
Of course the distinction between protostellar and protoplanetary discs is by no means sharp and we cannot exclude the possibility of planet formation in discs that do not meet our criteria of `well-behaved'. With the choice of our sample and the 1D framework we make some strong but necessary assumptions to study planet formation statistically.

We initialise our simulations $\SI{1}{kyr}$ after the formation of the protostar. 
During the first $\sim \SI{e3}{yr}$ to $\sim \SI{e5}{yr}$  the system's mass increases, typically by more than an order of magnitude, through infall from the MCC. 
We assume that discs are formed in a short period of high and constant infall rates. This is in good qualitative agreement with \citetalias{2018MNRAS.475.5618B} (see their Fig.~11) in the sense that most systems are dominated by very high infall rates at early times, followed by a rather sudden drop. 
The procedure for setting up the initial conditions is explained in Sect.~\ref{subsect:Initial}.

\subsubsection{Infall location}
\label{ssub:infloc}

An important, yet difficult question is the location where the infalling material is added to the disc. 
Simple analytic models for the collapse of MCCs (for example \citealt{shu1977b}) calculate the trajectories for the infalling material and therefore also their intersection with the disc's midplane.
But this treatment is too simple.
First, the material does not land in the midplane but interacts with the disc higher up, which requires  a treatment of the disc's boundary. Second, the infalling material  carries specific angular momentum different from the the disc's. Thus, a description of how the two components mix is required.

A treatment of these processes (albeit based on an obsolescent infall model) can be found in \citet{Visser2010}.
Since in our model angular momentum and trajectories of the infalling material are unknown, we resort to a simpler approach. 
At early times in the disc's evolution, the infall rates are very high. Therefore the disc's mean specific angular momentum is comparable to that of the material incoming from the MCC. 
Thus we simply add the infalling material near the location characterised by a specific angular momentum equal to the mean specific angular momentum of the entire disc. Assuming a Keplerian disc with a surface density profile of the form $\Sigma = \Sigma_0 \left(\frac{r}{r_0}\right)^{-1}$ (consistent with \citetalias{2018MNRAS.475.5618B}), this radius $R_\mathrm{i}$ is well-defined and it turns out (see Appendix \ref{app:infloc} for a derivation):
\begin{equation}
    R_\mathrm{i} \approx 0.7~R_{63.2},
\end{equation}
where $R_{63.2}$ is the disc radius we take from \citetalias{2018MNRAS.475.5618B} (defined as the radius containing $\num{63.2} \%$ of the disc mass).
In practice we add the material from the MCC to the disc using an infall source term of Gaussian shape centred at $R_\mathrm{i}$, with a standard deviation $\sigma_\mathrm{i} = R_\mathrm{i}/3$:
\begin{equation}
    S_\mathrm{inf} (r,t) = \frac{\dot{M}_\mathrm{in}}{2 \sqrt{2} \pi^{3/2} R_\mathrm{i} \sigma_\mathrm{i}} \exp{\left[ - \left( \frac{r - R_\mathrm{i}(t)}{\sqrt{2} \sigma_\mathrm{i}} \right) \right]}.
\end{equation}
This is a compromise between having a source term that is too concentrated in radius and one that is very wide. We note that, aside extremes, the choice of this width does not influence the disc's fragmentation and evolution significantly. $R_\mathrm{i}$ is assumed to grow at a constant rate $b_\mathrm{disc}$: $R_\mathrm{i} \equiv R_\mathrm{i}(t) = R_\mathrm{i,k} + (t - \SI{1}{kyr}) b_\mathrm{disc}$, where $R_\mathrm{i,k}$ is the initial value of $R_\mathrm{i}$ at $t = \SI{1}{kyr}$. $R_\mathrm{i,k}$ and $b_\mathrm{disc}$ are obtained from the results of \citetalias{2018MNRAS.475.5618B}. For systems where we try to mimic the effect of magnetic fields on the infall, we keep $R_\mathrm{i}$ constant. How we model this MHD-collapse, and how the initial conditions are determined in detail, is explained in Sect. \ref{sec:param}.

\section{Investigated parameter space and initial conditions}\label{sec:param}

We perform five runs to investigate the influence of different physical parameters on the inferred discs. Each run consists of $\num{10000}$ systems.
The runs are summarised in Table~\ref{tov}.

\begin{table}
\centering
\begin{tabular}{ccccccc}  
\hline\hline
\multicolumn{2}{c}{Run}                & \begin{tabular}[c]{@{}c@{}}Stellar\\accretion\\ heating\end{tabular} &
\begin{tabular}[c]{@{}c@{}}Infall\\heating\end{tabular} &
\begin{tabular}[c]{@{}c@{}}Initial\\fragment\\mass\end{tabular} &
\begin{tabular}[c]{@{}c@{}}Infall\\model\end{tabular} &
\begin{tabular}[c]{@{}c@{}}$\alpha_\mathrm{bg}$\end{tabular}\\
\hline
1 & ``hydro''   & \Haek & \Haek      & $M_\mathrm{f}$     & ``hydro'' & \num{e-2} \\
2 & ``MHD''     & \Haek & \Haek      & $M_\mathrm{f}$     & ``MHD''   & \num{e-2} \\
3 & lowalpha  & \Haek & \Haek      & $M_\mathrm{f}$     & ``hydro'' & \num{e-3} \\
4 & noheat    & --- & ---      & $M_\mathrm{f}$     & ``hydro'' & \num{e-2} \\
5 & MJ        & --- & ---      & $M_\mathrm{J,FR}$  & ``hydro'' & \num{e-2}\\
\hline
\end{tabular}
\caption[]{Overview of the runs. $M_\mathrm{f}$ and $M_\mathrm{J,FR}$ denote the different initial fragment masses (see Sect.~\ref{ssub:mfrag}).}
\label{tov}
\end{table}

RUN\nobreakdash-1 is the default run in this study. 
It is based on the initial conditions obtained from hydrodynamical simulations where we include heating from infalling material of the MCC on the disc as well as accretion heating from disc material that is accreted onto the star. 
These heating mechanisms are also active in RUN\nobreakdash-2 and RUN\nobreakdash-3, but not in RUN\nobreakdash-4 and RUN\nobreakdash-5. RUN\nobreakdash-2 is a modification of the first run where we aim to mimic the effect of magnetohydrodynamics (MHD).
\citet{Hennebelle2016} calculate the early disc radius $r_\mathrm{H16}$ resulting from a magnetised collapse:
\begin{equation}\label{eq:rhen}
    r_\mathrm{H16} = \SI{18}{au} ~ \left( \frac{A}{\SI{0.1}{s}} \right)^{2/9} \left( \frac{B_z}{\SI{0.1}{G}} \right)^{-4/9} \left( \frac{M_{*\mathrm{d}}}{\SI{0.1}{\msun}} \right)^{1/3}.
\end{equation}
Here, $A$ is a measure for the ambipolar diffusivity\footnote{\citet{Hennebelle2016} use a non-standard notation in which their ambipolar diffusion coefficient ``$\eta_\mathrm{AD}$'' in their Eq.~(13) has units of time rather than length squared per time. We use instead $A$.} and  $B_z$ the magnetic field in the inner part of the core.
We use this expression, setting $A = \SI{0.1}{s}$ and $B_z = \SI{0.1}{G}$ to determine $R_\mathrm{i}$ for RUN\nobreakdash-2.
RUN\nobreakdash-2 therefore differs from RUN\nobreakdash-1 only in the infall location, which we choose very close to the star (\SIrange{1.7}{9}{\astronomicalunit}) and constant in time throughout the simulation. The locations are simply chosen such that the disc radius agrees with the analytic result from Eq.~\ref{eq:rhen} at the end of the infall phase. In practice finding the ``correct'' radii is an iterative process, but it works well (see the middle left panel in Fig.~\ref{fig:disc_2} in the Results section).
The precise infall locations we use are given in Appendix~\ref{app:infallradii}. RUN\nobreakdash-3 is equal to RUN\nobreakdash-1 with the exception of a lower assumed $\alpha_\mathrm{bg}$. In RUN\nobreakdash-4 we turned off both infall and accretion heating. RUN\nobreakdash-5 is similar to RUN\nobreakdash-4 except for the choice of the initial fragment mass, where we use the larger $M_\mathrm{J,FR}$ (see Sect.
~\ref{ssub:mfrag}).

Our model can cover a large range in stellar mass.
We use a range of final (at the end of the simulation) stellar masses from \SIrange{0.05}{5}{\msun} and divide this interval into $100$ logarithmically spaced bins. We compute the evolution of $100$ systems in each of these bins to give a total of $\num{10000}$ systems per run. The initial conditions are chosen such that the resulting population of stars has a mass distribution in agreement with the stellar initial mass function (IMF, see following sct.). 
The final stellar mass of a system with a given set of initial conditions is a priori unknown. We must therefore make an estimate, as we discuss in the following section.  

\subsection{Initial setup}\label{subsect:Initial}

We initiate each simulation at $\SI{1}{kyr}$ by setting up a disc with an initial mass $M_\mathrm{disc,i}$ and an initial radius $R_\mathrm{disc,i}$ along with a protostar of mass $M_{*,\mathrm{i}}$. We choose a power-law surface density profile with exponent $\num{-1}$, as expected by \citetalias{2018MNRAS.475.5618B}. The precise shape of the initial profile has little importance, since it is changed by  infall and viscous evolution immediately.
In order to choose suitable initial values, we select 35~systems from the single star sample in \citetalias{2018MNRAS.475.5618B} and construct probability distributions for $M_\mathrm{disc,i}, M_{*,\mathrm{i}}$ and $R_\mathrm{disc,i}$. The (constant) infall rate is set as a probability distribution around $\dot{M}_\mathrm{in} = (M_{*,\mathrm{10k}} + M_\mathrm{disc,10k} - M_{*,\mathrm{i}}-M_\mathrm{disc,i}) / (\SI{9}{kyr})$, where $M_{*,\mathrm{10k}}$ and $M_\mathrm{disc,10k}$ are the stellar mass and disc mass after $\SI{10}{kyr}$, respectively. The time of formation of a protostar therefore agrees with that in \citetalias{2018MNRAS.475.5618B} by construction.
A similar procedure is used so set up the initial disc radius. We choose a sub-sample of 20 among the 35 previously chosen discs (some do not have well-defined radii during the first $\SI{10}{kyr}$) to generate a probability distribution for $R_\mathrm{i,k}$. We also use these discs to gauge the rate $b_\mathrm{disc}$ at which the infall radius expands in time (see \S~\ref{ssub:infloc}). We assume that protostellar mass, disc mass and infall rate are correlated, and that disc radius and its expansion rated are correlated. Mass and radius are assumed to be uncorrelated. The specific discs we selected, along with histograms showing the distributions of the initial parameters, can be found in Appendix~\ref{app:initdist}.

\begin{figure}
  \includegraphics[width=\linewidth]{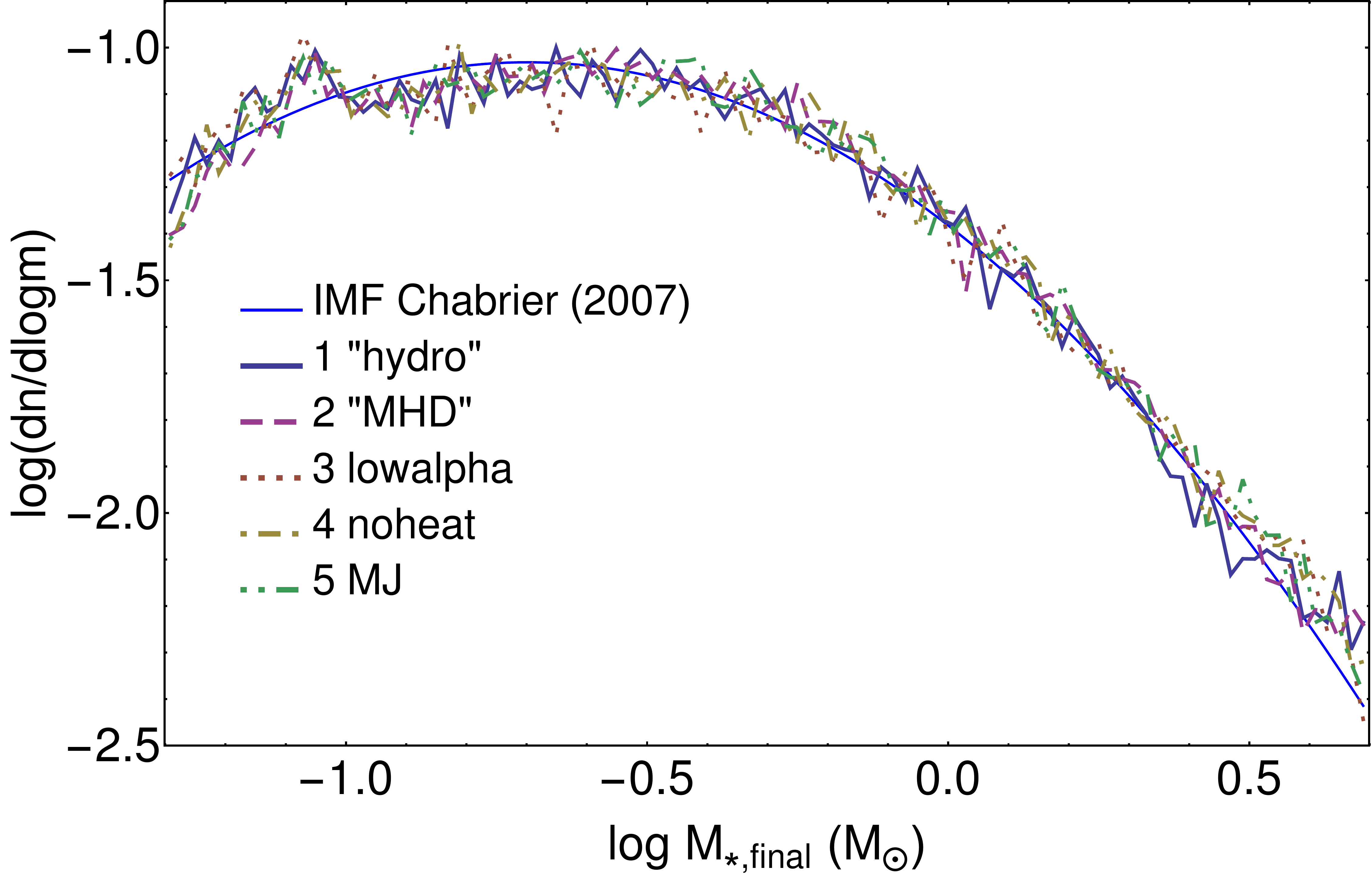}
  \caption{Stellar masses at the end of the simulation for all runs (see Table~\ref{tov}), compared to the \citet{Chabrier2007} IMF.}  
  \label{fig:mstar}
\end{figure}

At this stage, there remains one important unknown: the infall duration $t_\mathrm{infall}$. We cannot obtain this quantity from \citetalias{2018MNRAS.475.5618B}, since their simulations do not run long enough for all systems to reach the end of the main accretion phase. Therefore, we choose a different approach\footnote{In principle, it would be desirable to obtain this quantity from observations. However, while some observational estimates for the duration of the main infall phase exist (see for example the introduction of \citealt{Vorobyov2010a}), they are highly uncertain.}. At the end of the simulations, when the discs have dispersed, we expect the resulting distribution of stellar masses to obey the initial mass function. We use the IMF from \citet{Chabrier2007} for individual stars:
\begin{equation}
    \frac{\mathrm{d}n}{\mathrm{d} \log m} =
    \begin{cases}
    0.093 \times \exp \left\{ - \frac{1}{2}\frac{(\log \tilde{m} - \log 0.2)^2}{0.55^2} \right\}, & m \le \SI{1}{\msun}\\
    0.041 \tilde{m}^{-1.35}, & m \ge \SI{1}{\msun},
    \end{cases}
\end{equation}
where $\mathrm{d} n / \mathrm{d} \log m$ denotes the stellar number density in $\si{pc^{-3}}$ per logarithmic interval of mass and $\tilde{m}\equiv m/(1~\msun)$.
We choose the distribution for $t_\mathrm{infall}$ in such a way that in each run the resulting distribution of final stellar masses in our simulation agrees with this observed distribution. Figure~\ref{fig:mstar} shows this works quite well for all our five runs. The distributions for $t_\mathrm{infall}$ can be found in Appendix~\ref{app:initdist}.

\section{Formation and evolution of an example system}\label{sec:example}

\begin{figure*}[pt]
  \begin{subfigure}[pt]{0.49\textwidth}
  \includegraphics[width=\linewidth]{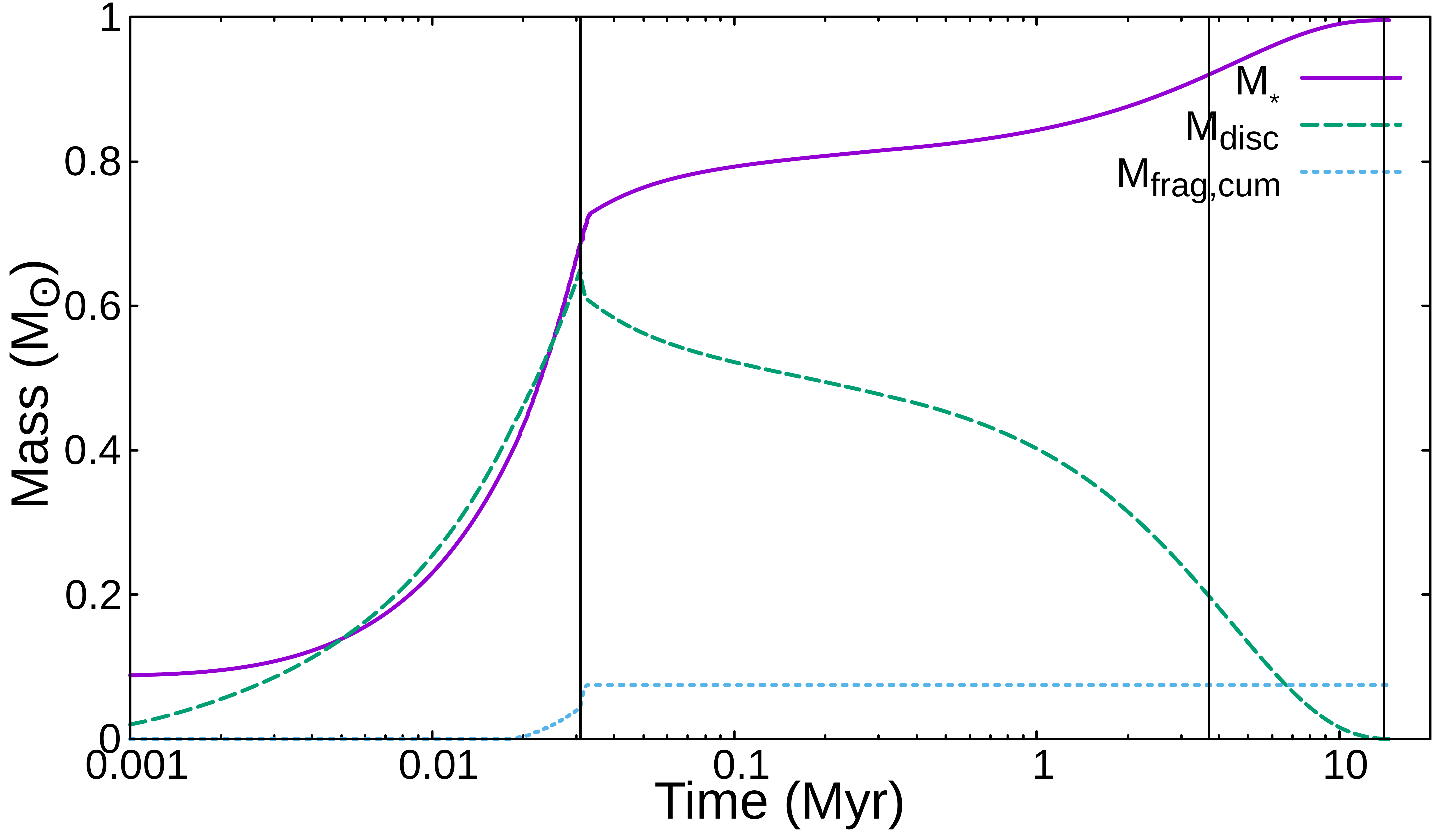}
  \end{subfigure}
  \begin{subfigure}[pt]{0.49\textwidth}
  \includegraphics[width=\linewidth]{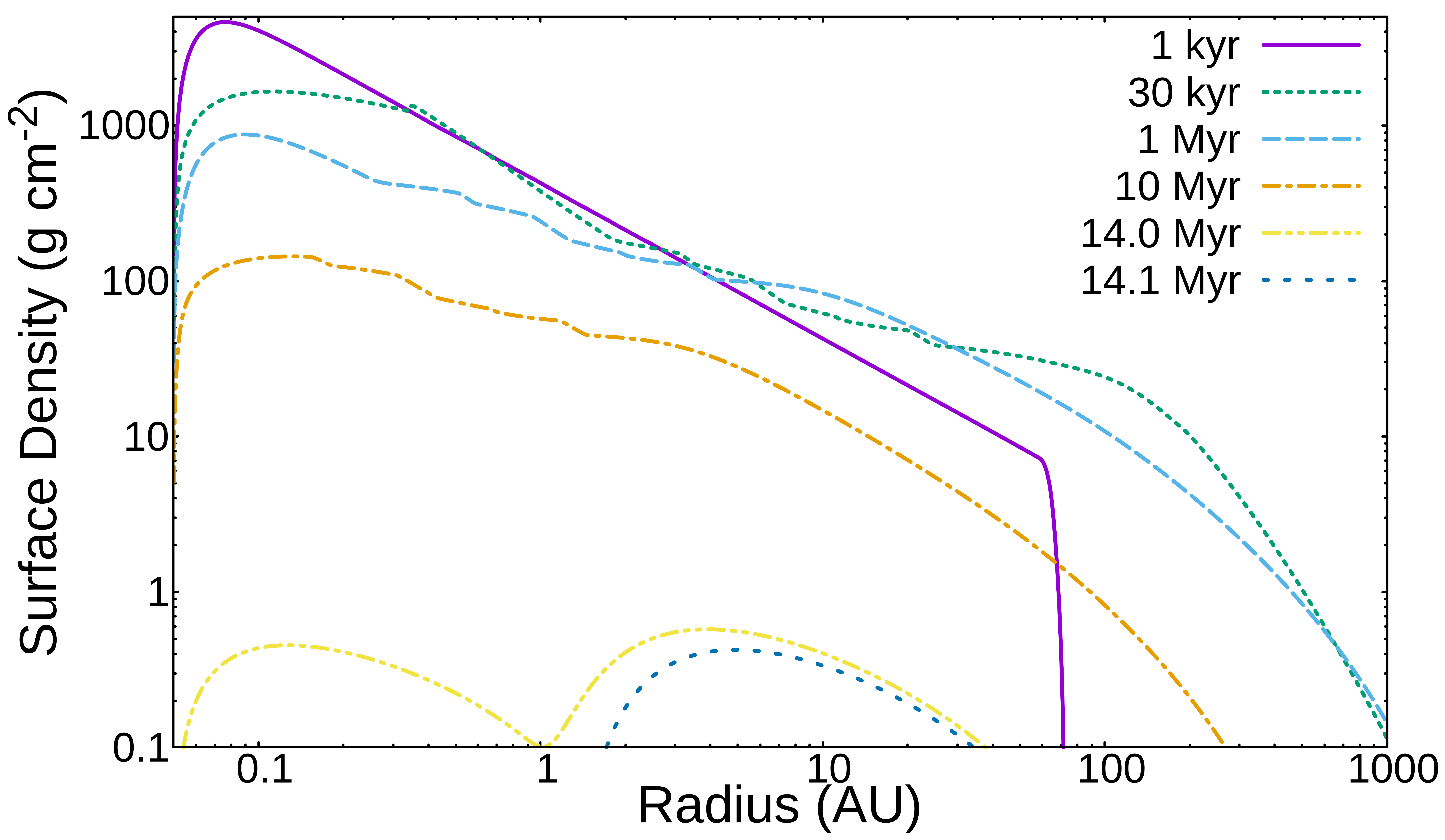}
  \end{subfigure}
  \begin{subfigure}[pt]{0.49\textwidth}
  \includegraphics[width=\linewidth]{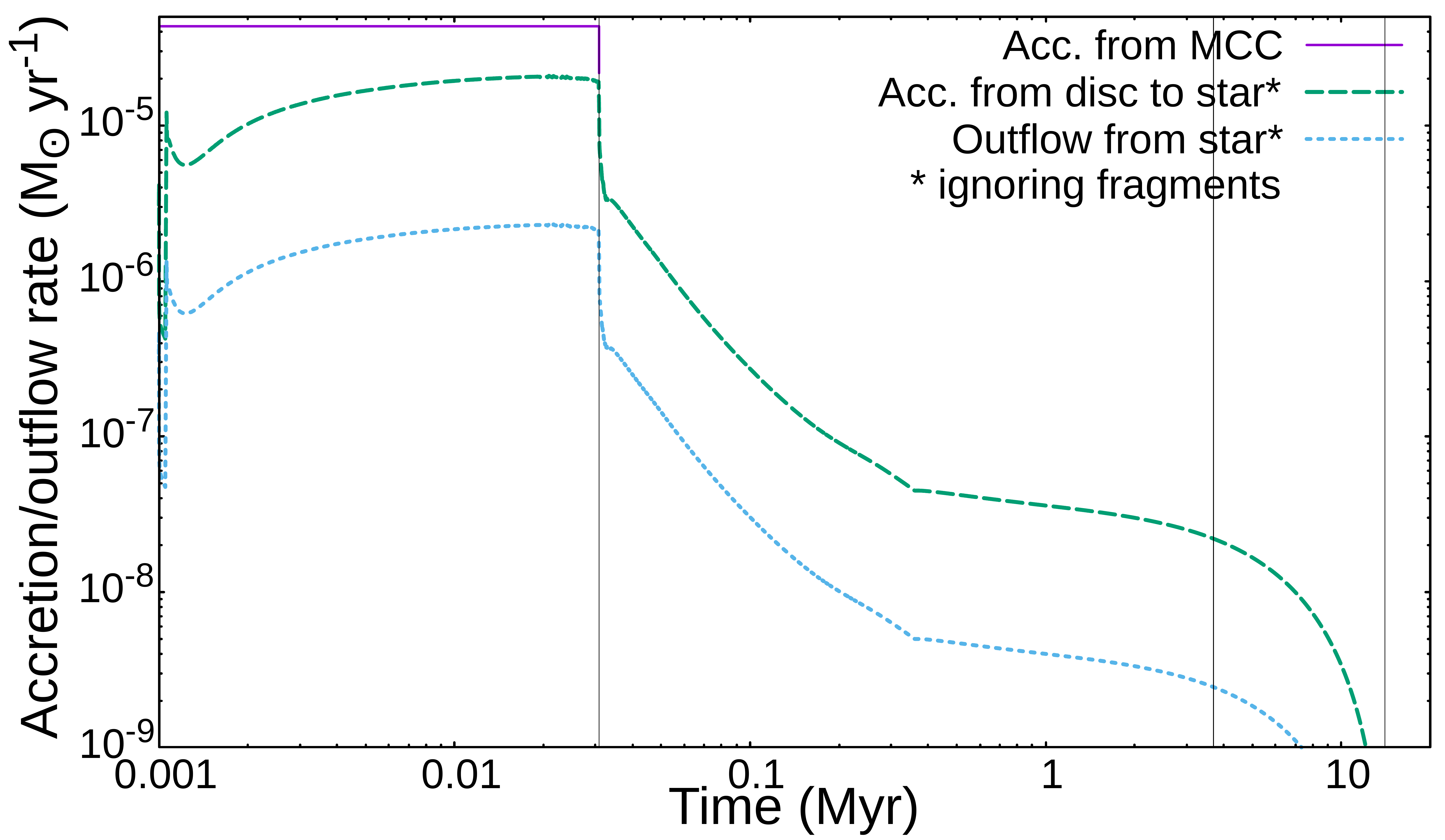}
  \end{subfigure}
  \begin{subfigure}[pt]{0.49\textwidth}
  \includegraphics[width=\linewidth]{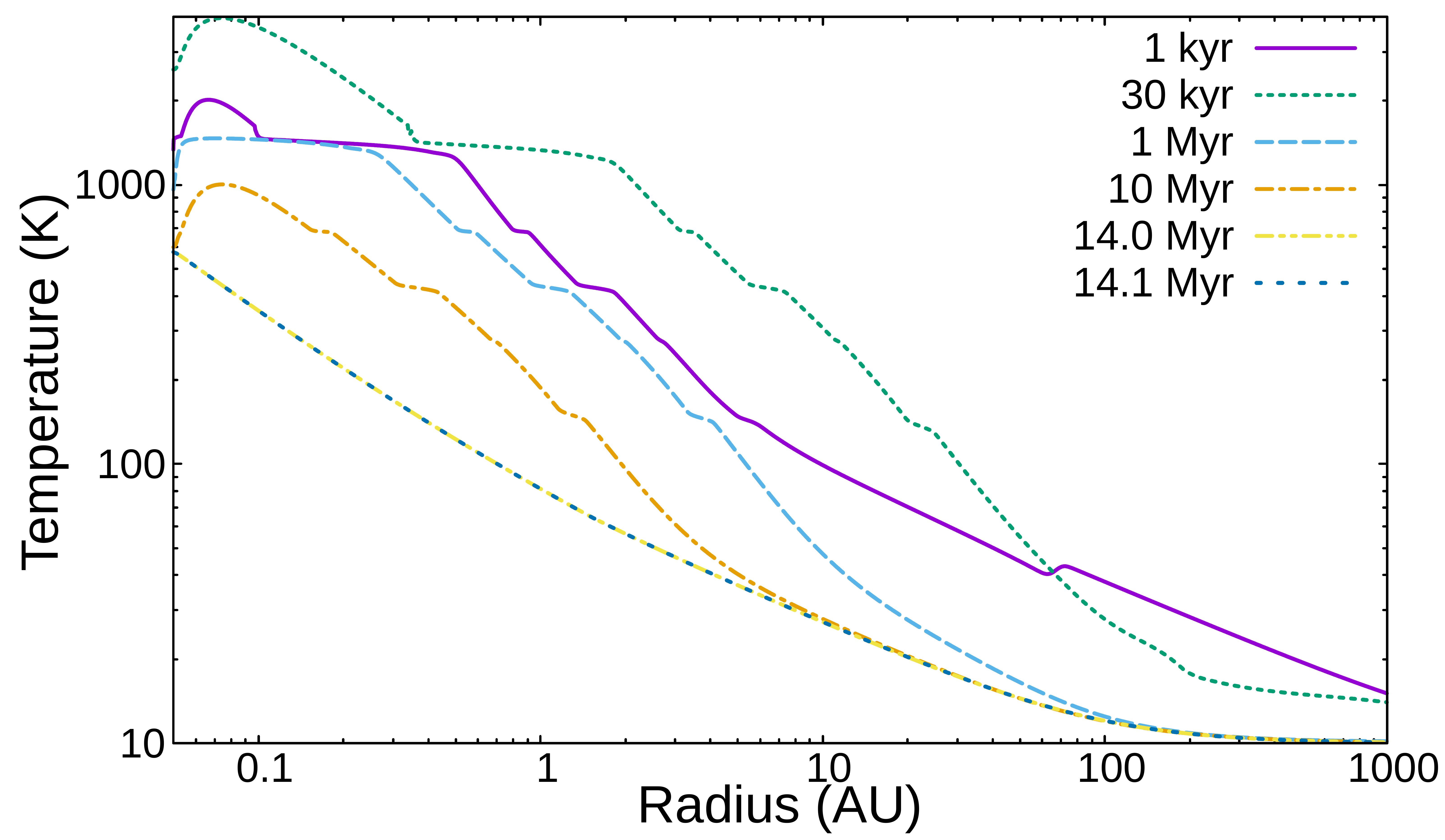}
  \end{subfigure}
  \begin{subfigure}[pt]{0.49\textwidth}
  \includegraphics[width=\linewidth]{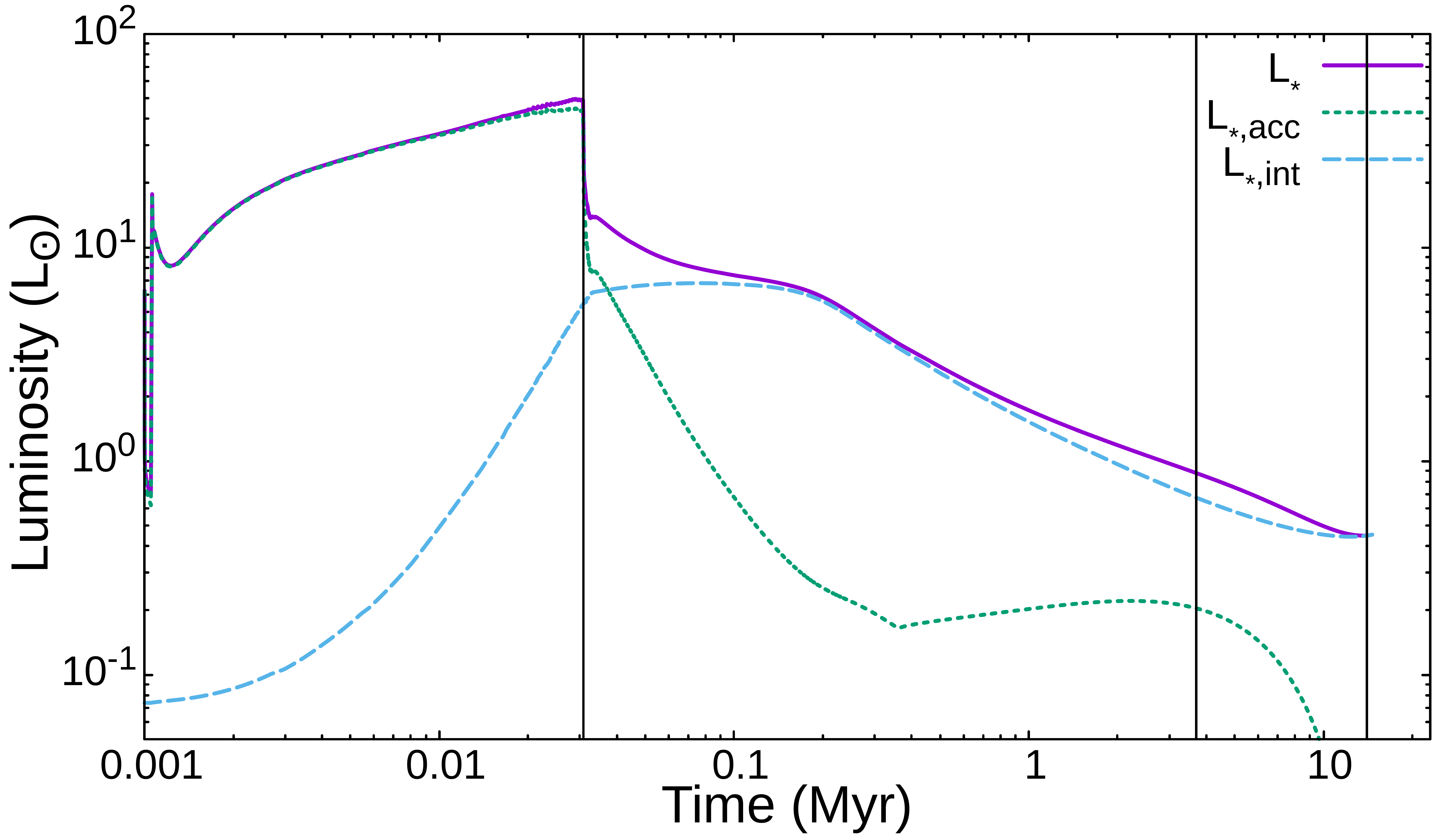}
  \end{subfigure}
  \begin{subfigure}[pt]{0.49\textwidth}
  \includegraphics[width=\linewidth]{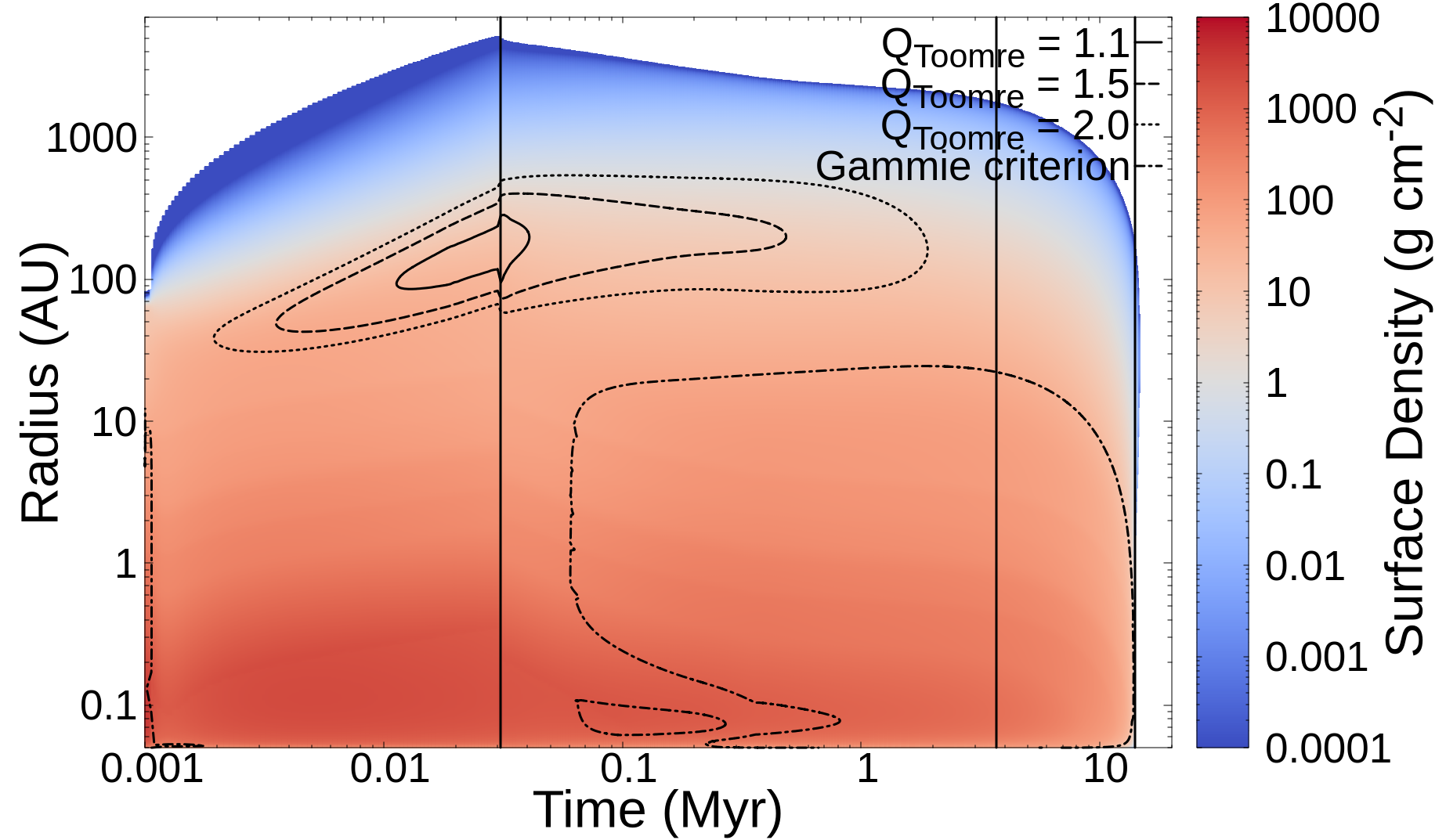}
  \end{subfigure}
  \caption{Time evolution of System~6410 from RUN\nobreakdash-1. Top left panel: stellar mass, disc mass and cumulative mass removed from the disc by fragmentation. Middle left panel: accretion and outflow rates. Bottom left panel: stellar luminosity. Top right panel: surface density at different times. Middle right panel: midplane temperature at different times. Bottom left panel: stellar luminosity (accretion and intrinsic). Bottom right panel: contour plot of the surface density with fragmentation criteria (see text). The black vertical lines denote, in order of increasing time:
  $t_\mathrm{infall}\approx30$~kyr, $t_\mathrm{pms}\approx 3.7$~Myr and $t_\mathrm{NIR}\approx 14$~Myr (see Sect.~\ref{sec:example}).}
  \label{fig:example1}
\end{figure*}

In this section, we demonstrate how a star-disc system forms and evolves by describing the time-evolution of a specific system (System~6410 from RUN\nobreakdash-1, see Table~\ref{tov}). This is a typical system forming an $\sim \SI{1}{\msun}$ star.
The system is initialised with a protostar of mass $\SI{0.09}{\msun}$ and a disc of mass $\SI{0.02}{\msun}$ with a radius of $\SI{41}{au}$.
The top left panel of Fig.~\ref{fig:example1} shows the evolution of the stellar mass, disc mass, and cumulative mass removed from the disc by fragmentation. The middle left panel depicts as a function of time the infall rate from the MCC, the accretion of disc material on the star, as well as the fraction of this material considered to flow out of the system. The bottom left panel shows the stellar luminosity and its two contributors, the stars intrinsic luminosity and the luminosity due to accretion of material onto the star.
The right top and right middle panels show the surface density and temperature, respectively, as a function of radius at selected times. The contour plot in the right bottom panel color-codes the radial distribution of the surface density as a function of time. The fragmentation criteria are also shown. Fragmentation is forbidden because of the Gammie cooling criterion (Eq.~\ref{gammie}) in regions enclosed by the dash-dotted lines (inside of about $\SI{20}{au}$ for $t > \SI{0.1}{Myr}$). The Gammie criterion thus never limits fragmentation in this simulation. It should be noted that the condition for the infall-dominated regime of fragmentation (Eq.~\ref{regime}) is always satisfied during the infall phase and is therefore not shown in the figure.

The evolution of the system can be divided into four phases.
The first is the infall phase. In this example system, it lasts $\sim \SI{30}{kyr}$. In this phase, the system is dominated by high infall rates from the MCC and fast transport of angular momentum due to the global instability of the disc (Sect.~\ref{subsubsect:gobaltrans}). The ratio of disc mass to stellar mass, $q$, is high ($q\sim \num{1}$). The disc is globally unstable and fragments 16 times in the outer region. Both star and disc grow in mass by a factor of a few.
The second phase is short ($\approx \SI{2}{kyr}$) and barely seen in Fig.~\ref{fig:example1} as a sudden drop in disc mass, accretion rate and luminosity. Temperatures in the disc decrease quickly due to the reduced heating after infall has ceased. Consequently, the disc fragments another 15 times. $q$ drops by $\sim \SI{10}{\%}$ due to fast transport of angular momentum. A total of $\SI{0.07}{\msun}$ of matter is removed from the disc by fragmentation.
During the infall phase, on the order of one fragment per $\SI{}{kyr}$ is formed, and afterwards it is roughly ten times more. Both values are in the range of the number of fragments per time expected from hydrodynamic simulations of sufficiently high resolution (see \citet{Boley2009} and \citet{2017MNRAS.464.3158S} for an example with and without infall, respectively). We discuss the comparison to other studies further in Sect.~\ref{subs:compfrag}.
Fragments are added to the star in one time step ($\sim \SI{1}{yr}$ during this phase). This is much longer than an orbital period at the truncation radius. Since we do not model the accretion process of the clumps onto the star, the accretion rates and luminosities in Fig.~\ref{fig:example1} are smoothed between $\sim \num{20} - \SI{40}{kyr}$.
The accretion of a fragment leads to a strong increase in stellar luminosity for a short amount of time. This may be observed as episodic accretion \citep{2014prpl.conf..387A} and may affect the thermodynamics of the disc for a short time \citep{2016Natur.535..258C}. Our model cannot predict the duration or luminosity of these outbursts. However the interval between outbursts is typically a few \SIrange{e2}{e3}{yr}.
The third phase lasts approximately $\SI{100}{kyr}$ and is characterised by a fast redistribution of angular momentum by spiral arms. The disc-to-star mass ratio is reduced to $\sim \num{0.6}$.
In the last phase the viscous evolution of the disc progresses much more slowly in the absence of gravitational instabilities until the disc disperses after $\SI{14}{Myr}$.
Figures showing the time-evolution of two more example systems, a low-mass system from RUN\nobreakdash-2 and a high-mass system from RUN\nobreakdash-5, can be found in Appendix~\ref{app:example}. Animations of all three systems are provided at \url{http://www.space.unibe.ch/research/research_groups/planets_in_time/numerical_data}.

\section{Results}\label{sect:Results}

We now consider all runs set up in Table~\ref{tov} and study the statistics of their outcomes.
%
Table~\ref{globeoi} gives a summary of the disc properties at the end of the infall phase and the lifetimes of all runs. Mean and standard deviation of the respective quantities are shown. We discuss these together with the figures that follow.
The systems' properties at the end of infall are of particular interest, since they represent the initial conditions for planet formation models, where the collapse of the molecular cloud core is not modelled.

The results are organised as follows:
In Section~\ref{subs:12},
we analyse the disc properties at the end of the infall phase as well as the disc lifetimes for RUN\nobreakdash-1 and RUN\nobreakdash-2. The same quantities are discussed in Section~\ref{subs:345} for RUN\nobreakdash-3, RUN\nobreakdash-4 and RUN\nobreakdash-5. The following subsection (Section~\ref{ssec:dep}) presents the dependency of these quantities on the final stellar mass. We then discuss the potential fragmentation for all runs (Section~\ref{subs:frag}).

Figures \ref{fig:disc_a} and \ref{fig:disc_b} present the end-of-infall distributions of disc mass, disc radius, and disc-to-star mass ratio $q_\mathrm{infall} = M_\mathrm{disc,infall} / M_\mathrm{*,infall}$, as well as the disc lifetimes for the entire runs. 
Thus, they are a combination of the simulations performed in $100$ mass bins (see Sect.~\ref{subsect:Initial}). 
The different relative contributions of the stellar mass bins are considered and the slight deviations from the IMF (Fig.~\ref{fig:mstar}) are corrected.
The results should therefore be representative of (unbiased) observations of actual star forming regions, to the degree that the assumed IMF is representative of the stellar population in the region.

\begin{table}
\centering
\begin{tabular}{cccccc}
\hline\hline
\multicolumn{2}{c}{name} &
\begin{tabular}[c]{@{}l@{}} $M_\mathrm{disc,infall}$\\$(\msun)$ \end{tabular} &
\begin{tabular}[c]{@{}l@{}} $R_\mathrm{disc,infall}$\\$(\mathrm{au})$ \end{tabular} &
\begin{tabular}[c]{@{}l@{}} $q_\mathrm{infall}$\\~ \end{tabular} &
\begin{tabular}[c]{@{}l@{}} $t_\mathrm{NIR}$\\$(\mathrm{Myr})$ \end{tabular} \\
\hline
1 & ``hydro''    & $\num{0.29(6)}$ & $\num{200(100)}$ &  $\num{0.98(30)}$ & $\num{7.3(5)}$   \\
2 & ``MHD''      & $\num{0.11(1)}$ & $~~\num{36(7)}$    &  $\num{0.35(10)}$ & $\num{4.5(1)}$    \\
3 & lowalpha & $\num{0.32(7)}$ & $\num{220(100)}$ &  $\num{0.98(29)}$ & $\num{33.0(20)}$ \\
4 & noheat   & $\num{0.25(8)}$ & $\num{190(100)}$ &  $\num{0.80(31)}$ & $\num{7.0(6)}$   \\
5 & MJ       & $\num{0.22(8)}$ & $\num{180(100)}$ &  $\num{0.72(31)}$ & $\num{6.9(7)}$ \\
\hline
\end{tabular}
\caption[]{Global disc properties ($q_\mathrm{infall} =M_\mathrm{disc,infall} / M_\mathrm{*,infall})$}
\label{globeoi}
\end{table}

\subsection{\emph{RUN-1} and \emph{RUN-2}: ``Hydro versus MHD''}\label{subs:12}

\begin{figure*}[pt]
  \begin{subfigure}[pt]{0.49\textwidth}
  \includegraphics[width=\linewidth]{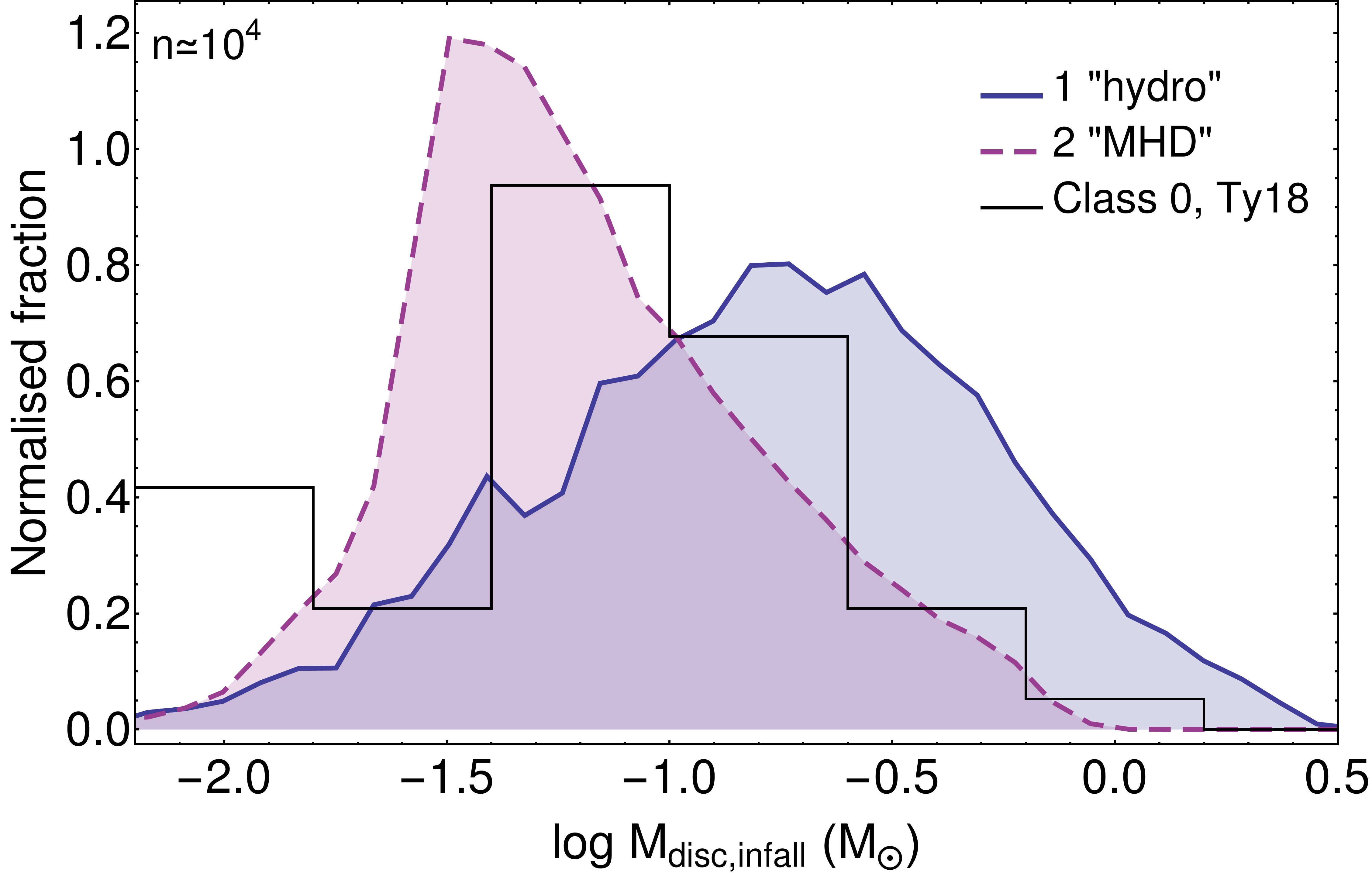}
  \end{subfigure}
  \begin{subfigure}[pt]{0.49\textwidth}
  \includegraphics[width=\linewidth]{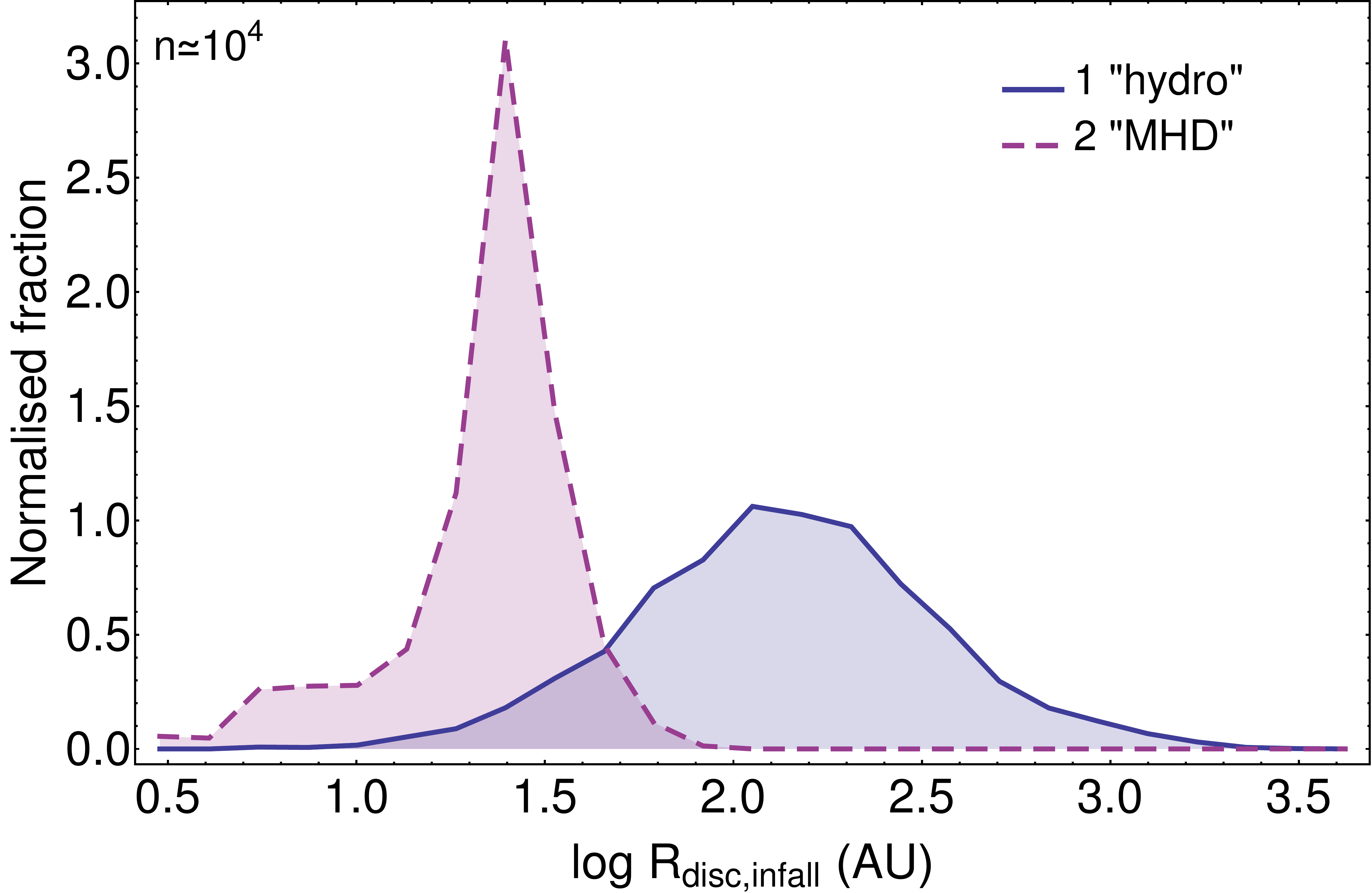}
  \end{subfigure}
  \begin{subfigure}[pt]{0.49\textwidth}
  \includegraphics[width=\linewidth]{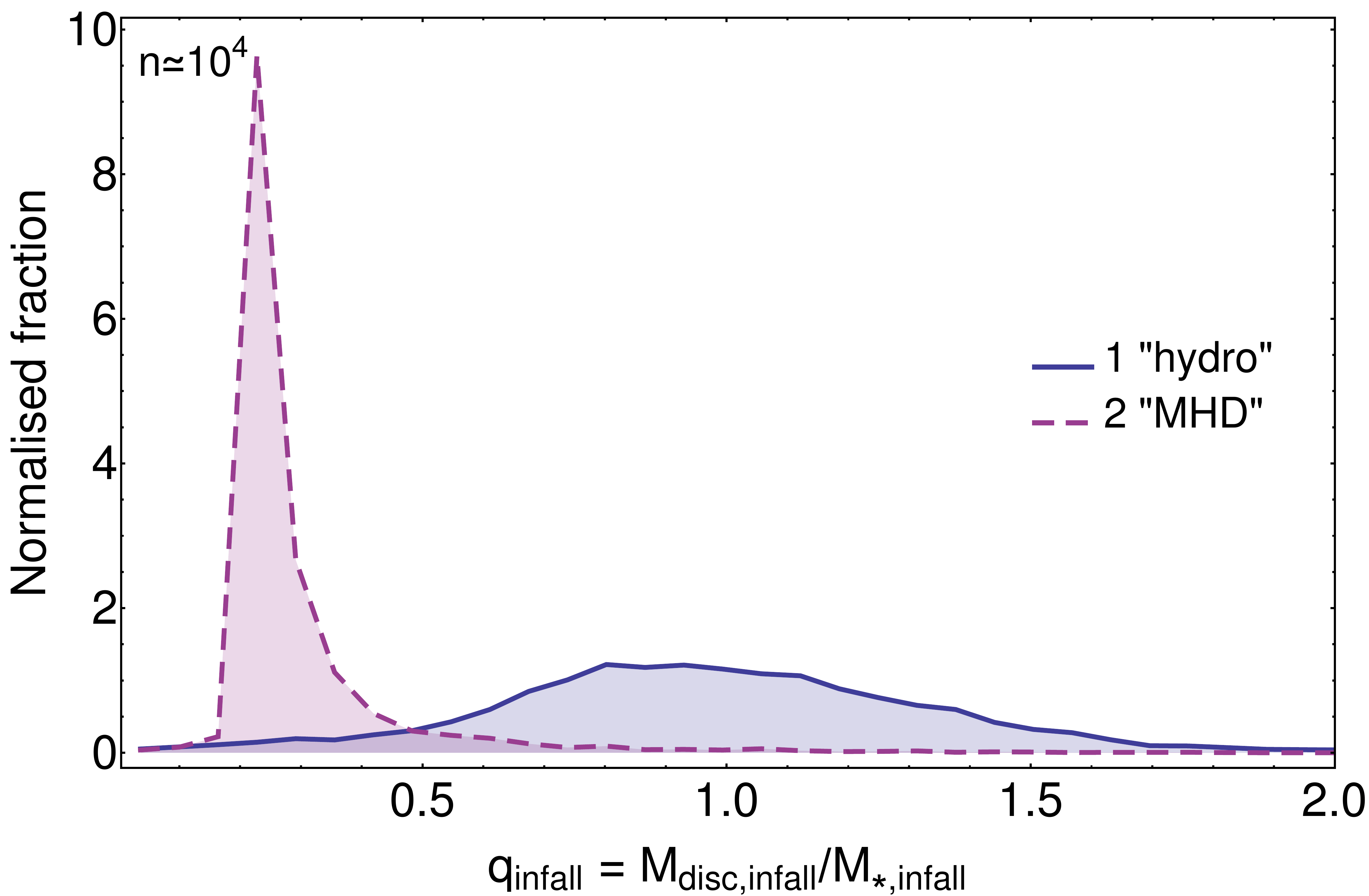}
  \end{subfigure}
  \begin{subfigure}[pt]{0.49\textwidth}
  \includegraphics[width=\linewidth]{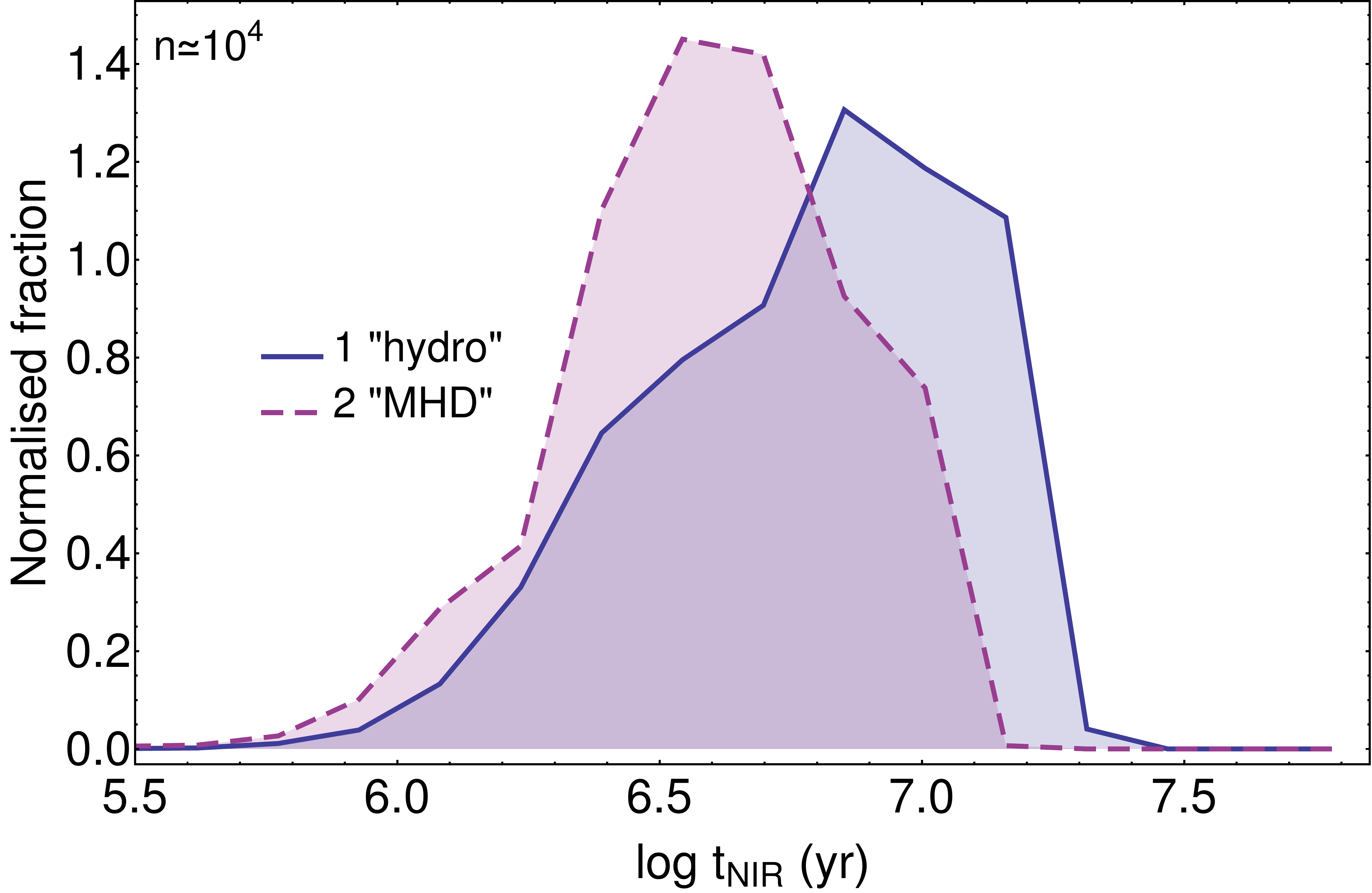}
  \end{subfigure}
  \caption{Distributions of the disc properties at the end of infall and lifetimes of the discs for RUN\nobreakdash-1 and RUN\nobreakdash-2. Top left: disc masses, top right: disc radii including the observational result of \citealt{2018ApJS..238...19T} (Ty18), bottom left: disc-to-star mass ratio $q_\mathrm{infall}$, bottom right: disc lifetimes. All stellar masses are included in this figure.}
  \label{fig:disc_a}
\end{figure*}

The disc properties at the end of the infall phase for  the ``hydro'' and the ``MHD'' runs and their inferred lifetimes are shown in Fig.~\ref{fig:disc_a}.

The mean disc mass in RUN\nobreakdash-1 is higher than that in RUN\nobreakdash-2 by almost a factor of three (top left panel of Fig. \ref{fig:disc_a}). 
This is expected since the infalling mass is deposited very close to the star throughout the simulation in RUN\nobreakdash-2. The top left panel of Fig.~\ref{fig:disc_a} also shows observed Class~0 disc masses from \citet{2018ApJS..238...19T} (see Section~\ref{subs:compdm} for a discussion).

The large difference between RUN\nobreakdash-1 and RUN\nobreakdash-2 is also visible in the disc radii (top right panel of Fig. \ref{fig:disc_a}), where a much wider distribution is seen and the discs are found to be larger by almost a factor six in RUN\nobreakdash-1 compared to RUN\nobreakdash-2. We use the same definition for the disc's radius as \citepalias{2018MNRAS.475.5618B} (the radius containing $\num{63.2} \%$ of the disc's mass).

The distributions of the ratios of disc masses to stellar masses at the end of the infall phase (bottom left panel) show the most striking difference between the two runs. In RUN\nobreakdash-1 the distribution is centered around $q_\mathrm{infall} \approx 1.0$ and very wide, while the mean value is roughly a factor three lower in RUN\nobreakdash-2, with a very narrow distribution. A consequence of the high values of $q_\mathrm{infall}$ is that the stellar mass at this time ($M_\mathrm{*,infall}$) differs significantly from the final stellar mass $M_\mathrm{*,final}$.

The shape of the distribution of disc lifetimes (bottom right panel of Fig. \ref{fig:disc_a}) is very similar in RUN\nobreakdash-1 compared to RUN\nobreakdash-2, but shifted to longer lifetimes by almost $50 \%$: a large fraction of the mass is transported to the star during the infall phase in RUN\nobreakdash-2 early on. The mean lifetimes of $\SI{7.3}{Myr}$ (RUN\nobreakdash-1) and $\SI{4.5}{Myr}$ (RUN\nobreakdash-2) are long, see Sect.~\ref{Sect:comp}. The quantity displayed, $t_\mathrm{NIR}$, corresponds to the disc's total lifetime as explained in Sect.~\ref{subs:disp} with no reduction due to the start of the pre-main sequence phase applied. Figures of $t_\mathrm{life}$ (with the reduction applied) can be found in Appendix~\ref{app:lifetimes}.

\subsection{\emph{RUN-3}, \emph{RUN-4} and \emph{RUN-5}: accretion heating and fragmentation}\label{subs:345}

Here we show the same figures as in Section~\ref{subs:12}, but this time for RUN\nobreakdash-3, RUN\nobreakdash-4 and RUN\nobreakdash-5. RUN\nobreakdash-1 is also shown for comparison.

\begin{figure*}[pt]
  \begin{subfigure}[pt]{0.49\textwidth}
  \includegraphics[width=\linewidth]{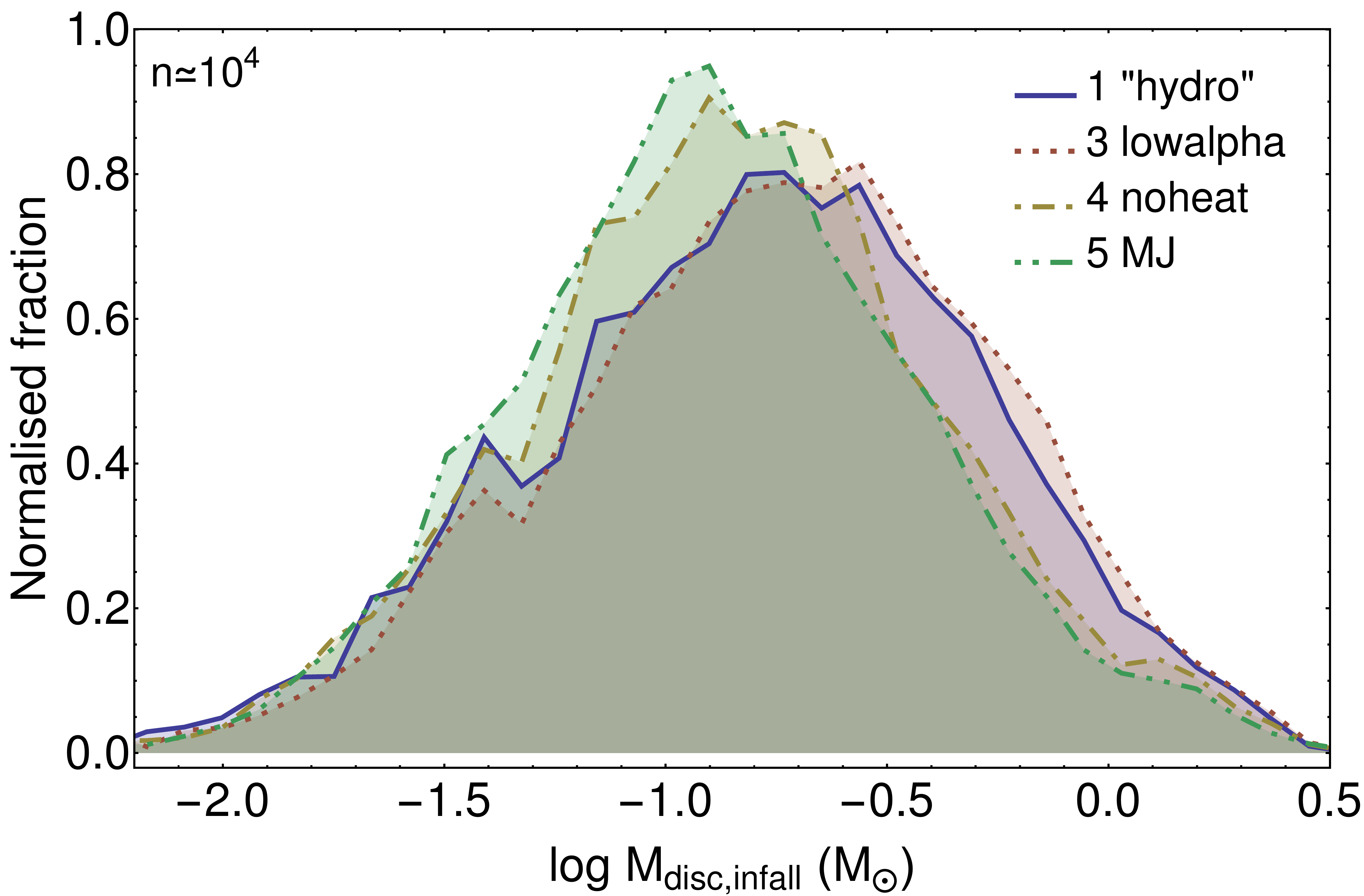}
  \end{subfigure}
  \begin{subfigure}[pt]{0.49\textwidth}
  \includegraphics[width=\linewidth]{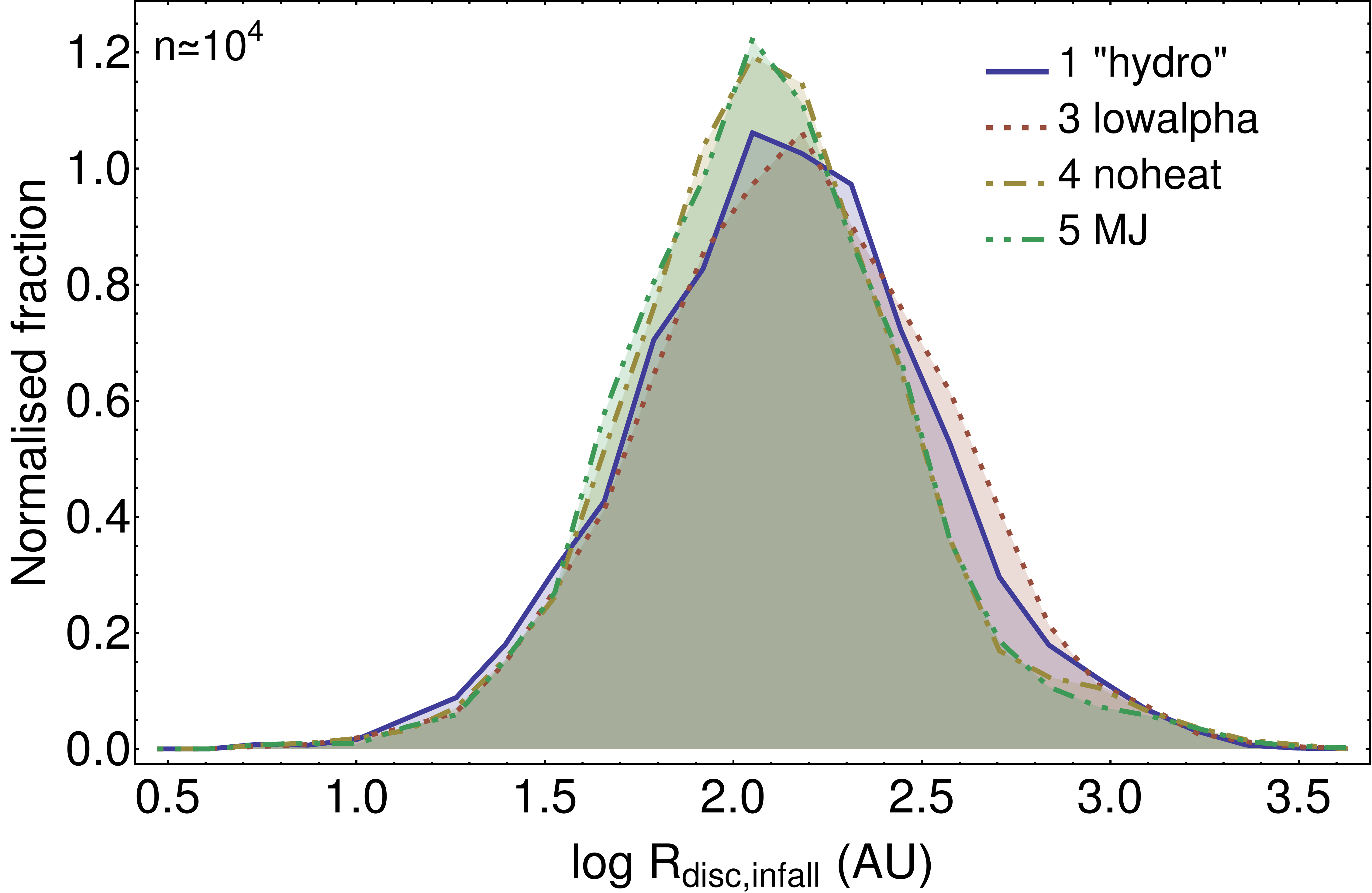}
  \end{subfigure}
  \begin{subfigure}[pt]{0.49\textwidth}
  \includegraphics[width=\linewidth]{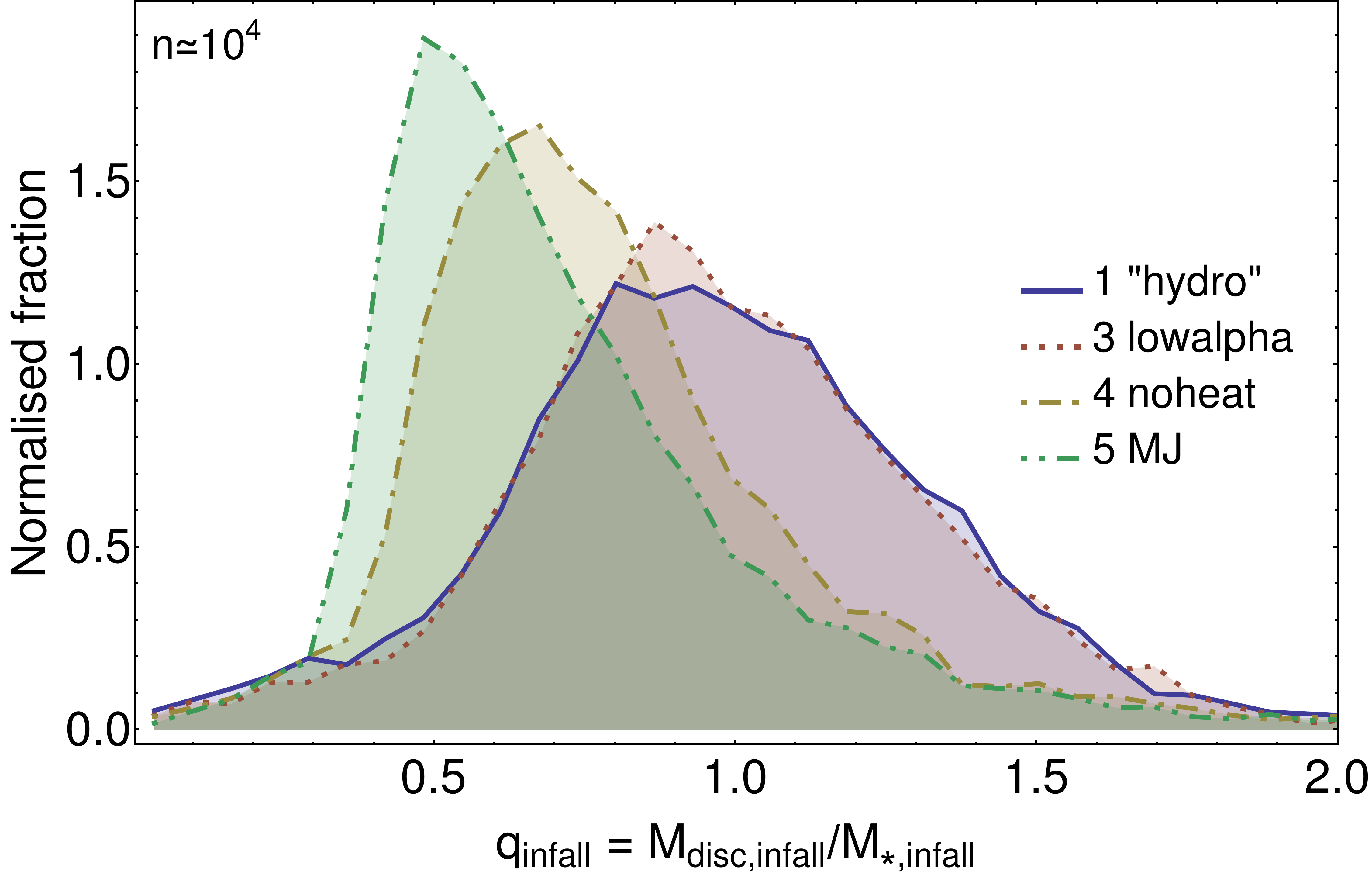}
  \end{subfigure}
  \begin{subfigure}[pt]{0.49\textwidth}
  \includegraphics[width=\linewidth]{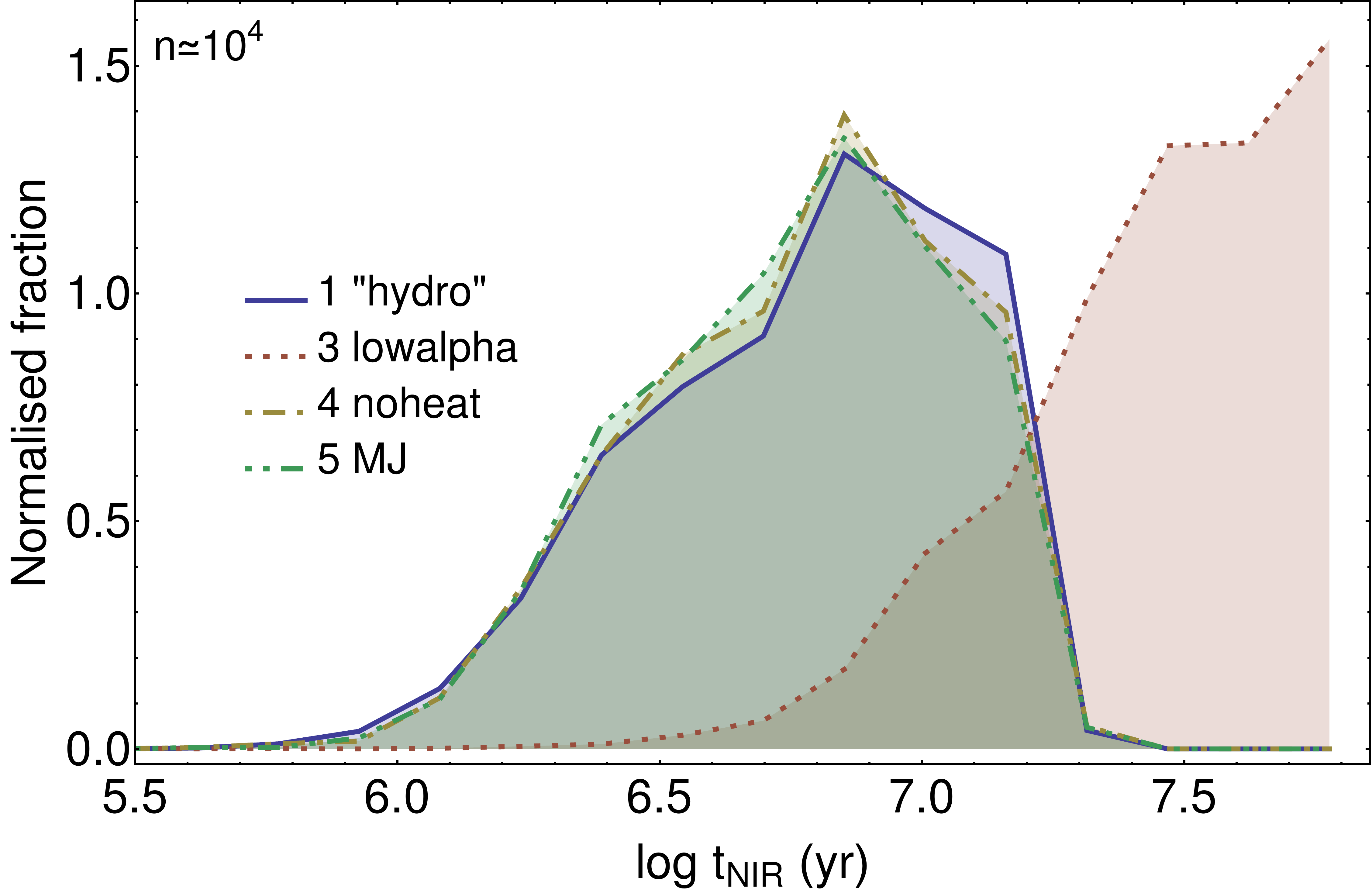}
  \end{subfigure}
  \caption{Properties at the end of the infall phase and lifetimes of the discs for RUN\nobreakdash-1, RUN\nobreakdash-3, RUN\nobreakdash-4 and RUN\nobreakdash-5. Top left: masses, top right: radii, bottom left: disc-to-star mass ratio $q_\mathrm{infall}$, bottom right: lifetimes.}
  \label{fig:disc_b}
\end{figure*}

There is almost no difference in the distribution of disc masses between RUN\nobreakdash-1 and RUN\nobreakdash-3 (top left panel of Fig.~\ref{fig:disc_b}). This is expected: the infall phase is dominated by the global gravitational instability of the discs, the lower background viscosity parameter $\alpha_\mathrm{bg}$ is only important later in the disc's evolution. The disc mass distributions of RUN\nobreakdash-4 and RUN\nobreakdash-5 are similar in shape to that of RUN\nobreakdash-1, but shifted to successively lower masses. Turning off accretion heating due to infalling matter from the MCC (infall heating) as well as accretion of disc material onto the star (stellar accretion heating) leads to lower temperatures in the disc and promotes fragmentation. In our model, fragmentation leads to the removal of mass from the disc to the star. This effect becomes more pronounced in RUN\nobreakdash-5, where much more mass is removed from the disc each time the conditions for fragmentation are satisfied (see also Sect.~\ref{subs:frag}). We note that the effect of stellar accretion heating is much more important than that of infall heating.

A similar effect on the disc radii is seen in the top right panel of Fig.~\ref{fig:disc_b}. It is less pronounced than the effect on the disc masses.

The most obvious differences between the runs that are based on hydrodynamic simulations (RUN\nobreakdash-1, RUN\nobreakdash-3, RUN\nobreakdash-4 and RUN\nobreakdash-5) are seen in the distribution of $q_\mathrm{infall}$ (bottom left panel of Fig.~\ref{fig:disc_b}). When the disc fragments more often due to lower temperatures in the disc, the disc mass is reduced, while the stellar mass grows. So $q_\mathrm{infall}$ exhibits the `double effect'. The distribution of $q_\mathrm{infall}$ is shifted to lower values in RUN\nobreakdash-4, and even more so in RUN\nobreakdash-5, compared to RUN\nobreakdash-1 and RUN\nobreakdash-3.

The distributions of lifetimes differ negligibly between RUN\nobreakdash-1, RUN\nobreakdash-4 and RUN\nobreakdash-5 (bottom right panel of Fig.~\ref{fig:disc_b}). Fragmentation can remove substantial amounts of mass from the disc. However, it leaves the inner disc (important for the determination of the lifetimes) mostly unchanged over long time scales. Lifetimes are longer by a factor $\sim \num{5}$ in RUN\nobreakdash-3. This is expected: during the vast majority of the disc's life, its evolution is dominated by the choice of $\alpha_\mathrm{bg}$.

\subsection{Dependency on the final stellar mass}\label{ssec:dep}

\begin{figure*}[pt]
  \begin{subfigure}[pt]{0.49\textwidth}
  \includegraphics[width=\linewidth]{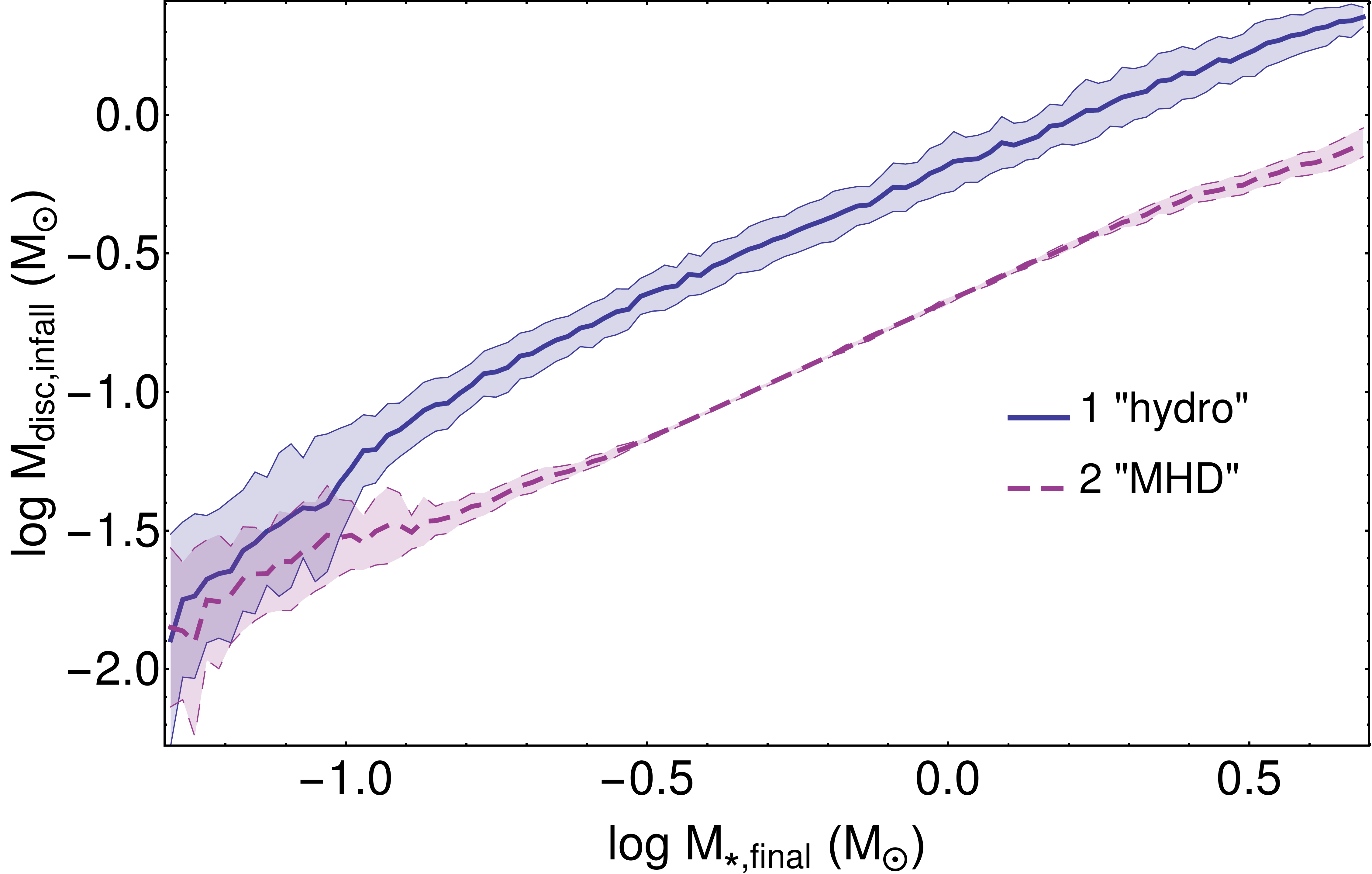}
  \end{subfigure}
  \begin{subfigure}[pt]{0.49\textwidth}
  \includegraphics[width=\linewidth]{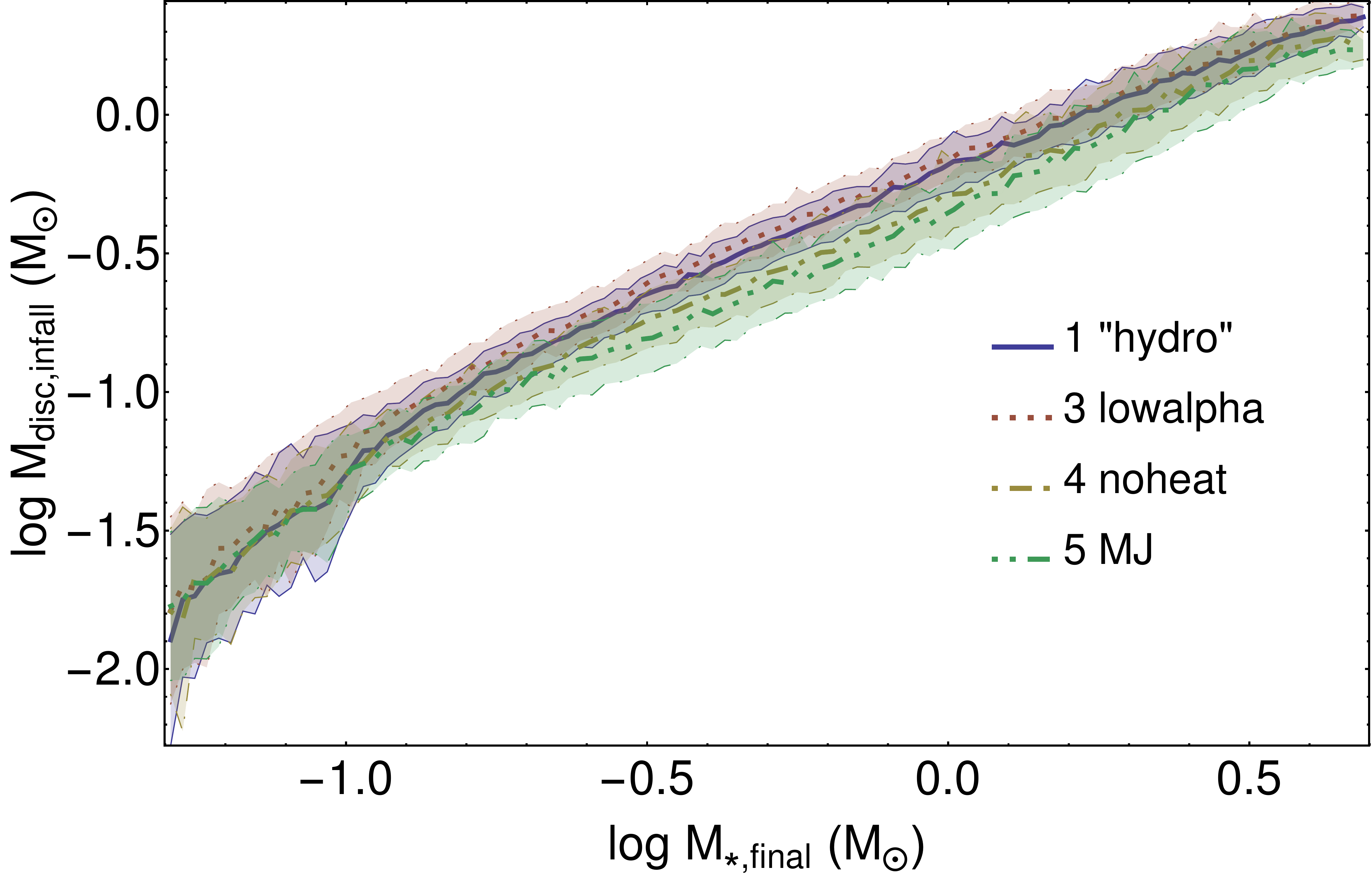}
  \end{subfigure}
  \begin{subfigure}[pt]{0.49\textwidth}
  \includegraphics[width=\linewidth]{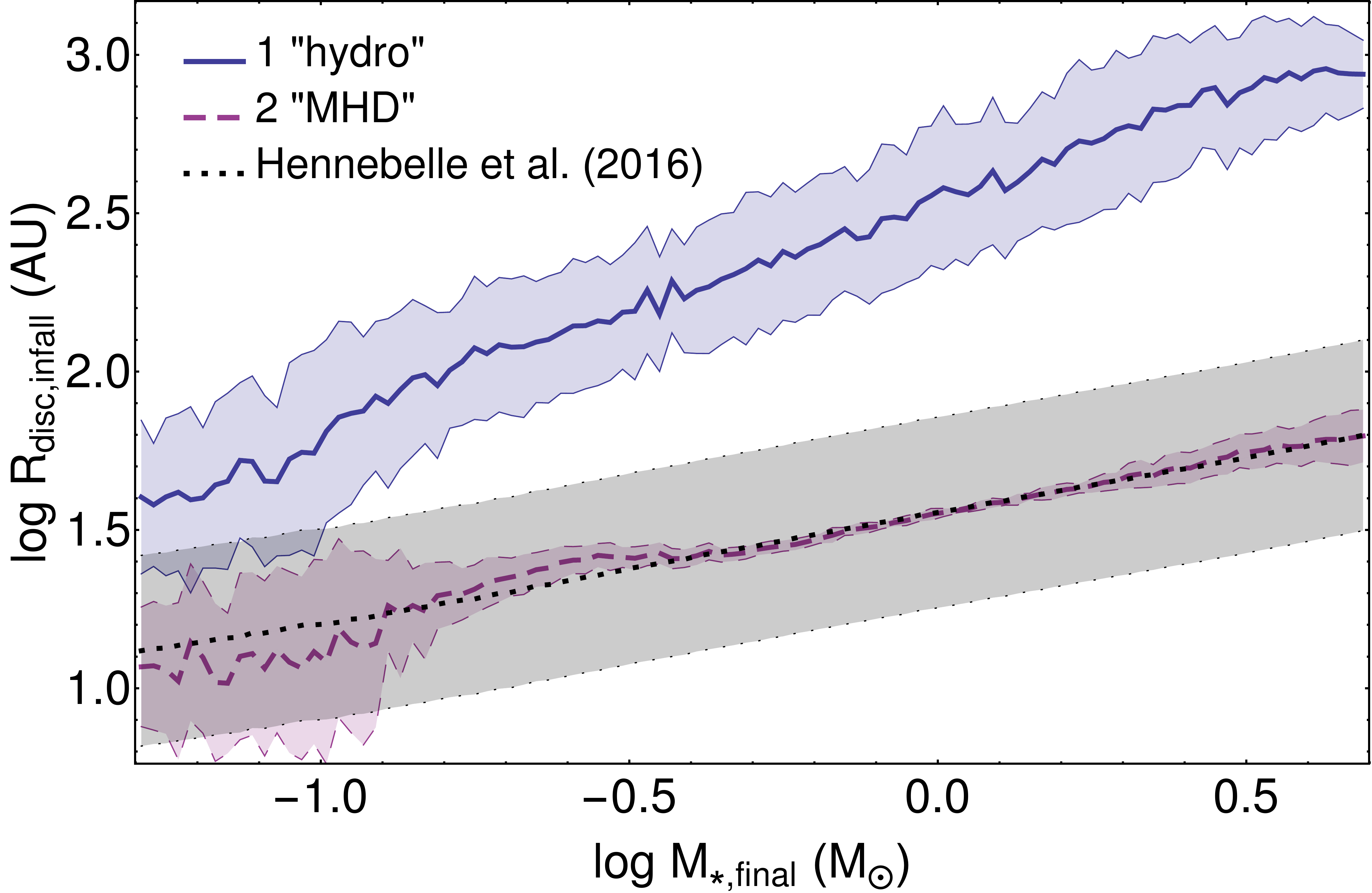}
  \end{subfigure}
  \begin{subfigure}[pt]{0.49\textwidth}
  \includegraphics[width=\linewidth]{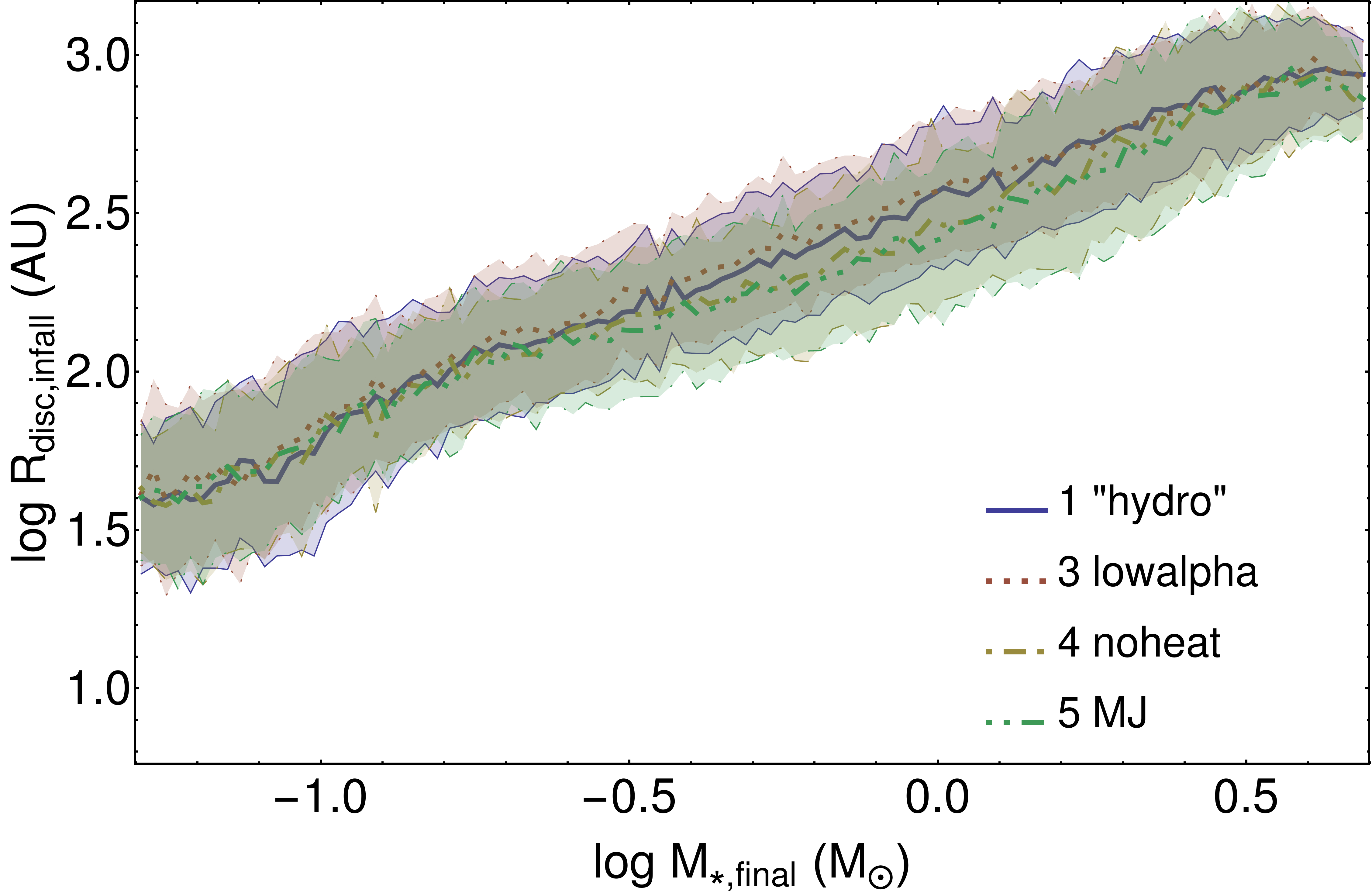}
  \end{subfigure}
  \begin{subfigure}[pt]{0.49\textwidth}
  \includegraphics[width=\linewidth]{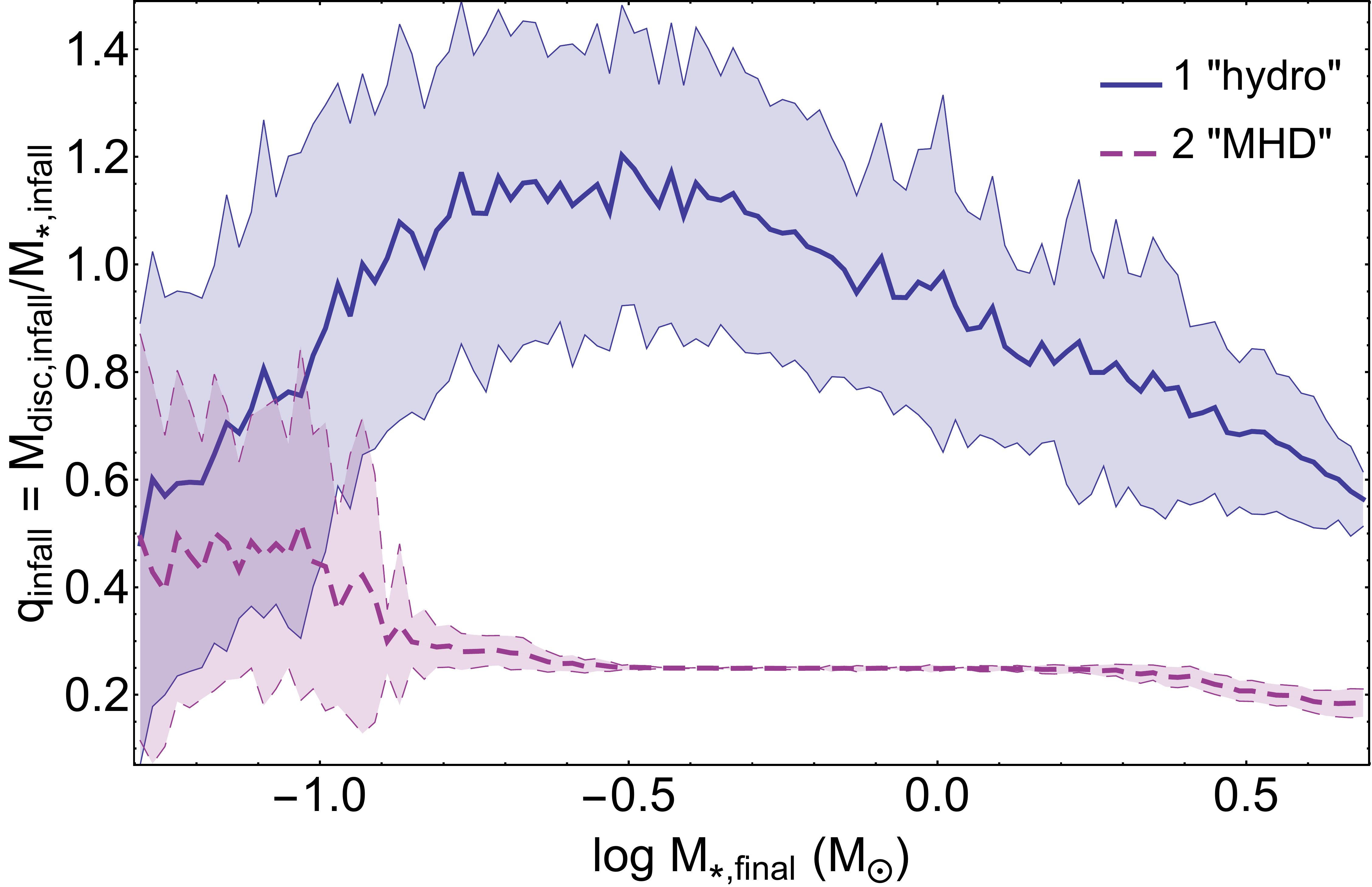}
  \end{subfigure}
  \begin{subfigure}[pt]{0.49\textwidth}
  \includegraphics[width=\linewidth]{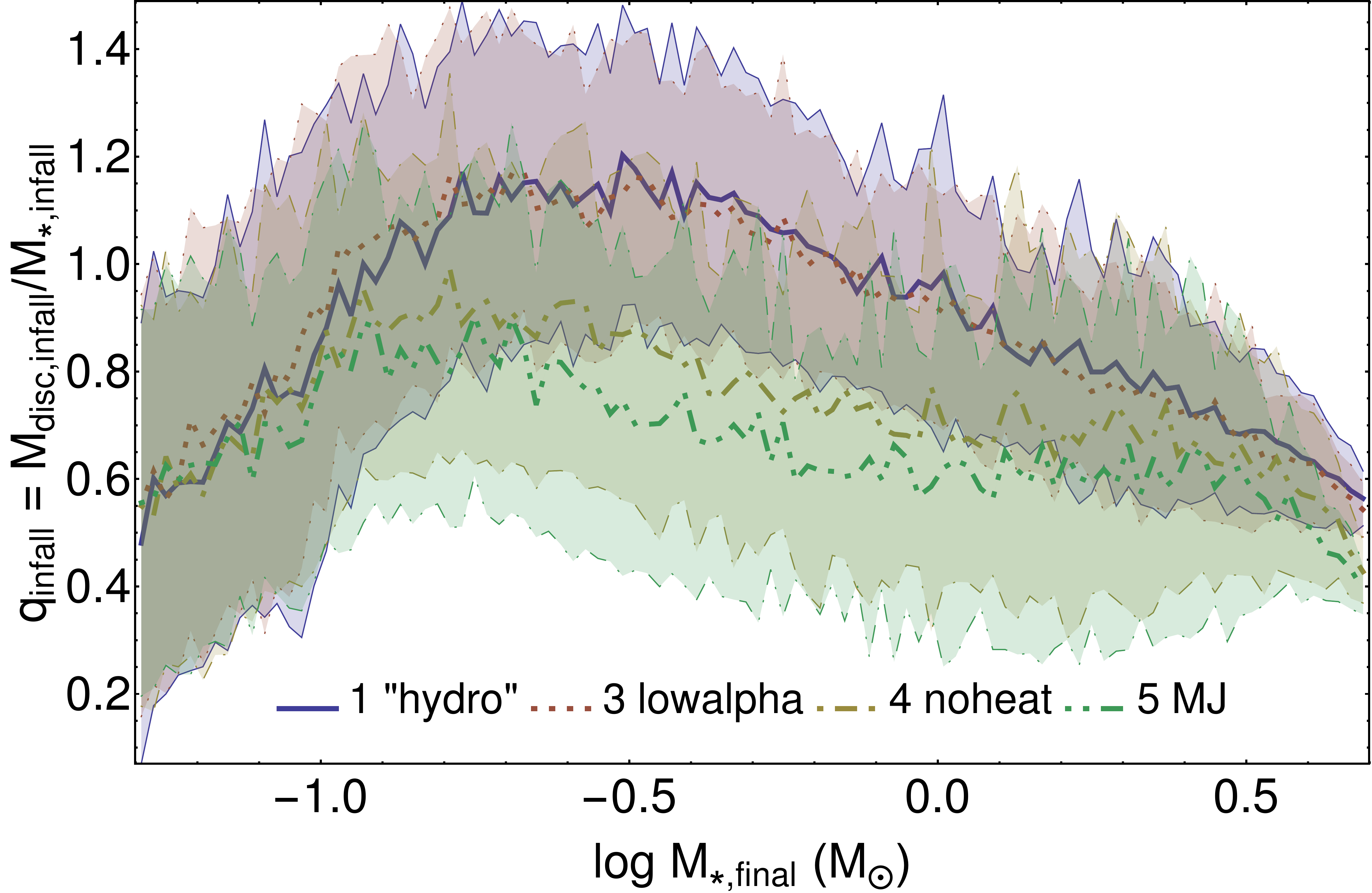}
  \end{subfigure}
  \begin{subfigure}[pt]{0.49\textwidth}
  \includegraphics[width=\linewidth]{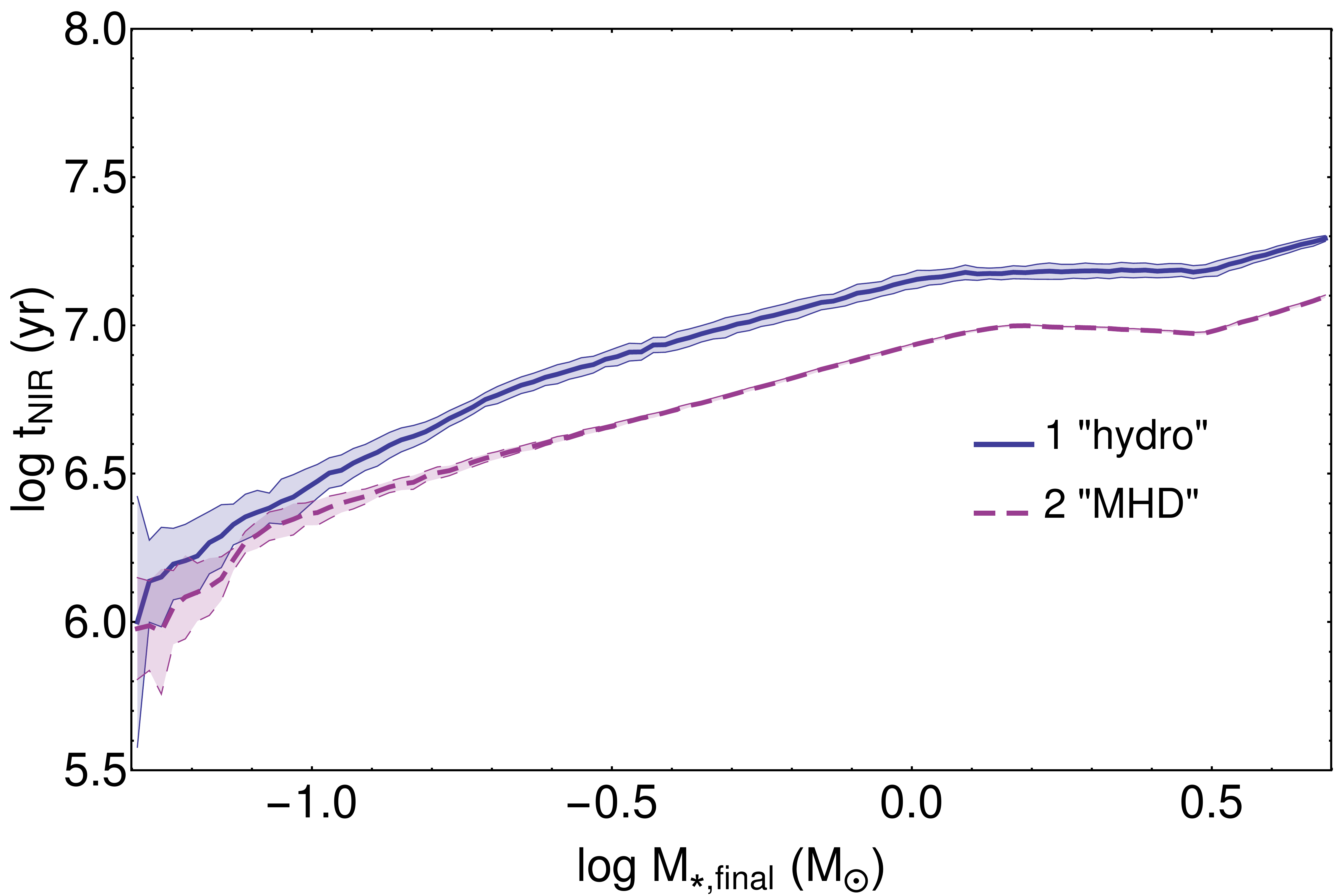}
  \end{subfigure}
  \begin{subfigure}[pt]{0.49\textwidth}
  \includegraphics[width=\linewidth]{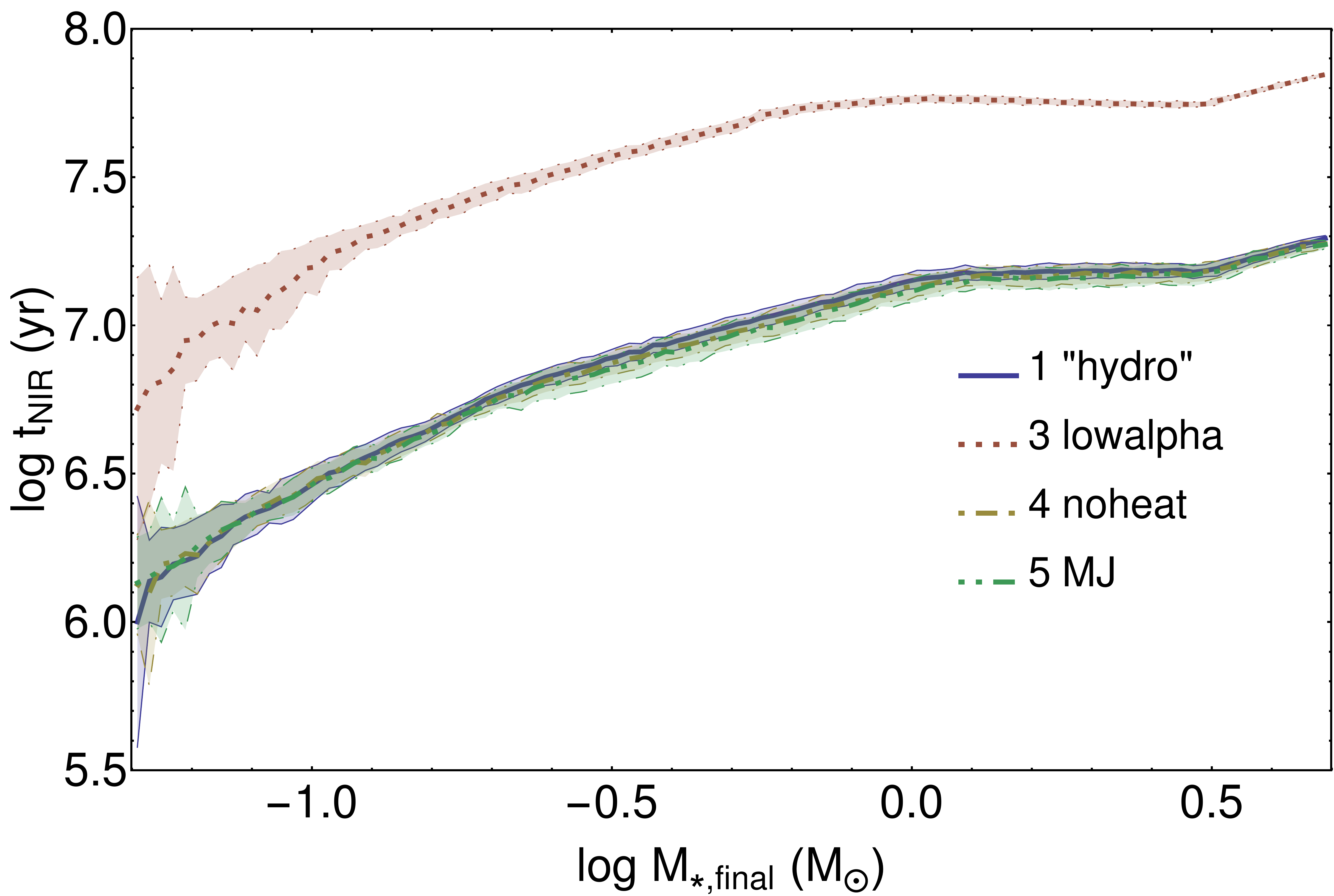}
  \end{subfigure}
  \caption{End-of-infall properties and lifetimes of the discs as a function of the final stellar mass $M_\mathrm{*,final}$. Left column: RUN\nobreakdash-1 and RUN\nobreakdash-2. Right column: RUN\nobreakdash-1, RUN\nobreakdash-3, RUN\nobreakdash-4 and RUN\nobreakdash-5. From top to bottom: masses, radii, disc-to-star mass ratios $q_\mathrm{infall}$ at the end of the infall phase and lifetimes.}
  \label{fig:disc_2}
\end{figure*}

Here we show the same quantities as in the preceding section, but this time as a function of the final stellar mass $M_\mathrm{*,final}$, because this quantity is more easily observed than the stellar mass at the end of the infall phase. Figure~\ref{fig:disc_2} shows the mean values of disc mass, disc radius and disc-to-star mass ratio for each of the $100$ mass bins. The spread in these parameters is also shown: the shaded region corresponds to $\pm$ one standard deviation. RUN\nobreakdash-1 and RUN\nobreakdash-2 are depicted in the left, RUN\nobreakdash-1, RUN\nobreakdash-3, RUN\nobreakdash-4 and RUN\nobreakdash-5 in the right panels of the figure.
Histograms for specific mass bins can be found in Appendix \ref{app:massbin}.

The dependency of $M_\mathrm{disc,infall}$ on $M_\mathrm{*,final}$ is roughly linear and very similar in shape for all runs. The spread is highest at low stellar masses. The reason is that these systems are more strongly dominated by the initial conditions: a larger fraction of the total mass is already in the system at the beginning of the simulation.

The discs' radii, $R_\mathrm{disc,infall}$, scale linearly with $M_\mathrm{*,final}$ to good approximation. The differences between the runs are very small, with exception of RUN\nobreakdash-2, as expected. The middle left panel of Fig.~\ref{fig:disc_2} also contains an analytic estimate for the early disc radius (see Eq.~\ref{eq:rhen} and Sect.~\ref{subs:compdr} for a discussion).

The ratio $M_\mathrm{disc,infall} / M_\mathrm{*,infall}$ varies strongly with $M_\mathrm{*,final}$ (third row in Fig.~\ref{fig:disc_2}). For all runs but RUN\nobreakdash-2, it rises from around $0.5$ at the lower end of the mass range considered to some maximum value at around $\SI{0.2}{\msun}$, then decreases again to $\approx 0.5$. The maximum value reached depends on the cooling and fragmentation mechanisms that are at play. When fragmentation is inhibited by stellar accretion heating, the mean values of $q_\mathrm{infall}$ reach $1.1$ (RUN\nobreakdash-1 and 3) while they stay below $0.8$ in most systems when heating is turned off (RUN\nobreakdash-4). They are further reduced when a larger initial fragment mass is used (RUN\nobreakdash-5). This shows that the amount of mass that is removed from the disc by fragmentation is limited by the initial fragment mass in runs 1-4. The influence of the heating mechanisms discussed above is, however, more important than that of the initial fragment mass.

The behaviour of $q_\mathrm{infall}$ is different in RUN\nobreakdash-2 compared to the other runs. Here $q_\mathrm{infall}$ depends only weakly on the final stellar mass. It is typically around $0.4$ at the lowest stellar masses and decreases to a nearly constant $\sim 0.25$ at stellar masses $\gtrsim \SI{0.2}{\msun}$.

The discs' lifetimes as a function of final stellar mass are depicted in the bottom row of Fig.~\ref{fig:disc_2}. Lifetimes exhibit a weak positive correlation with final stellar mass. We noted earlier, that runs 1, 4 and 5 show very little difference in disc lifetime. This remains true when looking at the dependence on stellar mass. The lifetimes in RUN\nobreakdash-2 are roughly $40 \%$ shorter than those in RUN\nobreakdash-1, due to the lower post-infall masses. Using a lower $\alpha_\mathrm{bg}$ produces lifetimes that are around a factor of $\num{5}$ longer except at the lowest masses. We  compare the disc lifetimes to observations in Sect.
~\ref{subs:complt}.

\subsection{Fragmentation}\label{subs:frag}
Here we look at the fragmentation behaviour of the discs. An overview is given in Table~\ref{table:frag}, where the global fragmentation properties of all runs are shown. Some of the simulated discs fragment, producing anywhere between $\sim 1$ and $\sim \SI{e3}{}$ fragments, while others do not fragment at all. The formation of several hundred fragments in a single disc has not been reported in the literature to our knowledge. Such large numbers of fragments are likely an overestimate related to our assumptions. We discuss this in Sect.~\ref{Sect:comp}. The fraction of fragmenting discs is given in the first column of Table~\ref{table:frag}. The following columns show the mean number of fragments, the mean initial mass of the fragments and the mean fragmentation location, respectively. The values in these three columns only take into account the discs that do fragment. For instance, if in one run every other disc fragments, with five fragments each, the mean number of fragments is $5$.

\begin{table}[ht]
\begin{tabular}{cccccc}
\hline\hline
\multicolumn{2}{c}{Run} & \begin{tabular}[c]{@{}l@{}}(a)\end{tabular} & \begin{tabular}[c]{@{}l@{}}(b)\end{tabular} & \begin{tabular}[c]{@{}l@{}}(c)\end{tabular} & \begin{tabular}[c]{@{}l@{}}(d)\end{tabular} \\
\hline
1 & ``hydro'' & $\num{0.45}$  & $\num{24(15)}$ & $\num{1.60(45)}$ & $\num{110(39)}$  \\
2 & ``MHD''   & $\num{0}$     & $0$            & ---                & ---  \\
3 & lowalpha  & $\num{0.49}$  & $\num{24(15)}$ & $\num{1.40(48)}$ & $\num{100(41)}$  \\
4 & noheat    & $\num{0.76}$  & $\num{130(60)}$& $\num{0.68(21)}$ & $\num{63(21)}$  \\
5 & MJ        & $\num{0.77}$  & $\num{19(9)}$  & $\num{6.7(20)}$  & $\num{58(20)}$ \\
\hline
\end{tabular}
\caption{ Fragmentation characteristics of each run. 
(a): the fraction of discs that fragment, (b): the mean number of fragments, (c): the mean initial fragments mass in $\mj$, (d): the mean fragmentation location in $\SI{}{au}$.}
\label{table:frag}
\end{table}

\begin{figure}[ht]
  \includegraphics[width=\linewidth]{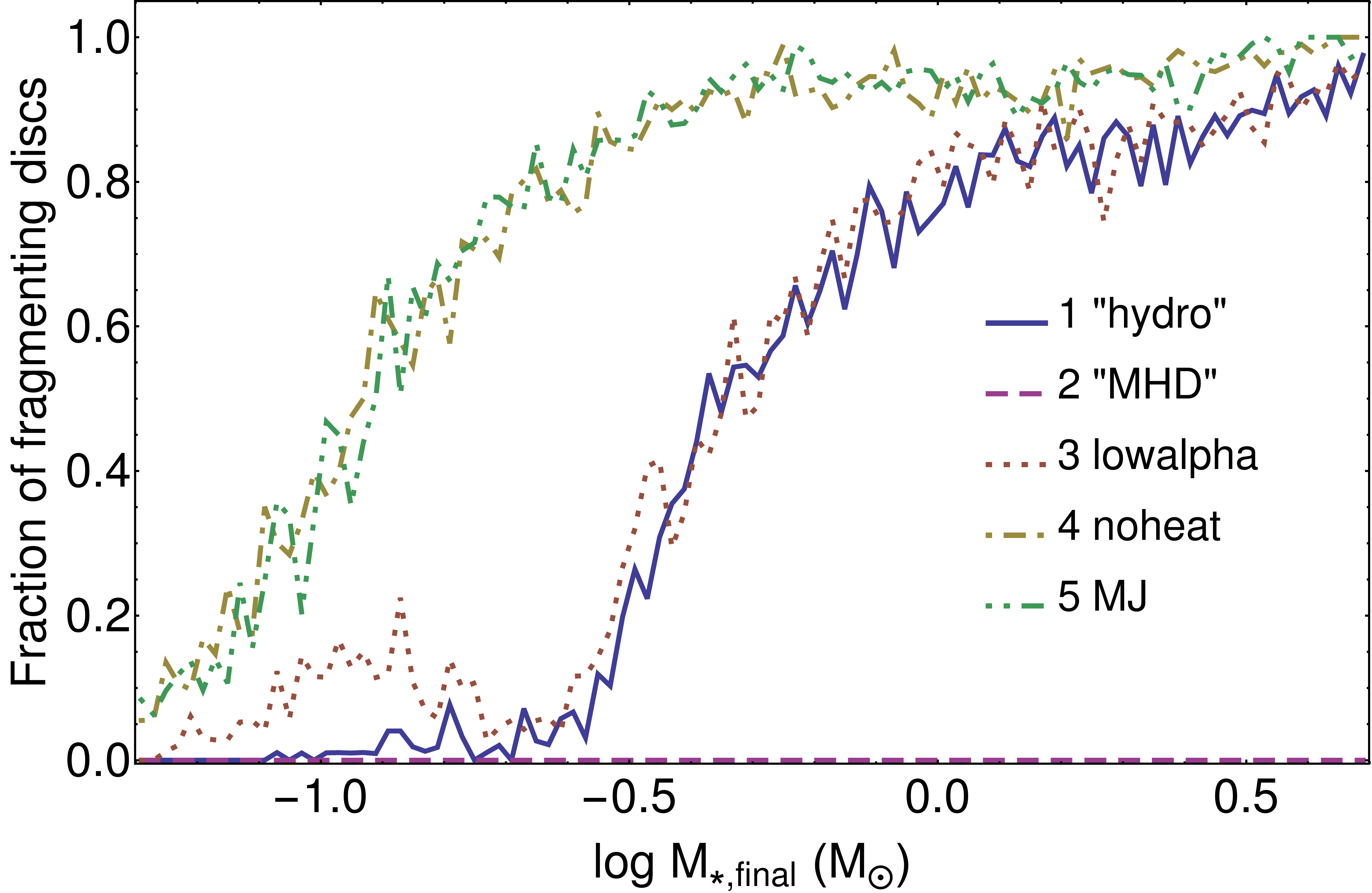}
  \caption{Fraction of discs that fragment versus final stellar mass for all runs. The ``MHD'' run does not produce any fragments.}
  \label{fig:ff}
\end{figure}

Figure~\ref{fig:ff} displays the fraction of discs that fragment as a function of final stellar mass.

Of the $\num{10000}$ discs in RUN\nobreakdash-2, not a single fragments. This demonstrates the enormous importance of magnetic fields and magnetized collapse for the question whether protostellar and protoplanetary discs can be gravitationally unstable, and thus whether giant planets may form by gravitational instability.
Fig.~\ref{fig:ff} shows the background viscosity has very little influence on the fraction of discs fragmenting. There is only a slight difference at low stellar masses where only few discs fragment in RUN\nobreakdash-1 and RUN\nobreakdash-3. Our statistics are therefore not reliable in this region of parameter space and we do not deem this difference significant. There is however a significant difference when stellar accretion heating is turned off. In this case, the fraction of discs that fragment is increased by \SI{50}{\percent} as seen in Table
~\ref{table:frag}. Stellar accretion heating heats the discs keeping them gravitationally stable, thus inhibiting fragmentation. Figure~\ref{fig:ff} reveals this happens mainly in systems with final stellar masses $< \SI{1}{\msun}$.

For the remainder of this section, we concentrate on the discs that do fragment. RUN\nobreakdash-2 is not included in the following figures.

\subsubsection{Global fragmentation properties}\label{globfrag}

\begin{figure*}[pt]
  \begin{subfigure}[pt]{0.49\textwidth}
  \includegraphics[width=\linewidth]{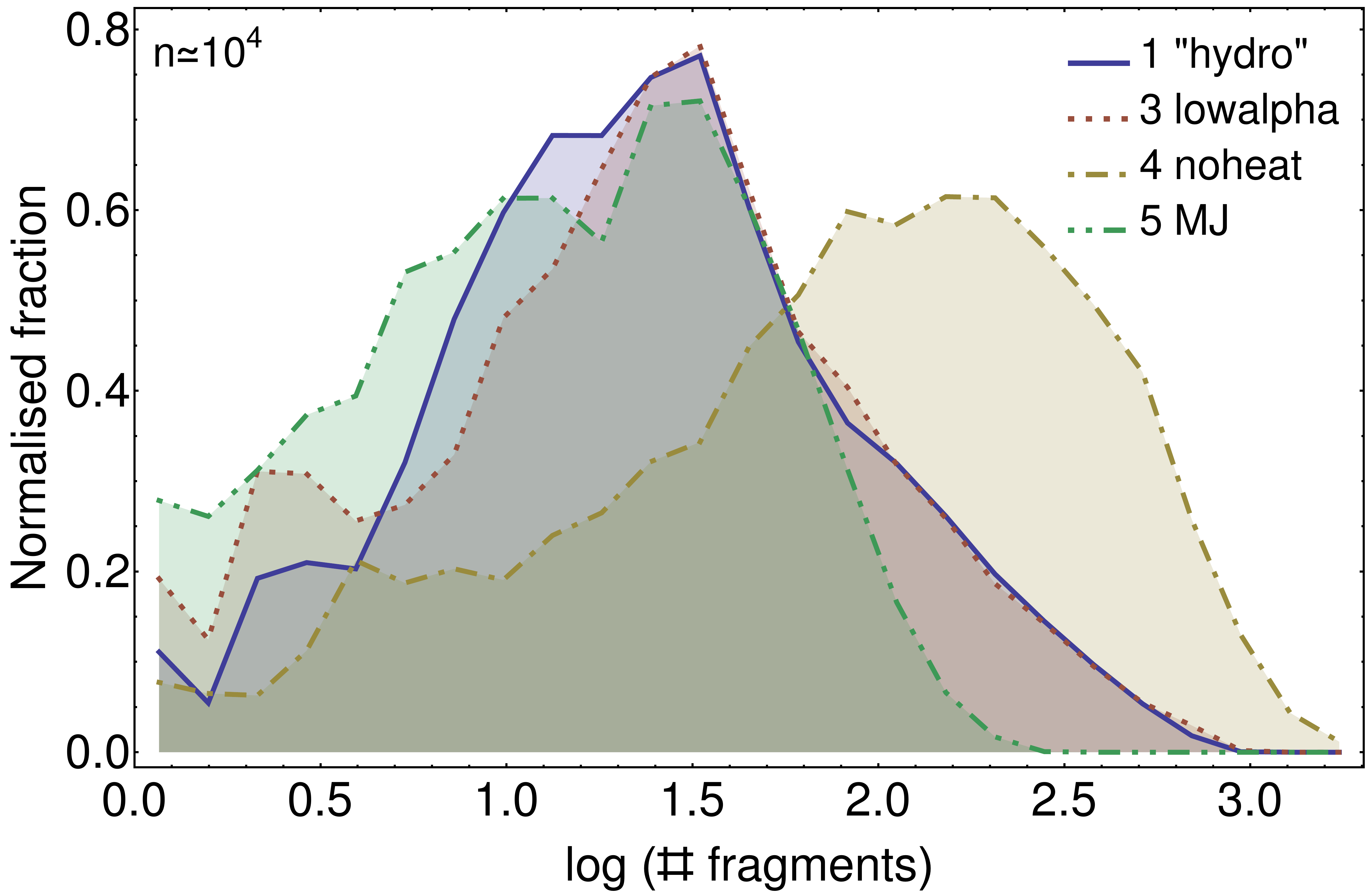}
  \end{subfigure}
  \begin{subfigure}[pt]{0.49\textwidth}
  \includegraphics[width=\linewidth]{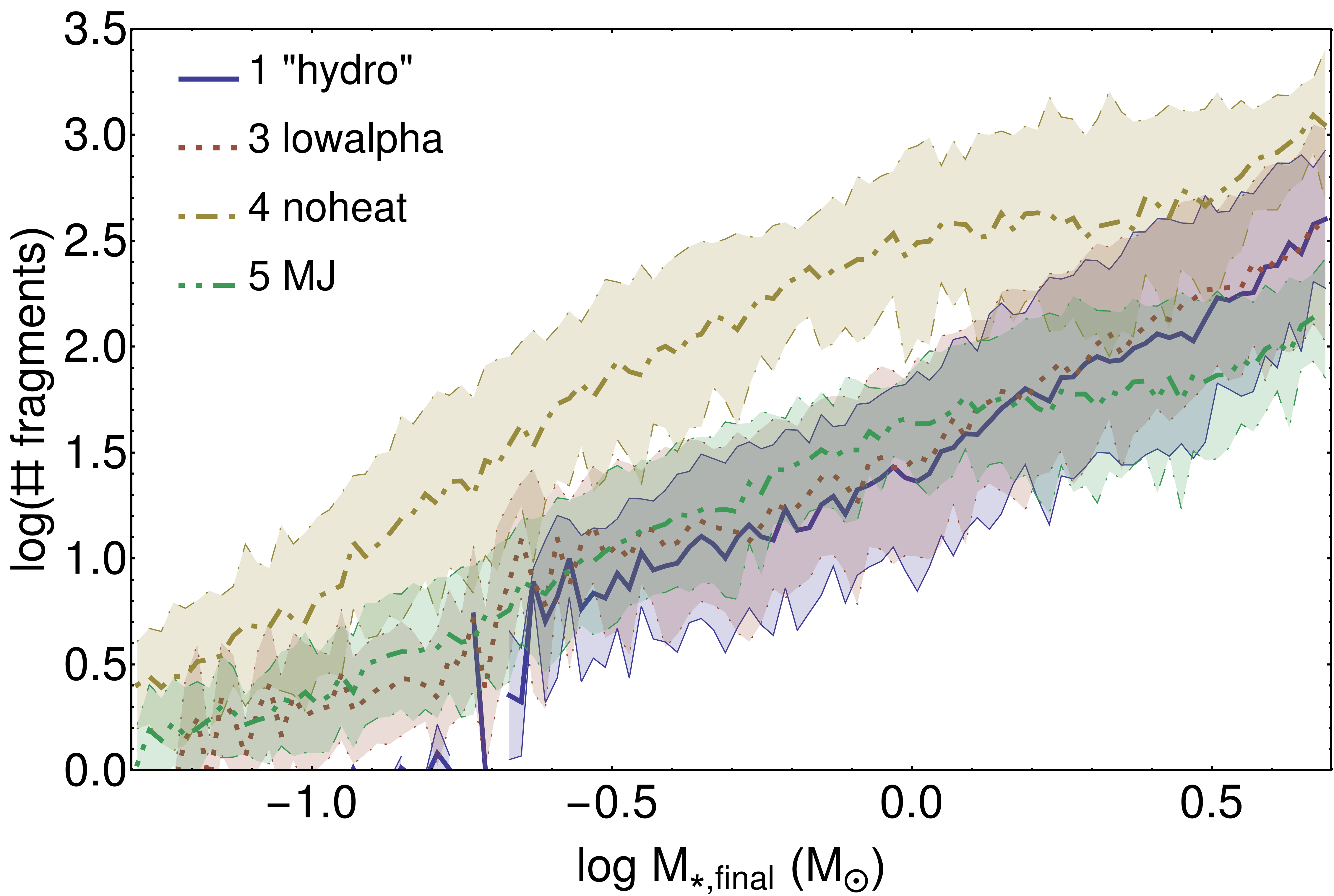}
  \end{subfigure}
  \begin{subfigure}[pt]{0.49\textwidth}
  \includegraphics[width=\linewidth]{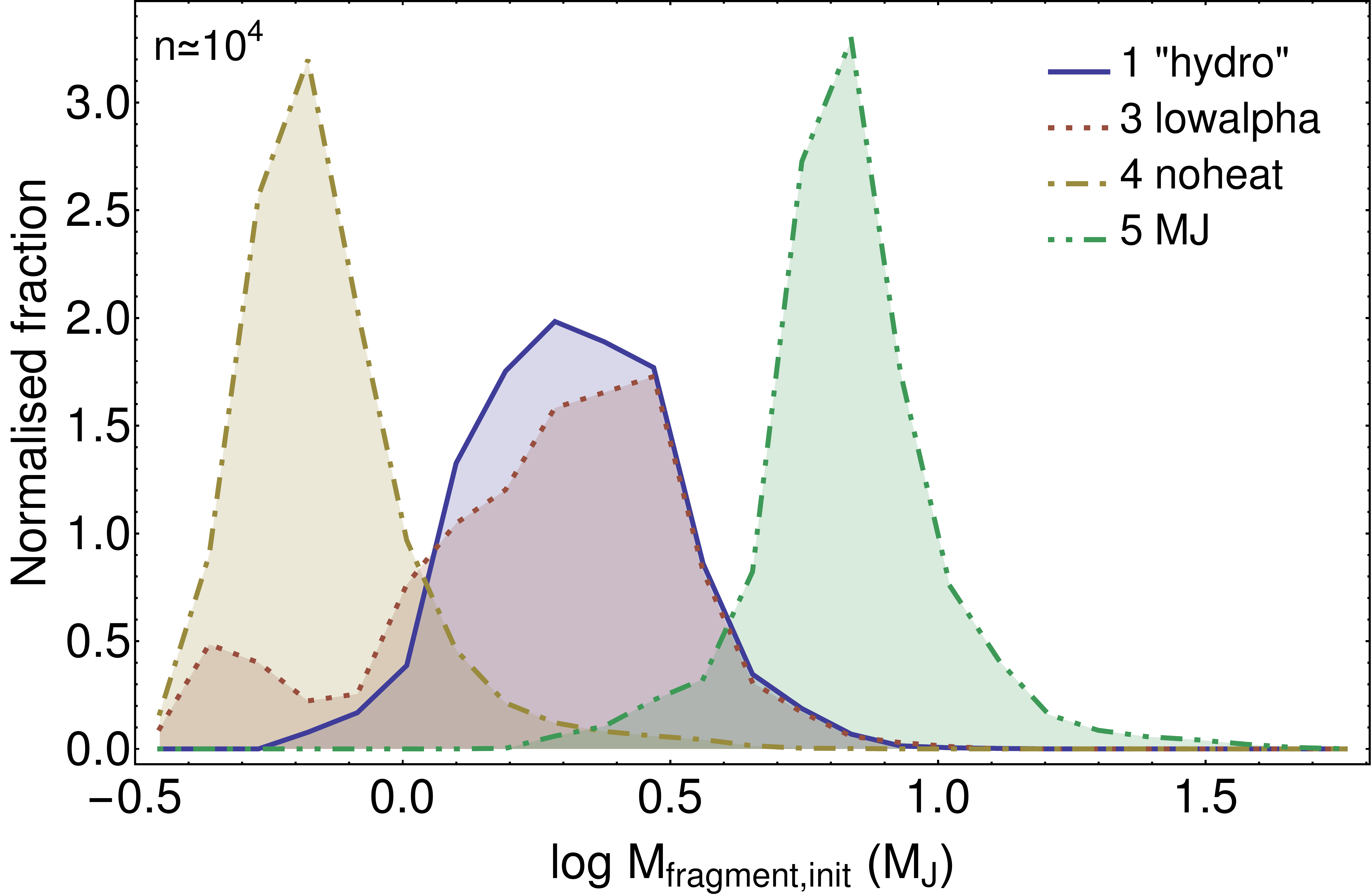}
  \end{subfigure}
  \begin{subfigure}[pt]{0.49\textwidth}
  \includegraphics[width=\linewidth]{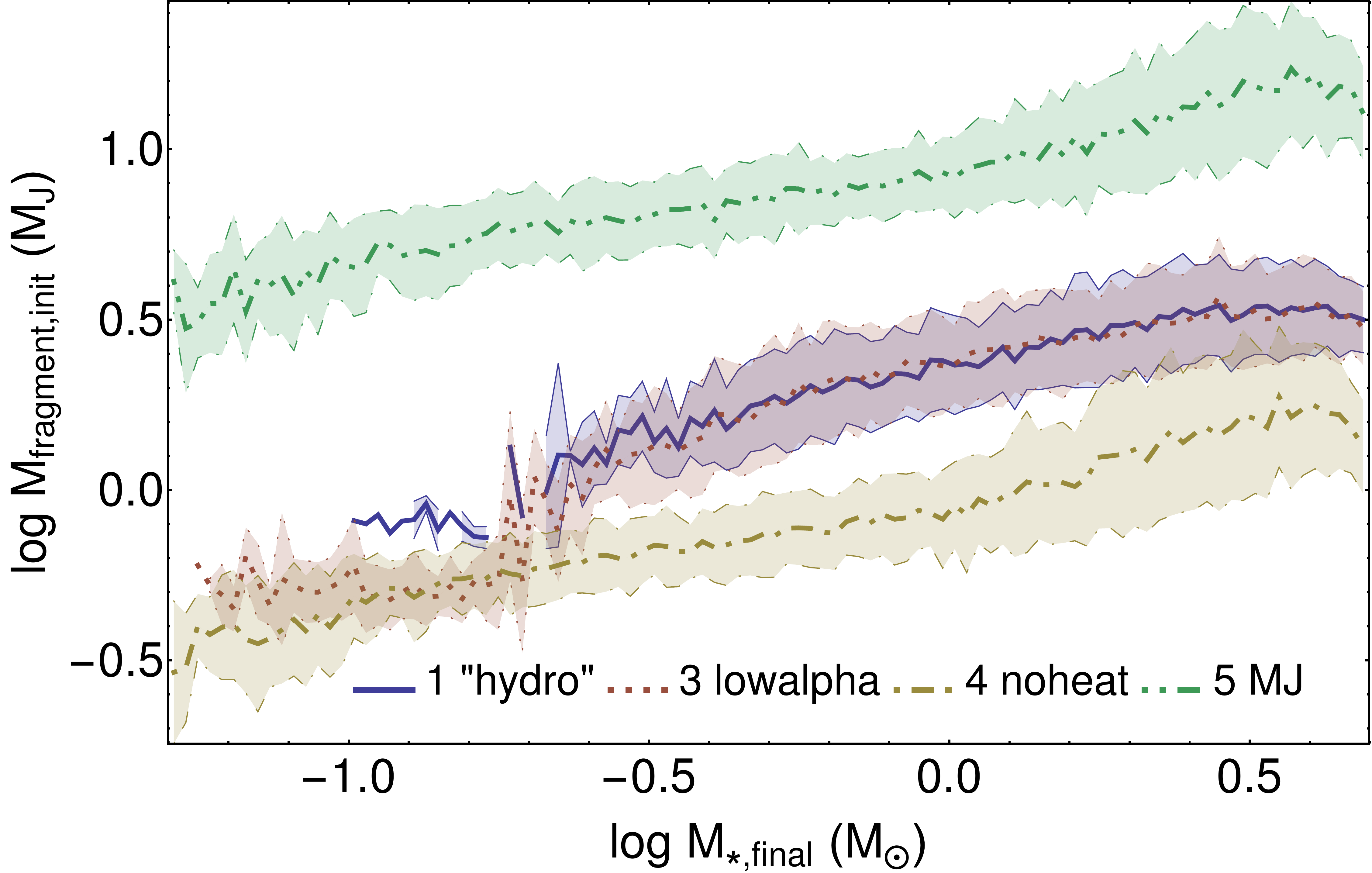}
  \end{subfigure}
  \begin{subfigure}[pt]{0.49\textwidth}
  \includegraphics[width=\linewidth]{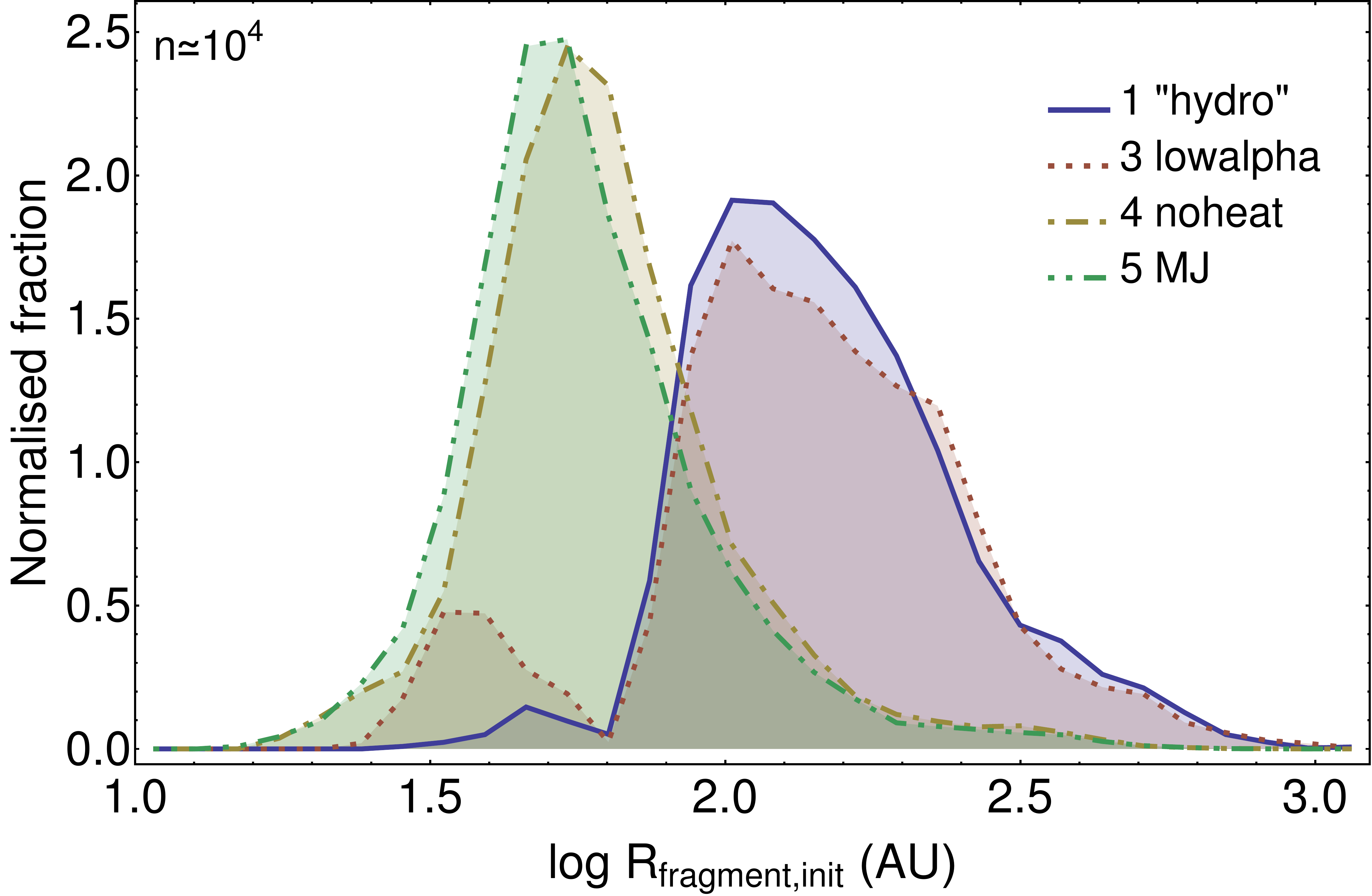}
  \end{subfigure}
  \begin{subfigure}[pt]{0.49\textwidth}
  \includegraphics[width=\linewidth]{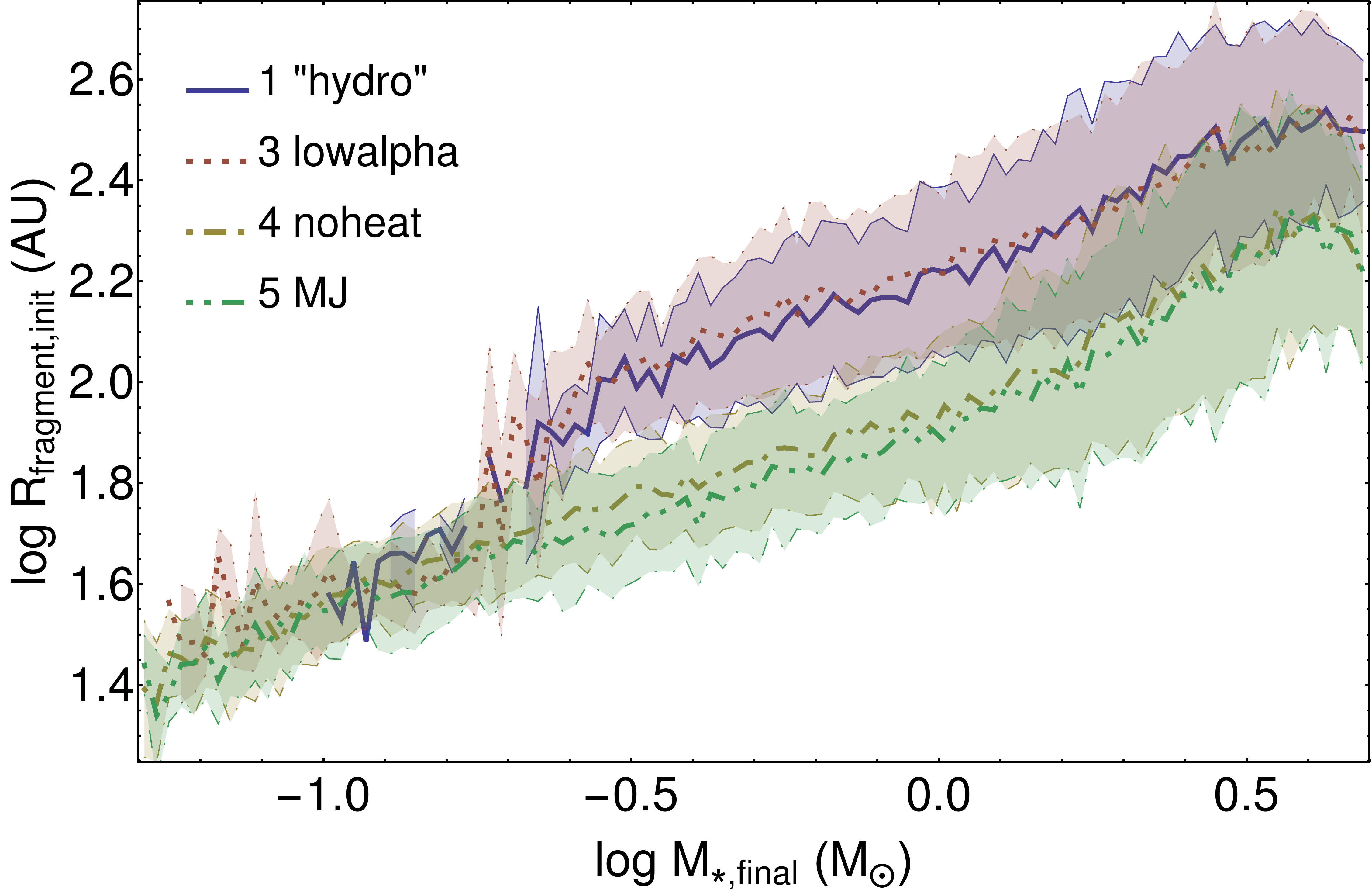}
  \end{subfigure}
  \caption{Fragmentation properties of all runs. Left column, from top to bottom: distribution of the number of fragments, mean initial fragment mass and mean fragmentation location. Right column: same properties as a function of final stellar mass.}
  \label{fig:frag}
\end{figure*}

Here, we focus on the fragmentation properties of the different runs as a whole. We discuss how many fragments are formed, their initial mass and the location in the disc at which the fragments form. We will refer to this location as ``fragmentation location''. The top left panel in Fig.~\ref{fig:frag}, shows the distribution of the number of fragments in fragmenting discs. RUN\nobreakdash-1 and RUN\nobreakdash-3 show a very similar behaviour. This is due to the very similar progress of the disc evolution during the infall phase, where fragmentation predominantly happens. Turning off the heating mechanisms gives rise to a strong increase in the number of fragments (RUN\nobreakdash-4). The number of fragments is more than quadrupled. Using the local jeans mass criterion for the initial fragment mass reduces the number of fragments by around a factor of $\approx 8$ (RUN\nobreakdash-5 compared to RUN\nobreakdash-4).

The mass distribution of the fragments (middle left panel of Fig.~\ref{fig:frag}) is very similar in RUN\nobreakdash-1 and RUN\nobreakdash-3, for the same reason as above. The fragments formed in RUN\nobreakdash-5 are much more massive initially, as expected from the different criterion. The mean mass is, however, not a factor $\num{25}$ larger than that in RUN\nobreakdash4, as one might naively expect from the ratio of $M_\mathrm{J,FR}$ to $M_\mathrm{f}$ (see Sect.
~\ref{ssub:mfrag}). Instead, the difference is only around a factor $\num{10}$. This is because the initial fragment mass is calculated self-consistently from the local conditions in the disc each time the conditions for fragmentation are satisfied. To illustrate this, let us consider two systems with identical initial conditions. The initial mass for the first fragment is indeed a factor of $\sim 25$ apart when the two different criteria are used. However, the subsequent evolution of the system is different and the initial fragment mass becomes self-limited also with the Jeans mass criterion.

The locations in the disc, where fragmentation happens, is depicted in the bottom left panel of Fig.~\ref{fig:frag}. The main difference here is between the runs with stellar accretion heating (RUN\nobreakdash-1, RUN\nobreakdash-3) and the runs without (RUN\nobreakdash-4, RUN\nobreakdash-5): the former fragment predominantly at radii $< \SI{100}{au}$, the latter at radii $\gtrsim \SI{100}{au}$. This shows again the importance of this heating mechanism. It also gives an explanation for the difference in mass discussed above: initial fragment masses are different (and typically lower) closer to the star.

\subsubsection{Fragmentation properties as a function of final stellar mass}

The fraction of fragmenting discs around stars $\la 0.1 \msun$ is low, in RUN\nobreakdash-1 and RUN\nobreakdash-3 in particular. (see Fig.~\ref{fig:ff}). Therefore the statistics in the corresponding mass bins is not very robust. This should be kept in mind when looking at the following figures.

The right column of Fig.~\ref{fig:frag} shows the mean number of fragments per disc, the mean initial fragment mass and the mean fragmentation location as a function of $M_\mathrm{*,final}$, respectively. Masses and locations are averaged twice. For example, first, the mean fragment mass in one system (producing multiple fragments) is computed. Then, the mean of these is calculated in each mass bin.
As seen in the figure, all three quantities increase with increasing final stellar mass. The dependency of the number of fragments $n_\mathrm{fragment}$ is strongest: $n_\mathrm{fragment} \sim M_\mathrm{*,final}^{5/4}$ (top panel). For the initial fragment mass, the dependency is weak, $M_\mathrm{fragment,init} \sim M_\mathrm{*,final}^{1/2})$ (middle panel) and for the fragmentation radius we have: $R_\mathrm{fragment,init} 
\sim M_\mathrm{*,final}$ (bottom panel).


\section{Comparison to previous studies and to observations}\label{Sect:comp}
Clearly, it is interesting to compare our results to other theoretical studies as well as to observations. However, as we are going to elaborate in this section, such a comparison is difficult. This is due to limitations in existing studies (including our own) as well as observational uncertainties and biases.
The population of stars resulting from our simulations agrees with the IMF by construction, as discussed (see Fig.~\ref{fig:mstar}). In the following we concentrate on the masses, radii and lifetimes of the discs as well as on their fragmentation.

\begin{figure*}[pt]
  \begin{subfigure}[pt]{0.49\textwidth}
  \includegraphics[width=\linewidth]{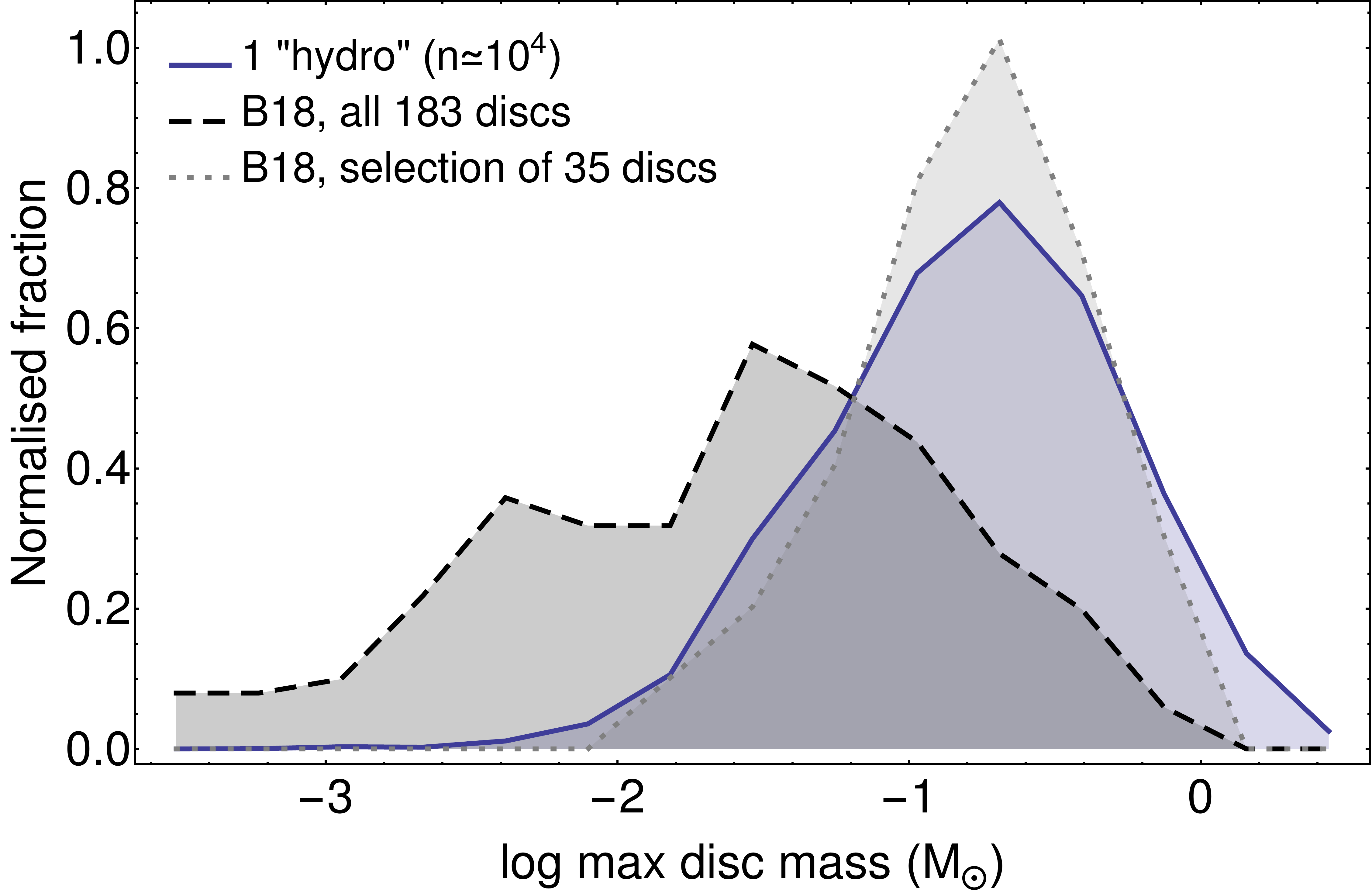}
  \end{subfigure}
  \begin{subfigure}[pt]{0.49\textwidth}
  \includegraphics[width=\linewidth]{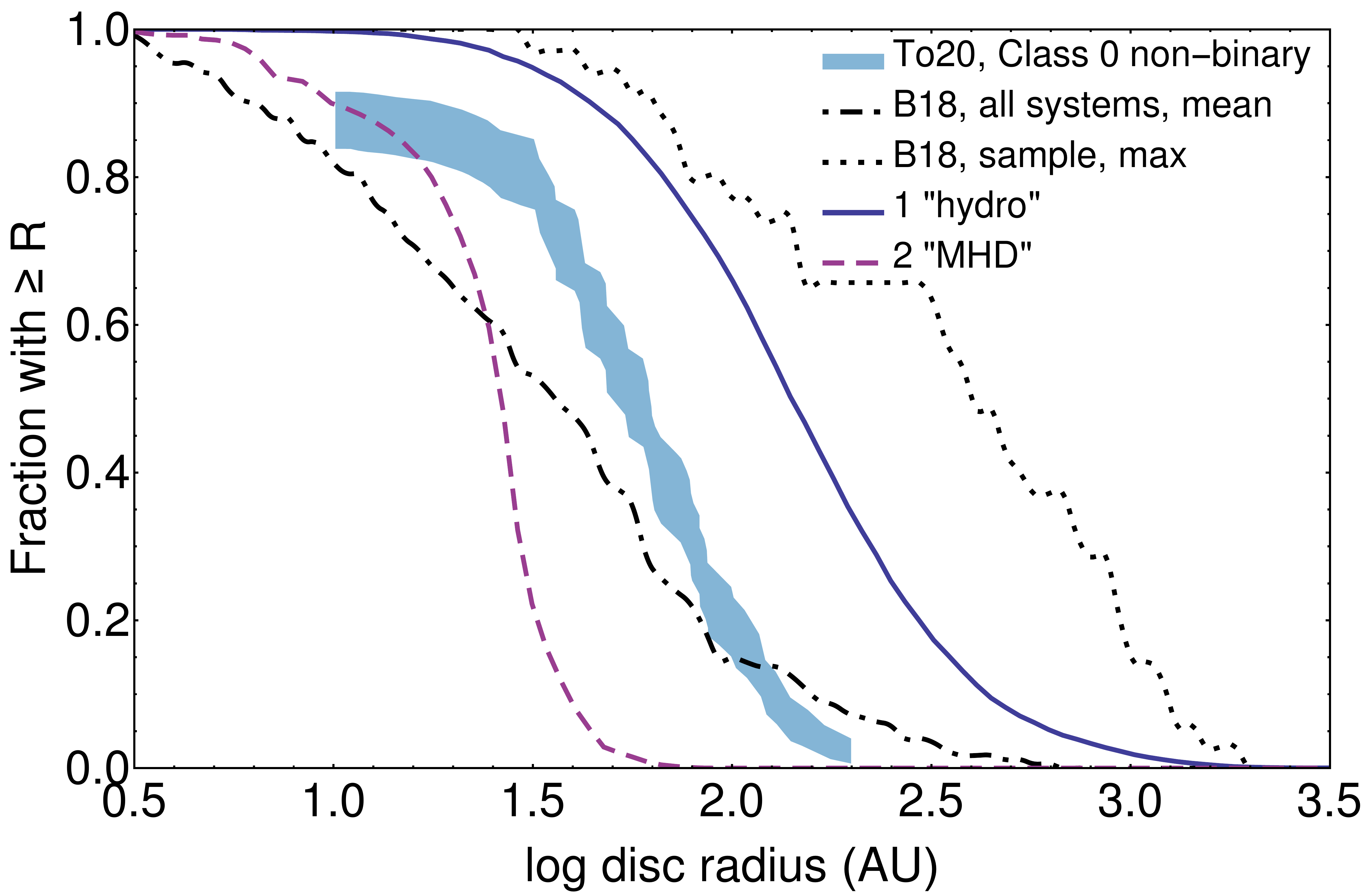}
  \end{subfigure}
  \begin{subfigure}[pt]{0.49\textwidth}
  \includegraphics[width=\linewidth]{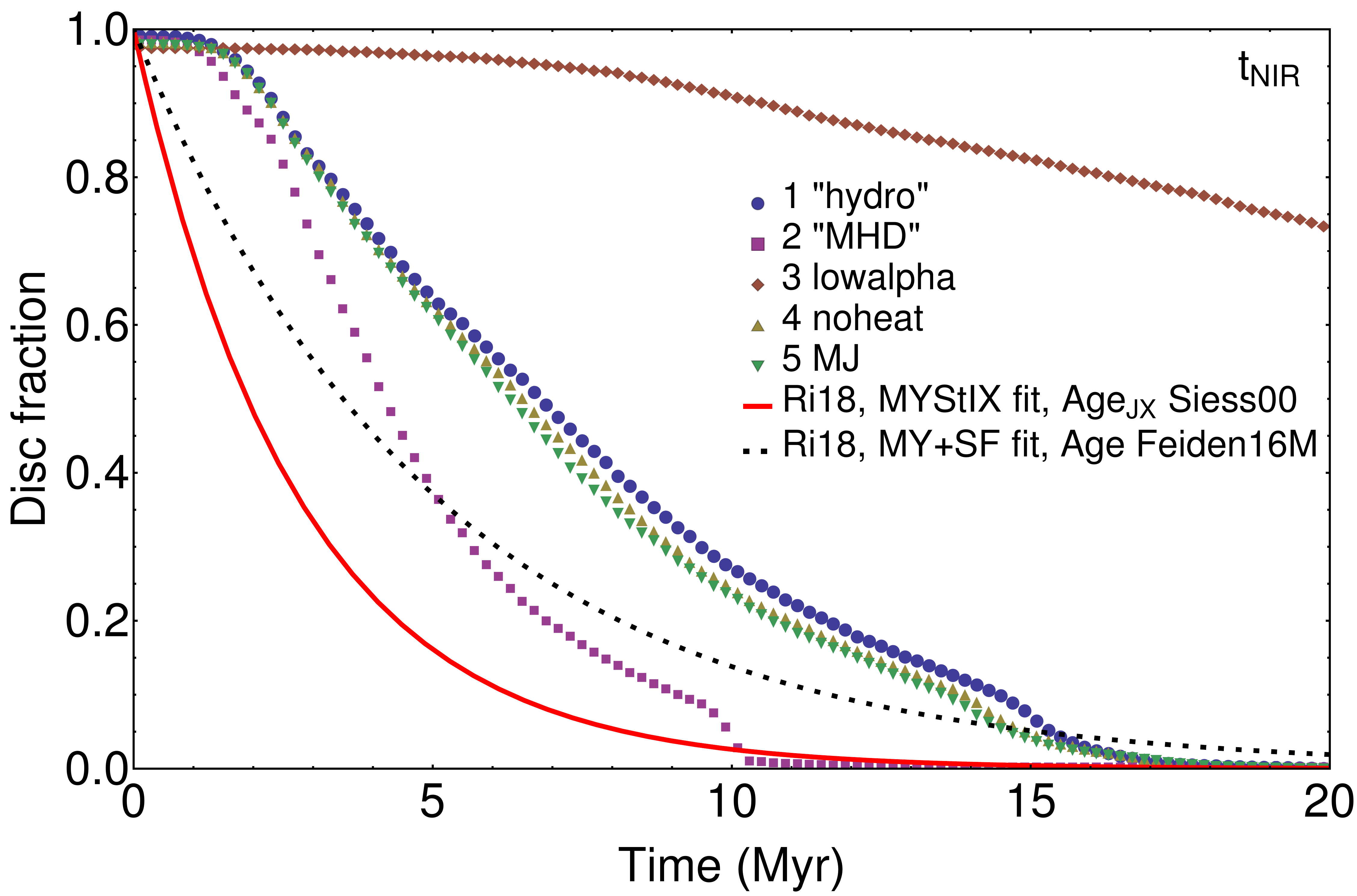}
  \end{subfigure}
  \begin{subfigure}[pt]{0.49\textwidth}
  \includegraphics[width=\linewidth]{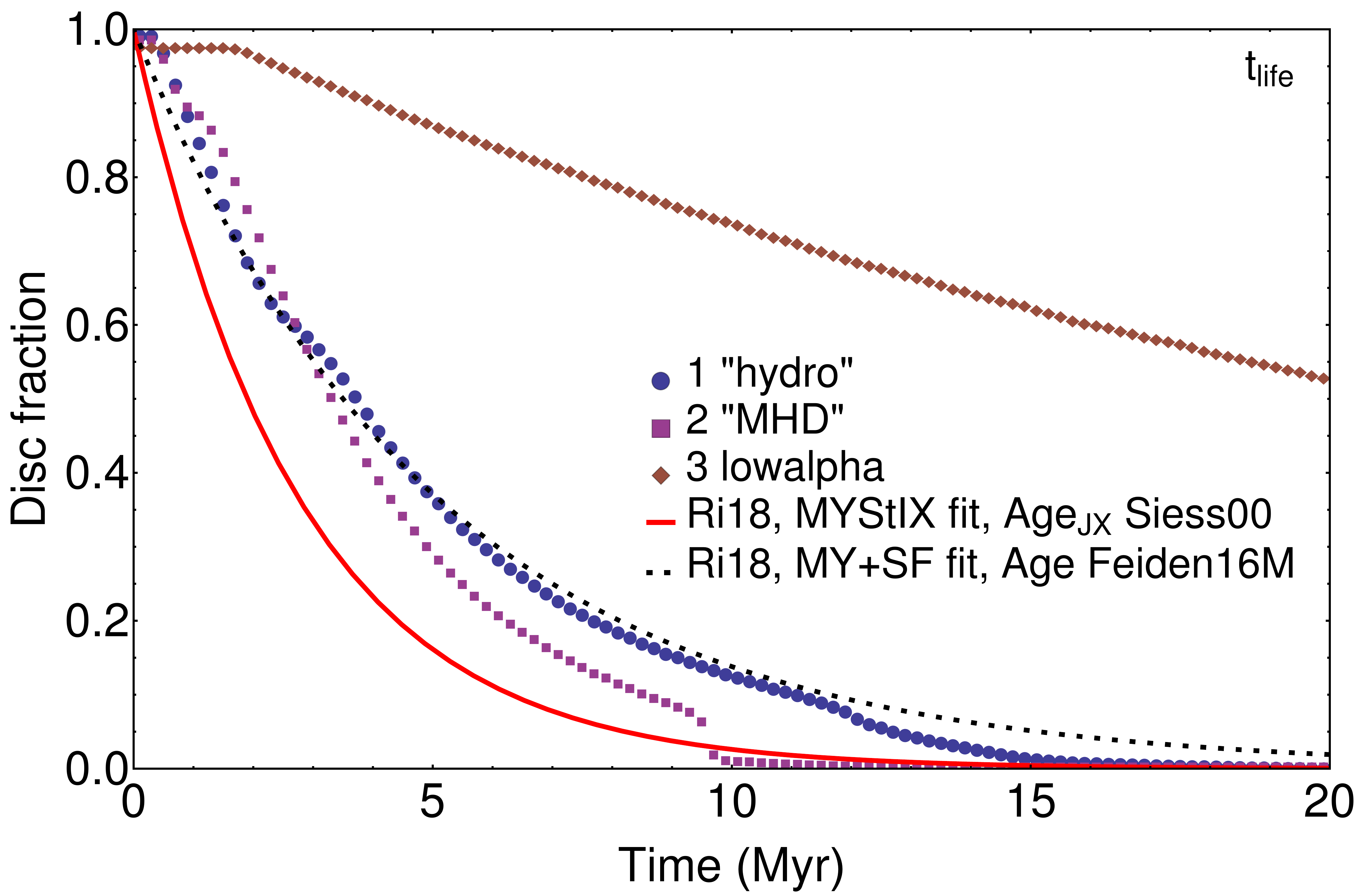}
  \end{subfigure}
  \caption{Comparisons of, top left: the maximum disc masses from RUN\nobreakdash-1 with those from the simulations of \citetalias{2018MNRAS.475.5618B}, top right: the disc radii at the end of the infall phase with observed \citet[][To20]{2020ApJ...890..130T}  and simulated \citepalias{2018MNRAS.475.5618B} Class 0 radii, bottom: the disc fractions as a function of time with fits from \citet[][Ri18]{Richert2018}; left: $t_\mathrm{NIR}$, right: $t_\mathrm{life}$ (reduced, see Sect.
 ~\ref{subs:disp}).  RUN\nobreakdash-4 and RUN\nobreakdash-5 are indistinguishable from RUN\nobreakdash-1 in the bottom right panel and omitted.}
  \label{fig:comp}
\end{figure*}

\subsection{Disc masses}\label{subs:compdm}

The most obvious study to compare our disc masses to is \citetalias{2018MNRAS.475.5618B}. However, the results cannot be directly compared  since we only used a sample to obtain initial conditions for our simulations. Nevertheless, we compare the results to identify the differences.

The top left panel of Fig.~\ref{fig:comp} shows the disc masses at the end of the infall phase from RUN\nobreakdash-1, together with the max. disc mass in two sets of data from \citetalias{2018MNRAS.475.5618B}. The first set comprises all the 183 ``discs'' in the simulation, the second includes only the 35 discs we chose to construct our initial distributions from (see \S~~\ref{subsect:Initial}). The representation  of RUN\nobreakdash-1 in Fig.~\ref{fig:comp} contains exactly the same data as the top left panel of  Fig.~\ref{fig:disc_a}, but with a different binning more appropriate to the other data displayed.

In our simulations, the end of the infall phase is a well defined point in time: infall rates are constant at the beginning, and zero later. The disc mass is always maximum (or very close to) at the end of this phase. This simple picture does not hold for the simulations in \citetalias{2018MNRAS.475.5618B}. There, infall rates are not constant and systems can stop accreting and restart later. Furthermore, some systems are still accreting at the end of the simulation, so the corresponding disc masses are still increasing by an unknown amount.

The mean value of the maximum disc masses of the full set shown in Fig.~\ref{fig:comp} is $\SI{0.08}{\msun}$, a little less than a third of that from RUN\nobreakdash-1. In the other set, containing only the 35 selected discs, it is $\SI{0.23}{\msun}$, or about $80 \%$ of that from RUN\nobreakdash-1. Given the difficulties explained, and considering we are comparing results from a 3D SPH calculation to those from an axis-symmetric 1D viscous evolution model, the agreement seems very reasonable.

\citet{2018ApJS..238...19T} performed an observational analysis of dust emission as part of the VLA Nascent Disk and Multiplicity (VANDAM) survey. They present estimates for the distribution of the protostellar disc masses in Perseus, both for Class 0 and Class I phases. We compare their Class~0 data to our RUN\nobreakdash-1 and RUN\nobreakdash-2 in the top left panel of Fig.~\ref{fig:disc_a}. \citet{2018ApJS..238...19T} give a mean disc mass in Class 0 of $\SI{0.075}{\msun}$, about $70 \%$ of our RUN\nobreakdash-2 or a quarter of our RUN\nobreakdash-1.

The similarity of the observed disc masses with our RUN\nobreakdash-2 should not be overstated. First, our data shows the maximum masses the discs reach during their life, while the VANDAM survey probes discs with some (unknown) distribution of Class-0 ages. Second, our sample of discs is one that covers a large range of stellar masses, without any observational bias applied. As discussed in Sect.~\ref{ssec:dep}, the distribution of disc masses strongly depends on that of the stellar masses. The masses of embedded protostars are very difficult to measure, and are not given in \citet{2018ApJS..238...19T}. If, for instance, systems with final stellar masses larger than $\approx \SI{0.3}{\msun}$ are favoured in the survey, this is indicative of a distribution of disc masses more massive than that of our RUN\nobreakdash-2.

\subsection{Disc radii}\label{subs:compdr}

In RUN\nobreakdash-2 (``MHD''), we chose an infall location close to the star and constant in time. This is an attempt to study the influence of magnetic fields on the formation and evolution of protoplanetary discs. As a consistency check, we show our disc radii together with the analytic expression from \citet{Hennebelle2016} in the second panel in the left column of Fig.~\ref{fig:disc_2}.
\citet{Hennebelle2016} compare their analytic estimate for the early disc radius to a large number of 3D, non-ideal MHD collapse simulations. They find agreement within a factor of two. This range is also shown (black shaded region). The disc radii from RUN\nobreakdash-2, including uncertainty, lie inside this region across almost the entire range of stellar masses. Indeed, we chose the infall location specifically to produce this agreement. Nevertheless, the outcome is not self-evident. During the infall phase, the viscosity in the disc is very high and the disc radii grow very quickly even when the infall location does not progress outwards.

\citet{2020ApJ...890..130T} (hereafter To20) performed a multiwavelength survey of hundreds of protostars. They used dust continuum emission to measure Class 0 dust disc radii. In the right panel of their Fig.~11, they compare the radii of discs around non-multiple protostars to non-binary systems from \citetalias{2018MNRAS.475.5618B}. We plot their data along with the end-of-infall radii from RUN\nobreakdash-1 and RUN\nobreakdash-2 in the top right panel of Fig.~\ref{fig:comp}. We also show two sets of data extracted from the online material of \citetalias{2018MNRAS.475.5618B}. The dash-dotted line depicts the temporal mean of the full sample, while the dotted line depicts the maximum radius the discs from our sample (see Sect. \ref{subsect:Initial}) reach during the simulation.

The observed disc radii lie in the middle between RUN\nobreakdash-1 and RUN\nobreakdash-2. The two representations of radii from \citetalias{2018MNRAS.475.5618B} span a huge range of disc radii and seem compatible both with the observed radii and those from our RUN\nobreakdash-1, but not with RUN\nobreakdash-2. However, we would like to stress again how difficult this comparison is. The data from our runs comes from a very specific point in time, while the observed data has an unknown distribution of underlying ages. Also, the observations are from dust continuum emission while our data (along with that from \citetalias{2018MNRAS.475.5618B}) results from the simulation of pure gas discs. The distributions of gas and dust may or may not be similar at early times. Therefore, we do not conclude discs stemming from a magnetised collapse are too small (these, too, grow quickly in time), or those from hydrodynamic collapse too large. An educated conclusion on this matter would necessitate the knowledge of the relationship between the distribution of dust of different sizes and that of the gas. Furthermore the ages of the observations would have to be well constrained. The uncertainties in the observations are also large. Nevertheless, disc radii somewhere between those from our RUN\nobreakdash-1 and RUN\nobreakdash-2 seem the most plausible.
A promising path to learn from the comparison of observational data with theoretical models is to include the evolution of heavy elements in models such as our own and use that to predict observables.

\subsection{Disc lifetimes}\label{subs:complt}

There is another theoretical study relating collapsing MCCs to disc lifetimes: \citet{Li2016} investigate the dependence of mass, angular velocity and temperature of MCCs in solid rotation on disc lifetimes across a large parameter space. They assume the mass distribution of the MCCs (not that of the resulting stars) is given by the IMF by \citet{Kroupa2001} and find a characteristic time of $\SI{3.7}{Myr}$ for the exponential decay of the disc fraction. They use a lower value for $\alpha_\mathrm{bg}$ (\num{1.5e-3}), however their collapse model is based on the Shu collapse and produces much less massive discs than we see in our work.

\citet{Richert2018} performed a study of the longevity of inner dust discs using data from $\num{69}$ young clusters. The work is based on data from two projects: ``Massive Young Star-Forming Complex Study in Infrared and X-ray'' (MYStIX, \citealt{2013ApJS..209...26F}) and ``Star Formation in Nearby Clouds'' (SFiNCs, \citealt{2017ApJS..229...28G}). They determine the ages of the clusters by combining the empirical $Age_{JX}$ method (described in \citealt{2014ApJ...787..108G}) with different PMS evolutionary models. In particular, they use the stellar evolutionary models from \citet{Siess2000} and the `magnetic' model from \citet{Feiden2016} which considers magnetic inhibition of convection. The corresponding ages are called $Age_{JX}$-Siess00 and $Age$-Feiden16M, respectively. We compare the disc fraction as a function of time from our work with their results. The bottom left panel of Fig.~\ref{fig:comp} compares the disc fractions (based on our criterion for the disc lifetime, see Sect.~\ref{ssec:dep}) to two of their fits. We note that these fits are based on data points that only go from $\approx \num{0.5}-\num{4}\SI{}{Myr}$.
In said range our disc fractions are much higher. Also, our disc fractions cannot be reasonably fit by an exponential decay of the form $\exp(-t/\tau)$ for some characteristic time $\tau$.

We also show the disc fractions as a function of time based on $t_\mathrm{life}$ (see Sect.~\ref{ssec:dep}) in the bottom right panel of Fig.~\ref{fig:comp}. The results from RUN\nobreakdash-1 are (likely coincidentally) almost identical to the \citet{Richert2018}-fit to their combined MYStIX+SFiNCs sample based on $Age$-Feiden16M. This neither indicates that the $Age$-Feiden16M disc fractions are more accurate than those based on, for example, $Age_{JX}$-Siess00, nor that the reduction proposed by \citetalias{Kimura2016} is correct, nor that our disc fractions are correct. However it does make one important point: the choice of time zero may be crucial when comparing the lifetimes of protoplanetary discs from simulations with those from observations. We conclude that the mean total lifetime of protoplanetary discs may well be a factor of a few higher than the oft-cited 1--3~Myr.
Values for the background viscosity coefficient much lower than $\alpha_\mathrm{bg}=\num{e-2}$, as in our RUN\nobreakdash-3, are clearly not compatible with observations (though we note that some discs with lifetimes of several $\SI{10}{Myr}$ have been observed, for example \citealt{2020MNRAS.494...62L}).
There has been a debate in the literature about the dependency of disc lifetimes on the host stellar mass. Some authors find longer disc lifetimes around low mass stars (for example \citealt{2015A&A...576A..52R}), which would be in disagreement with our results. \citet{Richert2018} find no evidence for such a dependency. The choice of the photoevaporation model and/or the strength of the evaporation rates (we use a constant value) also has a strong influence on the disc lifetimes. An example of such a model is \citet{2019MNRAS.487..691P}. We will study the influence of the photoevaporation model on our results in future work.

\subsection{Fragmentation}\label{subs:compfrag}
The numbers of fragments we find in RUN\nobreakdash-1 and RUN\nobreakdash-3 are substantial: $\num{24}$ when averaged across the whole population and up to several hundred around stars $\gtrsim \SI{1}{\msun}$. The number even approaches $\sim \num{1000}$ in some cases when stellar accretion heating is ignored (RUN\nobreakdash-4). We expect this to be an overestimate. The reason is that we ignore further gas accretion of the clumps once they have formed. If gas accretion were allowed, the surface density in the discs would be reduced and further fragmentation inhibited. It is, however, not guaranteed that mass accretion reduces the number of fragments by orders of magnitude in all cases. First, accretion competes with migration and/or gap formation, which means that fragments may migrate away from the gravitationally unstable region of the disc more quickly than they accrete \citep{Muller2018a}. Second, fragments may also be tidally disrupted \citep{Boley2010,2010MNRAS.408L..36N}, returning some or all of their mass back to the disc. The number of fragments is lower by almost an order of magnitude in RUN\nobreakdash-5 compared to RUN\nobreakdash-4 because we use a much higher initial fragment mass (see Sect.~\ref{globfrag}). A possible interpretation of the difference between the two initial fragment masses is as follows: while $M_\mathrm{F}$ is a measure of the fragment mass precisely at the time when the conditions for fragmentation are first satisfied, $M_\mathrm{J,FR}$ can be thought of as the fragment mass after accreting disc mass for a few orbits. Realistically, no more than a few dozen fragments around stars with $\gtrsim \SI{1}{\msun}$ are expected. Nevertheless, this estimate is still higher than what is found in some high resolution hydrodynamic simulations that typically find $\sim 1-10$ fragments (for example \citealt{Hall2017}). Such simulations, however, are so computationally expensive that they can only be run for a few~$\SI{}{kyr}$ ($\sim \num{1}$ orbit at $\SI{100}{au}$). Therefore, this could be one ``burst'' of fragmentation. Our discs are evolved for much longer and are replenished in mass by infall for several $\SI{10}{kyr}$ in the case of more massive systems, so many fragmentation ``bursts" are possible. Another difficulty with studying fragmentation in hydrodynamic simulations is that  fragmentation is only seen at sufficient numeric resolution \citep{2020arXiv200813653A}. \citetalias{2018MNRAS.475.5618B} find only $\num{10}$ out of the studied $\num{183}$ systems fragment. However, while the number of SPH particles in their simulation is very high ($\num{3.5e7}$), the simulation covers an entire star cluster, and the number of SPH particles in the discs is clearly insufficient to accurately study fragmentation.  \citet{2020arXiv200813653A} perform simulations of a forming massive star using a grid-based self-gravity-radiation-hydrodynamics simulations. Around $\num{26}$ million grid cells are used to study the disc physics. This represents one of the highest-resolution simulations on disc fragmentation performed so far. In their higher resolution runs, several dozen fragments are found around massive stars. At lower resolution fragmentation is suppressed.

\section{Discussion}\label{sect:Discussion}

In this study, we make a number of simplifying assumptions: Protoplanetary discs are axisymmetric, discs consist exclusively of gas (except for the opacity in the temperature calculation), there is one main infall phase, and finally when a disc fragments, the initial fragment mass is removed from the disc and added to the star.
Furthermore, the initial conditions are generated based  on only a fraction of the $\num{183}$ systems from \citetalias{2018MNRAS.475.5618B}.

Assuming an axisymmetric disc means we cannot reproduce per se the spiral structure and global instability of the disc. While we can self-consistently determine the \emph{location} where fragmentation happens, we cannot do this for the number of fragments. Furthermore, we cannot model the stochastic nature of fragmentation \citep{2012MNRAS.421.3286P}. Instead, we rely on parametrisations (as described in Sect.~\ref{subsect:viscosity} and \ref{subsect:Fragmentation}). 
As a result, individual simulations cannot reproduce the 2D/3D behaviour of discs accurately at early times. However, using parametrisations of the 2D/3D results, we perform a parameter study, and these assumptions still allow us to study the formation and evolution of protoplanetary discs statistically and systematically. At later times, the discs' evolution is no longer dominated by gravitational instabilities and the discs are expected to be closer to rotational symmetry as seen in many observations. The influence of our assumption concerning the evolution of the fragments has already been discussed in Sect.~\ref{subs:compfrag}.
In choosing a sample of `well behaved' discs from \citetalias{2018MNRAS.475.5618B}, we may miss some discs that, despite going through a very chaotic formation process, may still form planetary systems. This limitation should be kept in mind for future studies of planet formation. It can be overcome when more advanced hydrodynamic studies of disc formation become available.
In summary, our assumptions do not prevent reasonable estimates of our main results: the distributions of disc masses and radii at the end of the infall phase, numbers and properties of fragments and disc lifetimes.

\section{Summary and conclusions}\label{sect:Conclusions}

We study the formation and evolution of protoplanetary discs from their emergence in a collapsing molecular cloud core to their dispersal after a few million years. 
We perform five runs using different assumptions, each consisting of $\num{10000}$ systems.
The systems are initialised by means of six parameters: initial stellar mass, initial disc mass, (constant) infall rate, disc radius and the rate at which the latter increases with time. 
Distributions for these parameters are obtained from a selection of systems from the hydrodynamic disc population synthesis by \citet{2018MNRAS.475.5618B}.

We include the influence of the disc's autogravitation on the angular frequency and scale height as well as transport of angular momentum by global instability, self-gravity and turbulent viscosity. Mass loss by photoevaporation is also included. Furthermore we perform one run in which we investigate the effect magnetic fields might have on the infall and disc formation (RUN\nobreakdash-2 ``MHD'').

When comparing our default run (RUN\nobreakdash-1) with RUN\nobreakdash-2, we find that this produces very different discs. RUN\nobreakdash-2 produces systems that typically have disc radii a factor of 6 smaller, disc masses a factor of 2.6 lower, and disc-to-star mass ratios that are a factor of 2.8 lower, with much narrower distributions. Systems in RUN\nobreakdash-2 also have disc lifetimes $\approx 40 \%$ lower. Furthermore, none of the systems in RUN\nobreakdash-2 fragment.

Fragmentation is, however, common in all other runs. We find that approximately half of the systems fragment when shock heating from disc material accreting on the star (accretion heating) heats the disc. If this effect is neglected, even three quarters of the systems fragment.
Clearly, based on our current results we cannot make any robust statement about the feasibility of planet formation via gravitational instability yet. 
If at least some discs form in a way similar to our runs based on hydrodynamical simulations, many of them must fragment and would provide numerous bound clumps. It would then seem conceivable that at least some of these could survive and grow to form giant planets similar, for example, to the ones observed in the HR8799 \citep{2008Sci...322.1348M} or GJ~3512 \citep{Morales2019} system. We plan to study in more detail what happens to fragments once they have formed in future work. 
In summary we find:
\newline
\newline
If protoplanetary discs are formed in agreement with hydrodynamic simulations \citepalias{2018MNRAS.475.5618B} as discussed in our work,
\begin{enumerate}[series=summary]
    \item they are massive early on: \SI{0.29\pm0.06}{\msun} or \num{1\pm0.3} times their host stellar mass;
    \item in systems with a final stellar mass of $\sim \SI{1}{\msun}$ they are even more massive: $\SI{0.7}{\msun}$ at the end of the infall phase;
    \item they remain massive (a few tenths of their host stellar mass) for a large fraction of their lives;
    \item they are large: $\SI{200\pm100}{au}$ at the end of the infall phase
    \item their total lifetimes are long: \SI{7.3\pm0.5}{Myr}, despite choosing a  high value of $\num{e-2}$ for the background viscosity $\alpha$-parameter;
    \item half of the systems fragment. Neglecting stellar accretion heating raises this fraction to three quarters;
    \item fragmentation potentially removes a lot of mass from the disc, either through numerous fragments, through a high initial fragment mass, or (not modelled) through subsequent accretion; and
    \item the final stellar mass has a strong influence on fragmentation, with more massive systems fragmenting more often and producing more fragments of higher mass.
    \end{enumerate}
    
If discs are instead formed in a way expected by a magnetised collapse (\citealt{Hennebelle2016}), 
\begin{enumerate}[resume=summary]
    \item they are almost a factor of three less massive, though their masses are still substantial: $\SI{0.1}{\msun}$ at the end of the infall phase;
    \item in systems with a final stellar mass of $\sim \SI{1}{\msun}$ they have even a mean mass of $\sim \SI{0.2}{\msun}$;
    \item they are a factor of six smaller: $\SI{36\pm7}{au}$ at the end of the infall phase;
    \item their lifetimes are$\sim 40 \%$ shorter: $\SI{4.5\pm0.1}{Myr}$
    \item fragmentation is suppressed completely.
\end{enumerate}

The comparison with masses and radii from observed young discs favours systems that are somewhere between the two extreme cases ``hydro'' and ``MHD'' studied here.
In all cases the discs are found to be massive for a relatively long time. This could explain the apparent lack of planet-forming material \cite[e.g.,][]{Manara2018}.
The discrepancy with observed lifetimes may in part be explained by the choice of `time zero'. 

\begin{acknowledgements}
We thank Lucio Mayer, Eduard Vorobyov, Ken Rice, Til Birnstiel and Matthew Bate for the insightful discussions. We also thank the anonymous referee for valuable comments. Calculations were performed on UBELIX (\url{http://www.id.unibe.ch/hpc}), the HPC cluster at the University of Bern. This work has been carried out within the framework of the National Centre of Competence in Research PlanetS supported by the Swiss National Science Foundation. The authors acknowledge the financial support of the SNSF. O.S., C.M.\ and G.-D.M.\ acknowledge the support from the Swiss National Science Foundation under grant BSSGI0$\_$155816 ``PlanetsInTime’’.
RH acknowledges support from SNSF grant \texttt{\detokenize{200020_188460}}. 
G.-D.M. acknowledges the support of the DFG priority program SPP~1992 ``Exploring the Diversity of Extrasolar Planets'' (KU~2849/7-1).
\end{acknowledgements}

\section*{ORCID iDs}
Oliver\,Schib\,\orcidicon{}\\ \url{https://orcid.org/0000-0001-6693-7910}\\
Nicolai\,Wenger\,\orcidicon{}\\ \url{https://orcid.org/0000-0002-8042-3796}\\
Gabriel-Dominique\,Marleau\,\orcidicon{}\\ \url{https://orcid.org/0000-0002-2919-7500}\\
Christoph\,Mordasini\,\orcidicon{}\\ \url{https://orcid.org/0000-0002-1013-2811}\\
Ravit\,Helled\,\orcidicon{}\\ \url{https://orcid.org/0000-0001-5555-2652}\\

\bibliographystyle{aa}
\bibliography{library.bib}

\begin{appendix}

\section{Photoevaporation}\label{app:evap}

Here we give the expressions for the sink terms used in our implementation of photoevaporation (see Sect.~\ref{subs:evap}).

External photoevaporation:
\begin{equation}
    S_\mathrm{ext}(r,t) =
    \begin{cases}
        S_\mathrm{wind} \left( 1 - \frac{1}{1+sm_\mathrm{ext}^{20}} \right),&\text{if } r > 0.1 \beta_\mathrm{M} r_\mathrm{mI}\\
        0, & \text{otherwise},
    \end{cases}
    \end{equation}
where
\begin{equation*}
    sm_\mathrm{ext} = \frac{r}{\beta_\mathrm{M} r_\mathrm{g,ext}},
\end{equation*}
$r_\mathrm{g,ext}(t) = G \mstar(t) / c_\mathrm{s,ext}^2$ the gravitational radius. We use a smoothing term for numerical reasons. We choose $c_\mathrm{s,ext}$ to be $\SI{2.5}{km.s^{-1}}$, $\beta_\mathrm{M} = 0.14$ and use $S_\mathrm{wind} = \SI{2.8e-8}{g.cm^{-1}.yr^{-1}}$ for all systems. This value corresponds to an evaporation rate of $\SI{e-8}{\msun.yr^{-1}}$ if the disc extends to $\SI{1000}{au}$ and is comparable to previous studies (e.g., \citealt{2003MNRAS.342.1139A,Mordasini2009}).

Internal photoevaporation:
\begin{equation}
    S_\mathrm{int}(r,t) = 2 c_\mathrm{s,int} d_r \mathrm{u} sm_\mathrm{int},
\end{equation}
where $c_\mathrm{s,int} = \SI{11.1}{km.s^{-1}}$,
\begin{equation*}
    d_\mathrm{r} = \num{1.8e4} (\mstar(t)/\msun)^{-0.25} \left(\frac{r_\mathrm{g,int}}{\num{e14}}\right)^{-1.5} \left(\frac{r}{r_\mathrm{g,int}}\right)^{-2.5}.
\end{equation*}
$sm_\mathrm{int}$ is a smoothing factor for which it holds:
\begin{equation}
    sm_\mathrm{int} =
    \begin{cases}
        1 - \left( 1 + \left(\frac{r}{0.14 r_\mathrm{g,int}}\right)^{20} \right) ^{-1}, & \text{if } r > r_\mathrm{int}\\
        0, & \text{otherwise},
    \end{cases}
\end{equation}
where $r_\mathrm{int} = 0.07 r_\mathrm{g,int}, r_\mathrm{g,int}(t) = G \mstar(t) / c_\mathrm{s,int}^2$.

\section{Infall location}\label{app:infloc}

Here we derive the relationship between $R_\mathrm{i}$ and $R_\mathrm{62.3}$ used in Sec~\ref{ssub:infloc}.

Consider a Keplerian disc with a surface density profile of the form $\Sigma_\mathrm{d}(r) = \Sigma_0 \left(\frac{r}{r_0}\right)^{-1}$, for some constants $\Sigma_0$ and $r_0$. The gas at radius $r$ will have specific angular momentum $j$:
\begin{equation}\label{app:infloc_j}
    j = r^2 \Omega_\mathrm{K},
\end{equation}
where $\Omega_\mathrm{K} = \sqrt{\frac{G \mstar}{r^3}}$ is the Keplerian angular frequency.
We can therefore calculate the mean specific angular momentum $\langle j_\mathrm{d} \rangle$ of this disc:
\begin{equation}\label{app:infloc_jd}
\begin{aligned}
	\langle j_\mathrm{d} \rangle = &\frac{2 \pi}{M_\mathrm{d,tot}}	\int_{R_\mathrm{in}}^{R_\mathrm{out}}\!r j \,\Sigma_\mathrm{d}(r)\,\mathrm{d}r \\
	     &\cong \frac{4 \pi \Sigma_0 \sqrt{G \mstar}}{3 M_\mathrm{d,tot}} R_\mathrm{out}^{3/2} \\
	     &\cong 2/3 \sqrt{G \mstar R_\mathrm{out}},
\end{aligned}
\end{equation}
where $R_\mathrm{in}$ and $R_\mathrm{out}$ are the disc's inner and outer edge, respectively. We used $R_\mathrm{in} \ll R_\mathrm{out}$ and the disc's total mass $M_\mathrm{d,tot} \cong 2 \pi r_0 \Sigma_0 R_\mathrm{out}$ above.

If all the disc material were concentrated at one radius, this would be the radius characterised by a specific angular momentum $\langle j_\mathrm{d} \rangle$. We choose this radius to be our infall radius $R_\mathrm{i}$. Combining (\ref{app:infloc_j}) and (\ref{app:infloc_jd}) yields:
\begin{equation}
    R_\mathrm{i} = \frac{4}{9} R_\mathrm{out}.
\end{equation}
Now we need to link $R_\mathrm{out}$ to $R_{63.2}$. We note that for the mass as a function of the radius in the disc it holds:
\begin{equation}
    M_\mathrm{d}(r) \cong 2 \pi \Sigma_0 r_0 r
\end{equation}
for $r \gg R_\mathrm{in}$.
For the mass contained inside of $R_{63.2}$ it must therefore hold:
\begin{equation}
\begin{aligned}
    2 \pi \Sigma_0 r_0 R_{63.2}
    &\cong M(R_\mathrm{63.2})
    \equiv 0.632~M_\mathrm{d,tot} \\
    &\cong0.632 \times 2 \pi \Sigma_0 r_0 R_\mathrm{out}
\end{aligned}
\end{equation}
which leads to:
\begin{equation}
    R_\mathrm{out} \cong R_{63.2} / 0.632
\end{equation}
Plugging this result into (\ref{app:infloc_j}) we finally find:
\begin{equation}
    R_\mathrm{i} \cong 0.7~R_{63.2}.
\end{equation}
\section{More examples of system evolution}\label{app:example}

\begin{figure*}[pt]
  \begin{subfigure}[pt]{0.49\textwidth}
  \includegraphics[width=\linewidth]{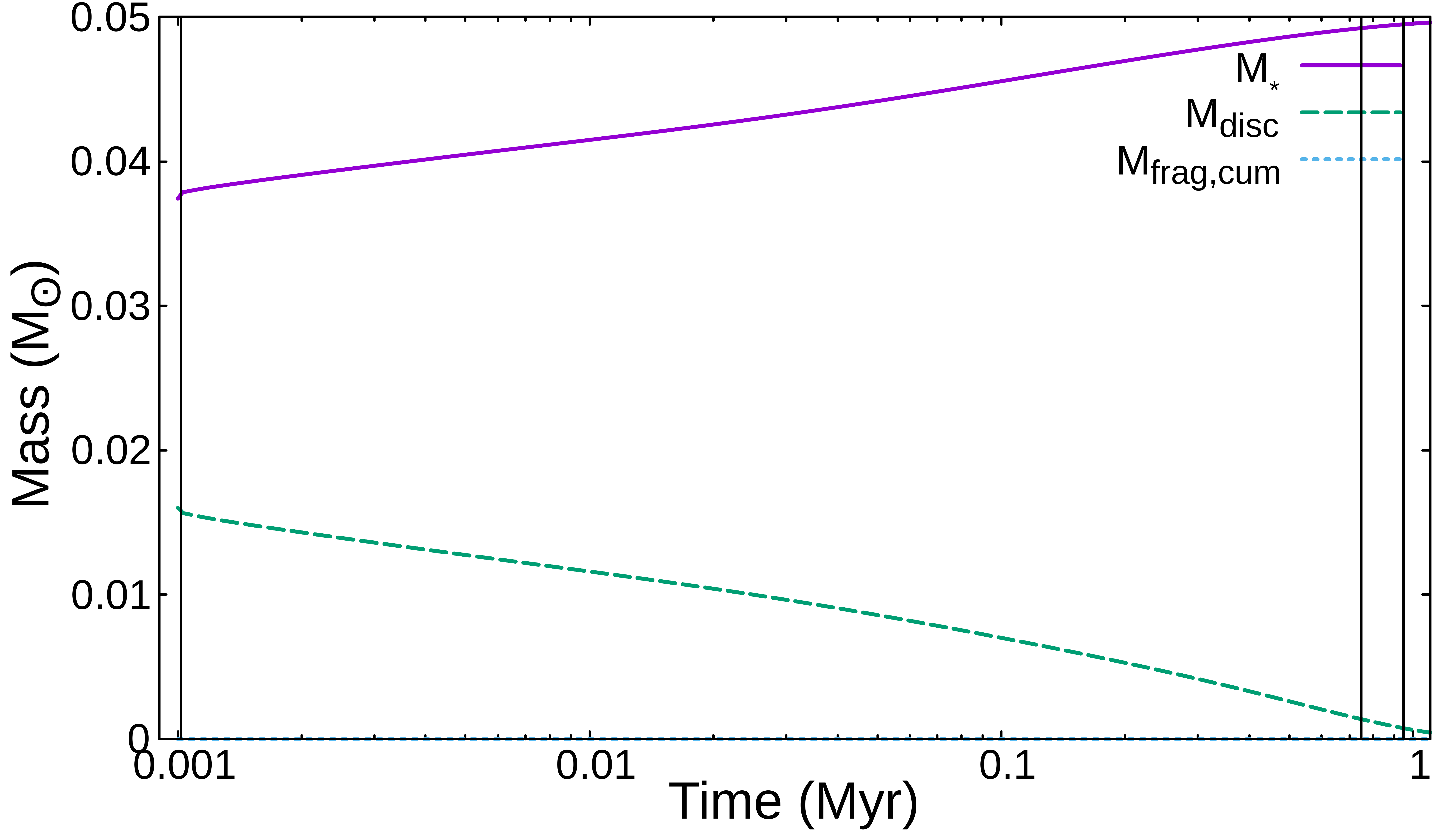}
  \end{subfigure}
  \begin{subfigure}[pt]{0.49\textwidth}
  \includegraphics[width=\linewidth]{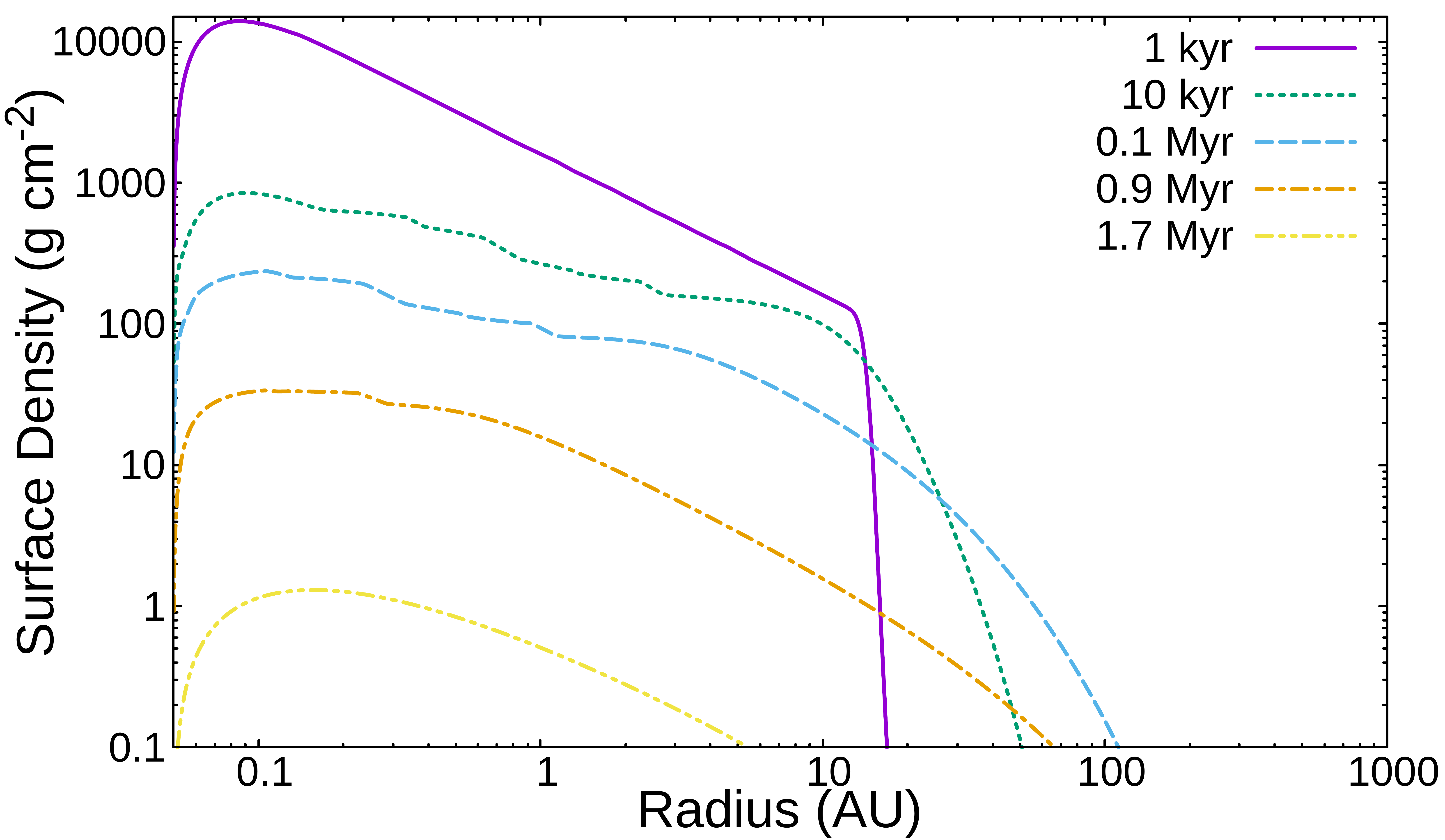}
  \end{subfigure}
  \begin{subfigure}[pt]{0.49\textwidth}
  \includegraphics[width=\linewidth]{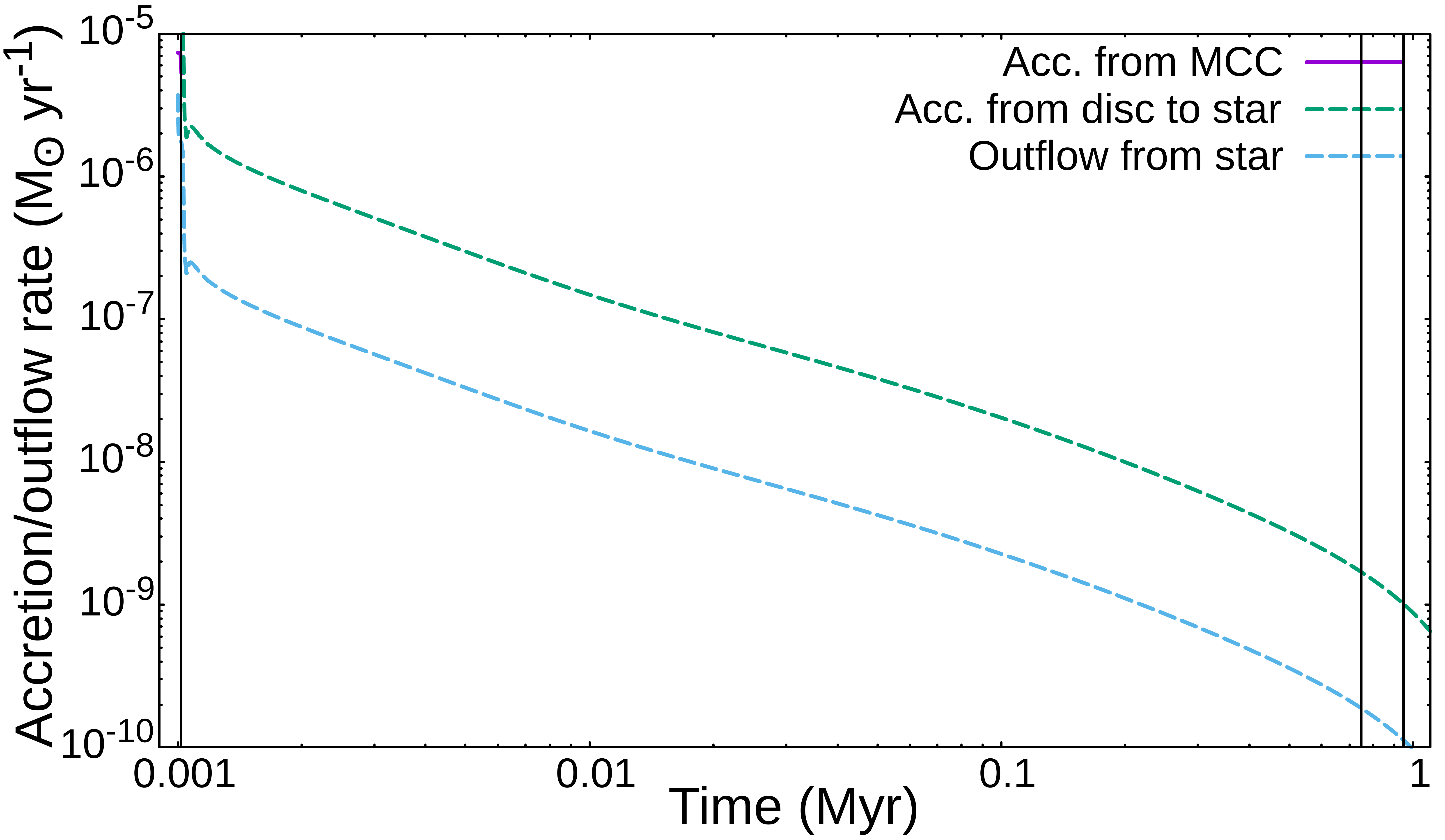}
  \end{subfigure}
  \begin{subfigure}[pt]{0.49\textwidth}
  \includegraphics[width=\linewidth]{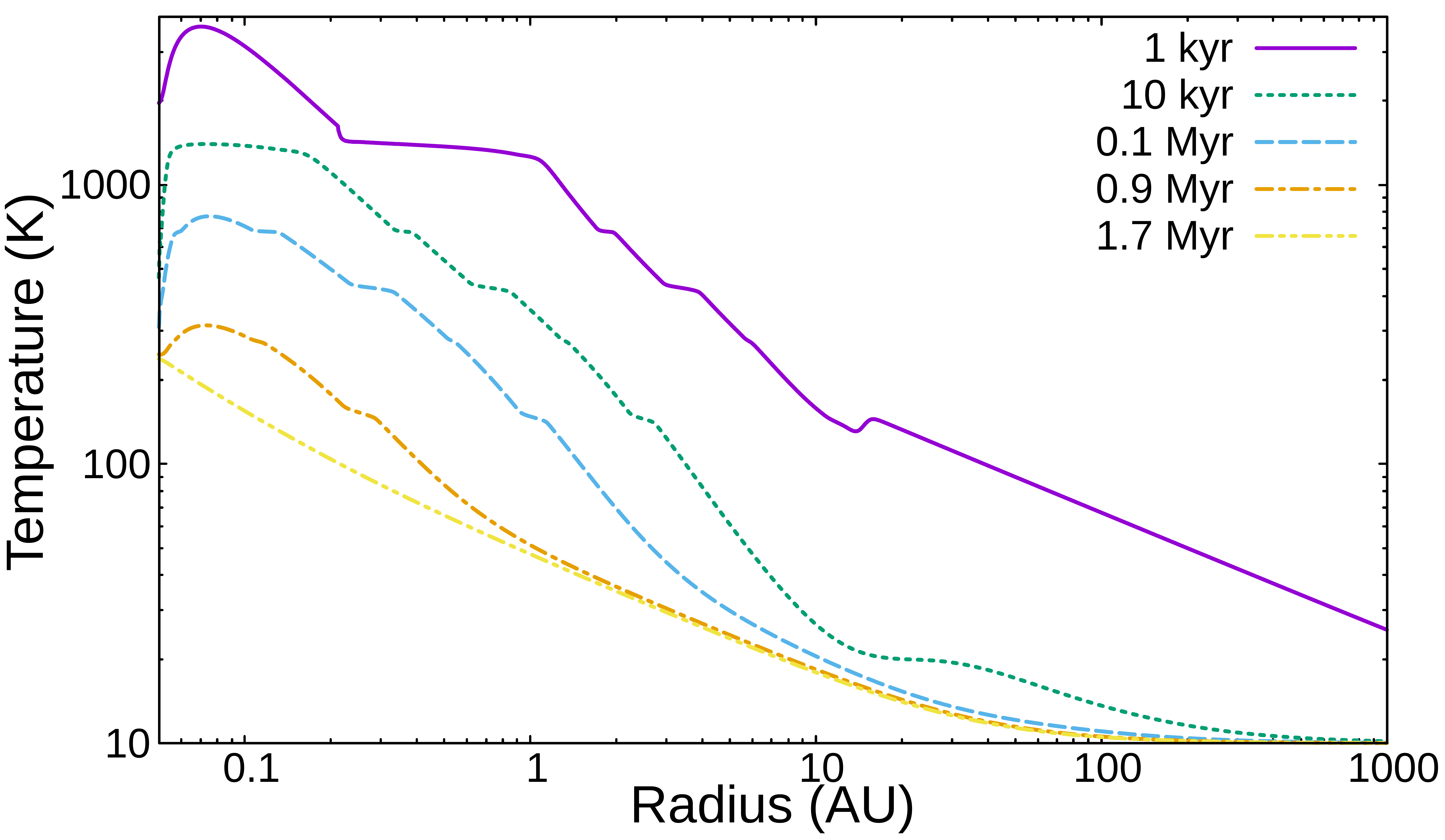}
  \end{subfigure}
  \begin{subfigure}[pt]{0.49\textwidth}
  \includegraphics[width=\linewidth]{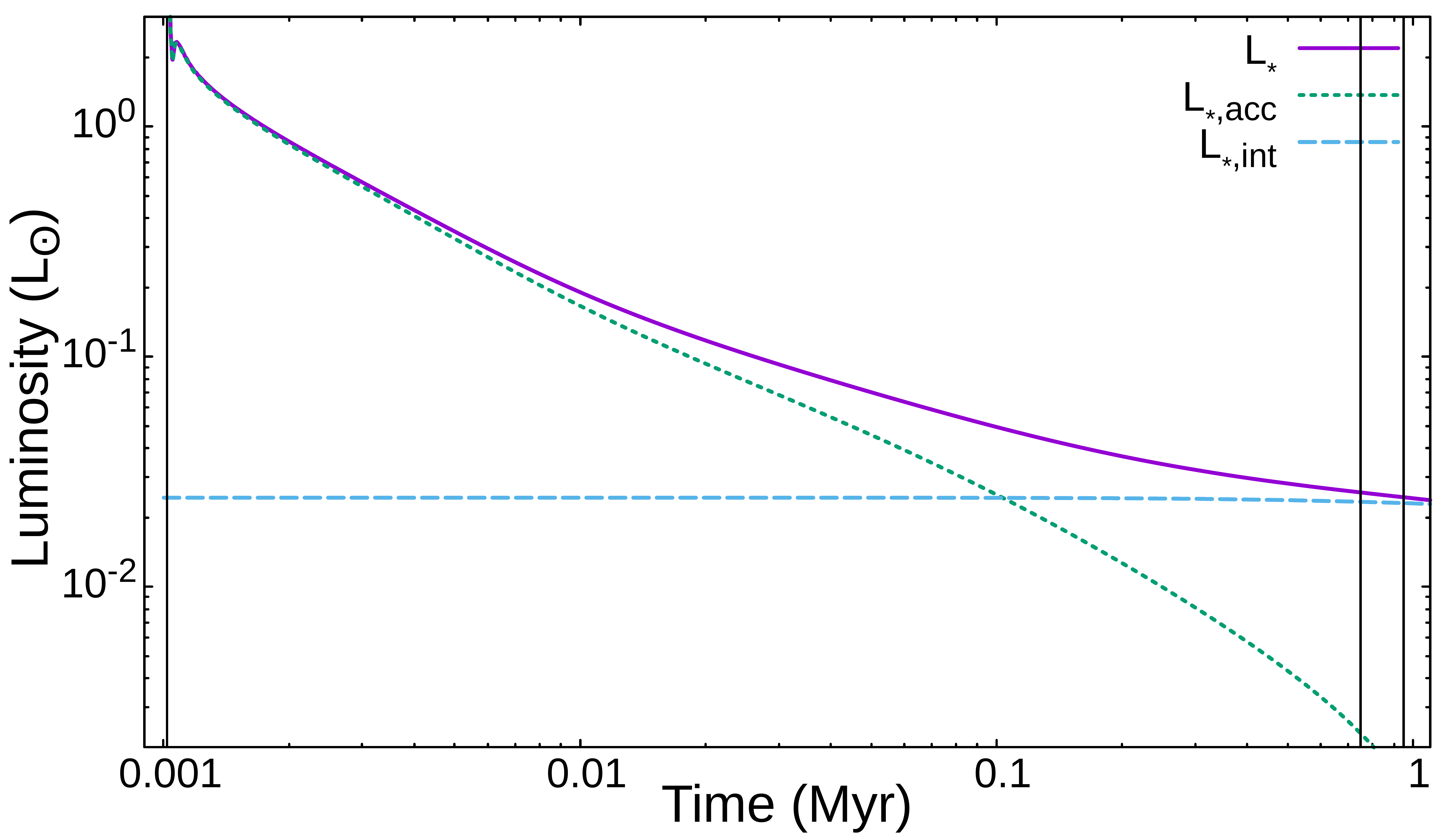}
  \end{subfigure}
  \begin{subfigure}[pt]{0.49\textwidth}
  \includegraphics[width=\linewidth]{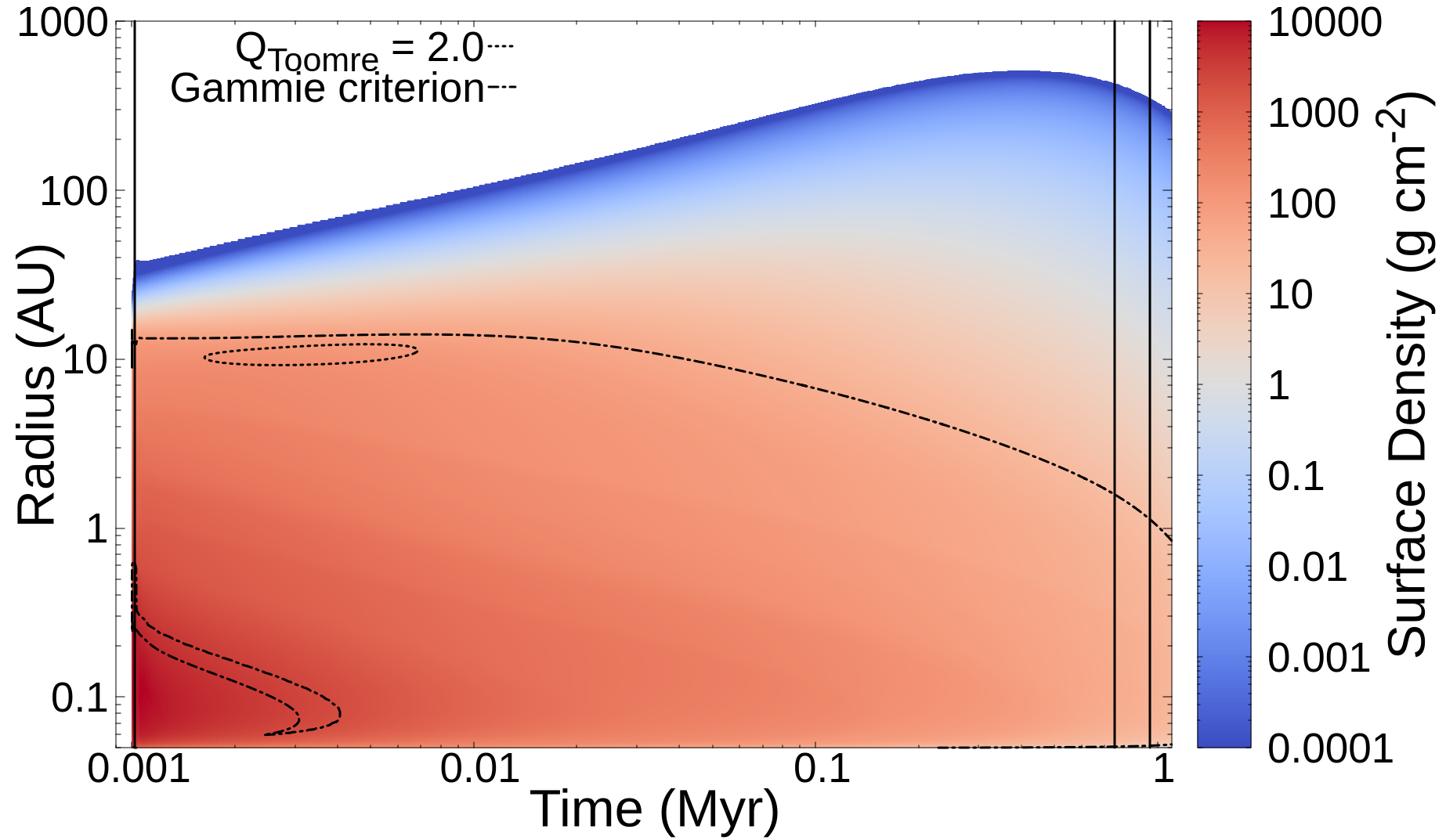}
  \end{subfigure}
  \caption{Time evolution of system 0004 from RUN\nobreakdash-2 ``MHD''. Top left panel: stellar mass, disc mass and cumulative mass removed from the disc by fragmentation. Middle left panel: accretion and outflow rates. Bottom left panel: stellar luminosity. Top right panel: surface density at different times. Middle right panel: midplane temperature at different times. Bottom left panel: stellar luminosity (accretion and intrinsic). Bottom Right panel: contour plot of the surface density with fragmentation criteria (see main text). The black vertical lines denote, in order of increasing time:
  $t_\mathrm{infall}\approx\SI{1.02}{kyr}$, $t_\mathrm{pms}\approx\SI{0.75}{Myr}$ and $t_\mathrm{NIR}\approx\SI{0.95}{Myr}$ (see Sect.~\ref{sec:example}).}
  \label{fig:example2}
\end{figure*}

Here we show two additional examples for the temporal evolution of star-and-disc systems (compare to Section~\ref{sec:example}). Figure~\ref{fig:example2} shows the evolution of system 0004 from RUN\nobreakdash-2 (``MHD''), an example of a very low mass system. The simulation starts with a $\SI{3.8e-2}{\msun}$ star and a $\SI{1.5e-2}{\msun}$ disc (top left panel). The stellar mass reaches $\SI{0.05}{\msun}$ at $t_\mathrm{NIR}$ ($\approx\SI{0.95}{Myr}$): a brown dwarf at the lower mass end considered in our simulations.
Nevertheless, the surface density (hence also the temperature, right top and middle panels) is already high ($\gg \SI{1000}{g cm^{-2}}$) at early stages. This is a consequence of the disc being very compact $(\sim\SI{10}{au}$ initially). The stellar luminosity is dominated by accretion during the first $\SI{100}{kyr}$ and decreases steadily after the end of the infall phase (as does the accretion rate, left bottom and middle panels). $Q_\mathrm{Toomre}$ never drops much below $\num{2}$ (right bottom panel), as a result  the disc does not fragment and gravitational instability plays a minor role in the system's evolution.

\begin{figure*}[pt]
  \begin{subfigure}[pt]{0.49\textwidth}
  \includegraphics[width=\linewidth]{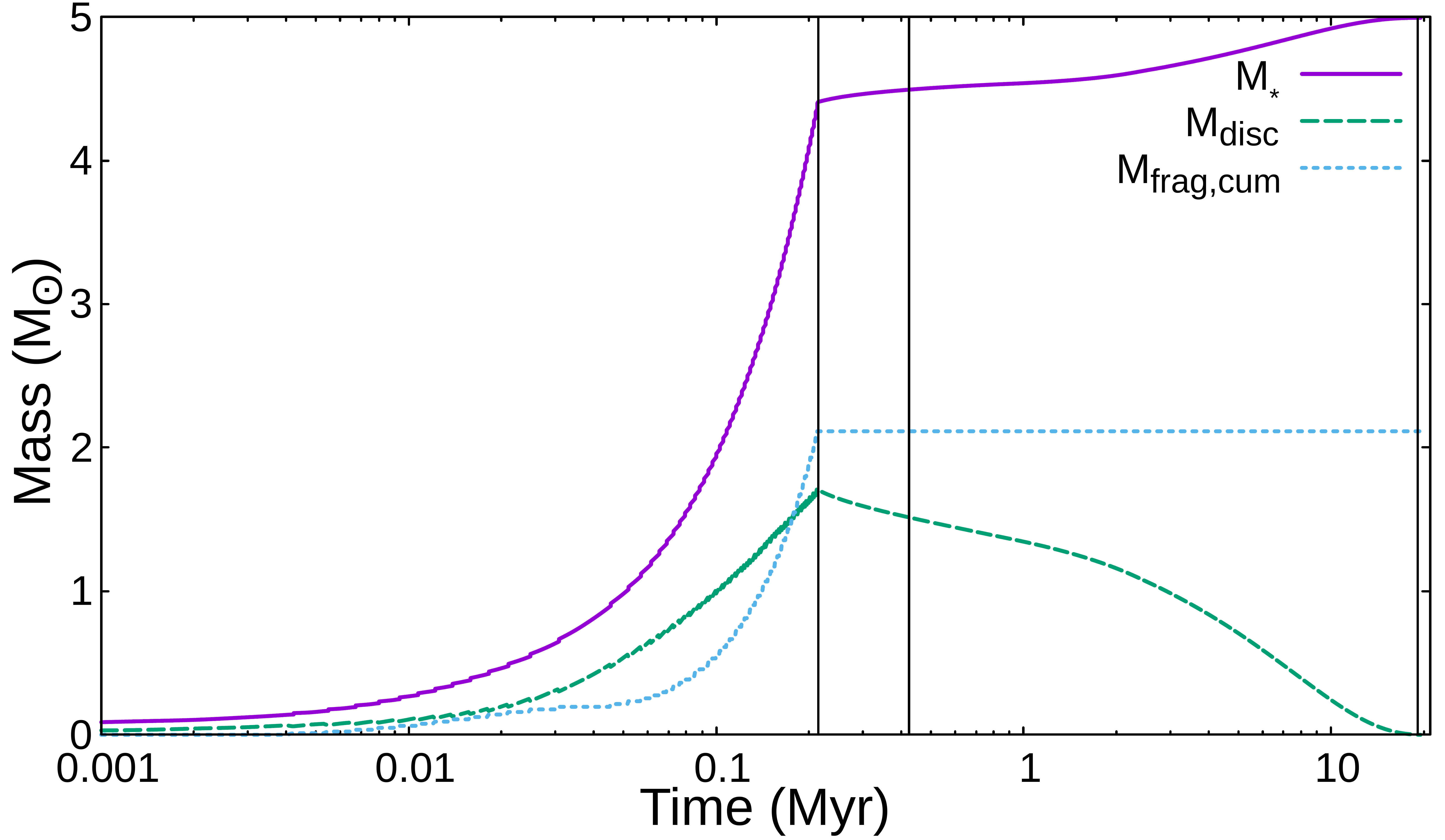}
  \end{subfigure}
  \begin{subfigure}[pt]{0.49\textwidth}
  \includegraphics[width=\linewidth]{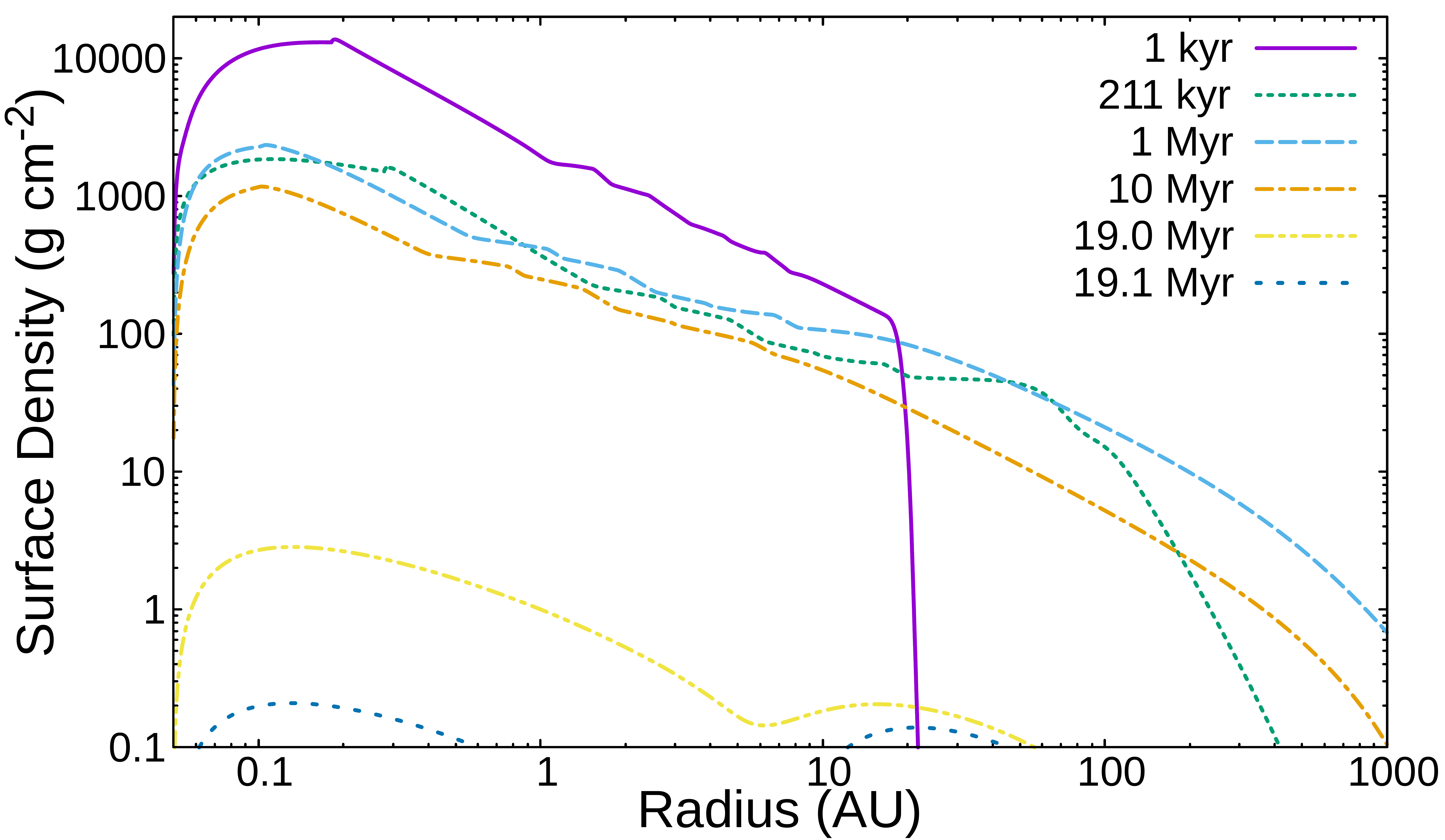}
  \end{subfigure}
  \begin{subfigure}[pt]{0.49\textwidth}
  \includegraphics[width=\linewidth]{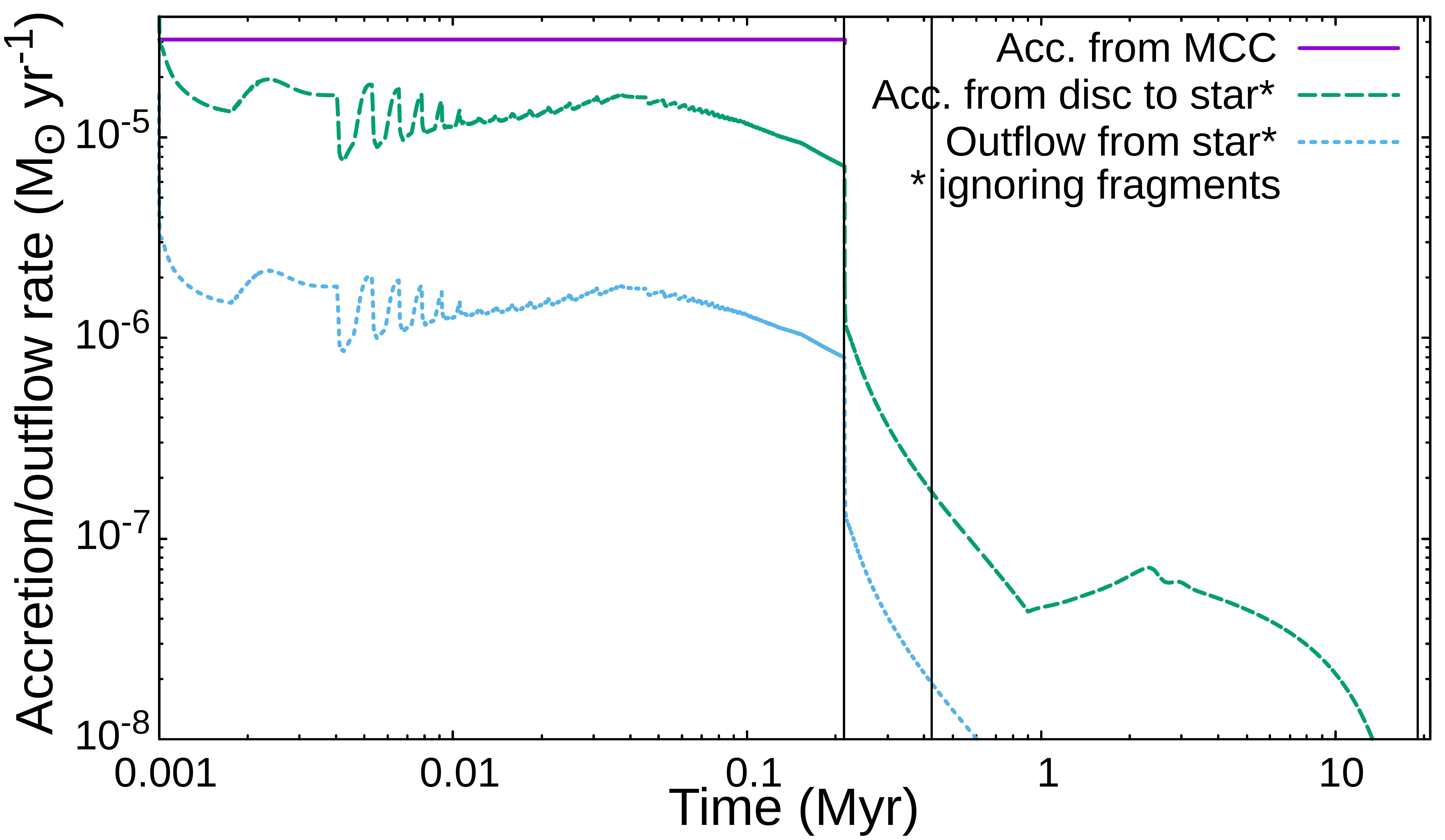}
  \end{subfigure}
  \begin{subfigure}[pt]{0.49\textwidth}
  \includegraphics[width=\linewidth]{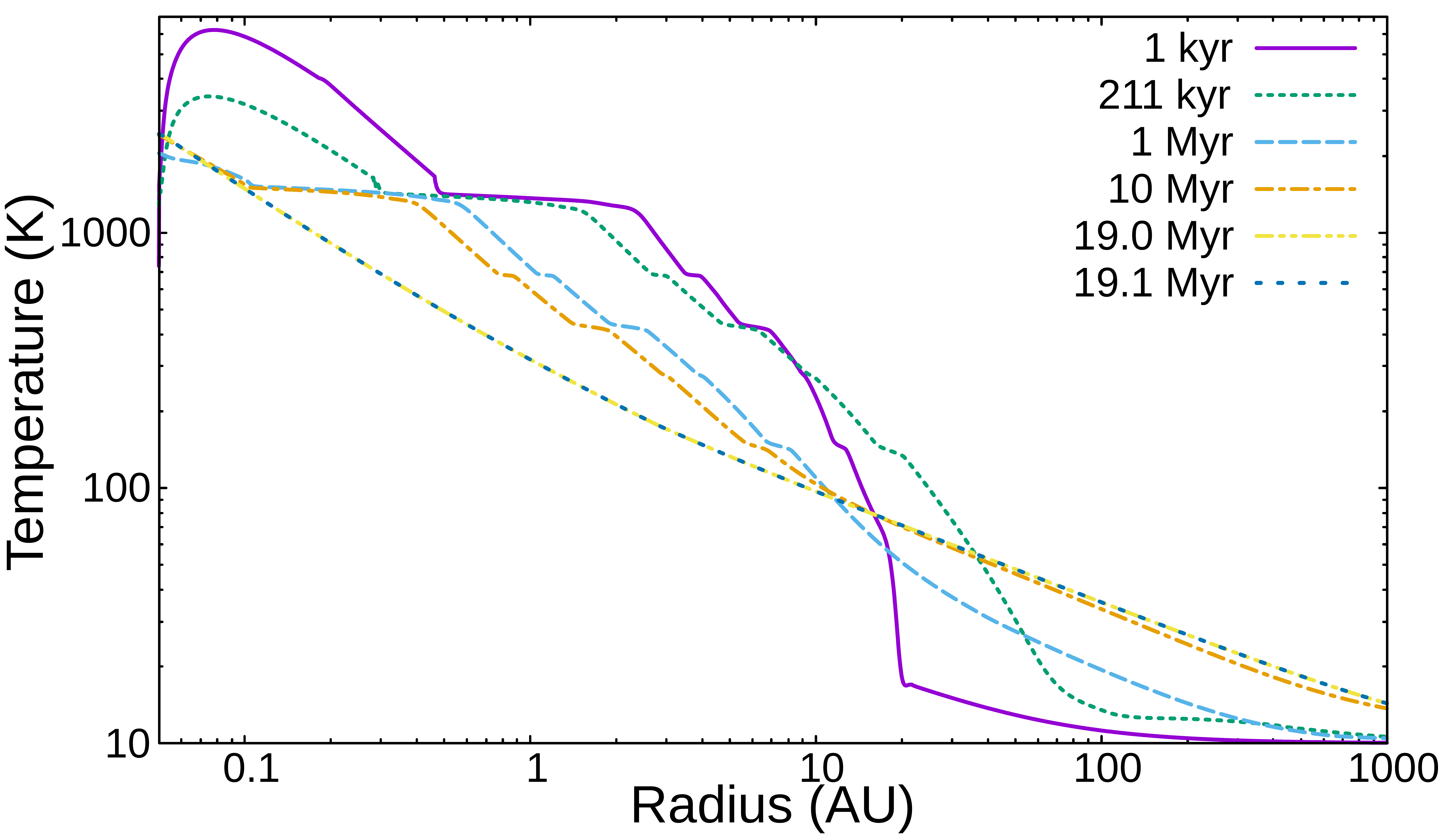}
  \end{subfigure}
  \begin{subfigure}[pt]{0.49\textwidth}
  \includegraphics[width=\linewidth]{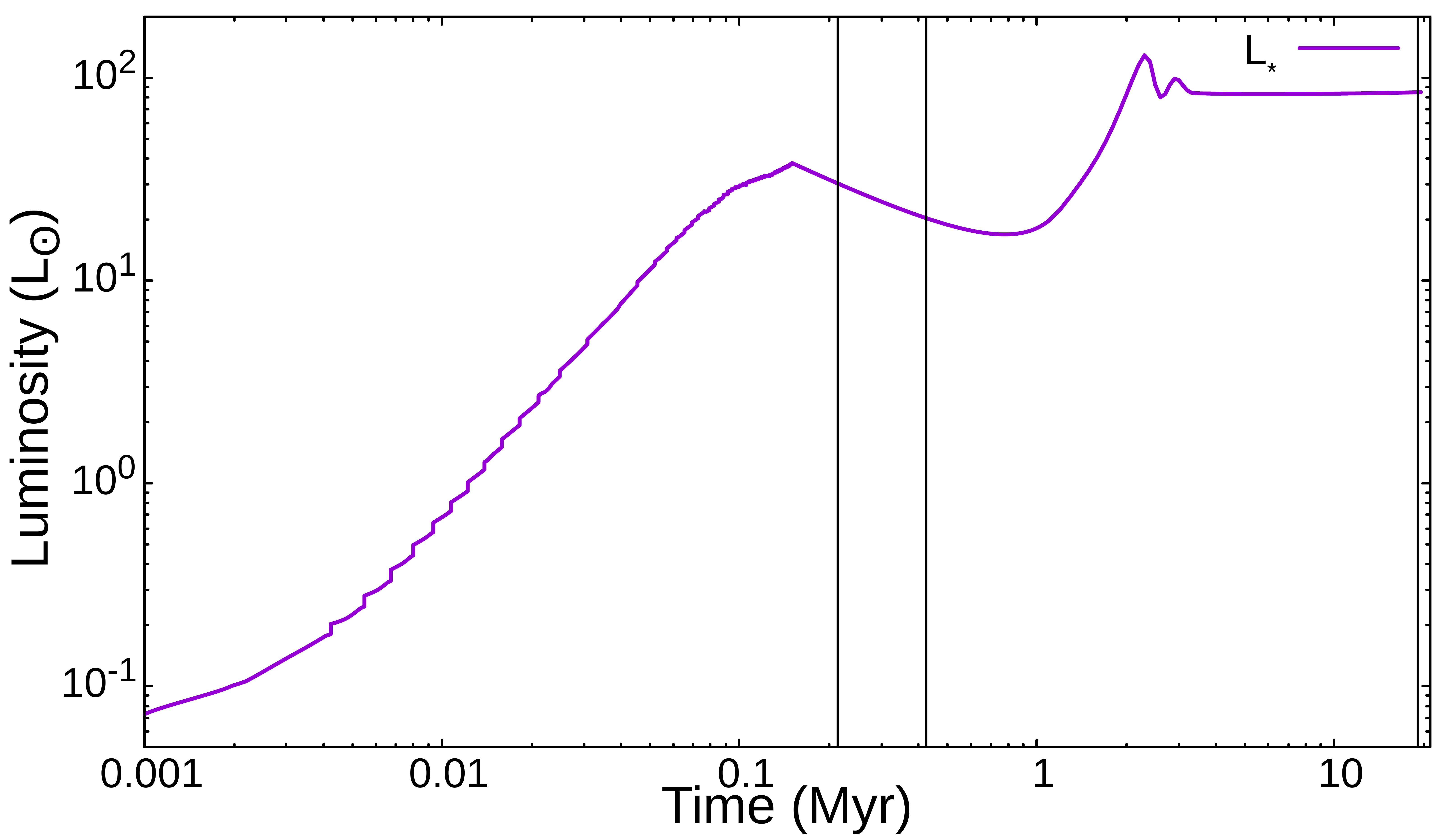}
  \end{subfigure}
  \begin{subfigure}[pt]{0.49\textwidth}
  \includegraphics[width=\linewidth]{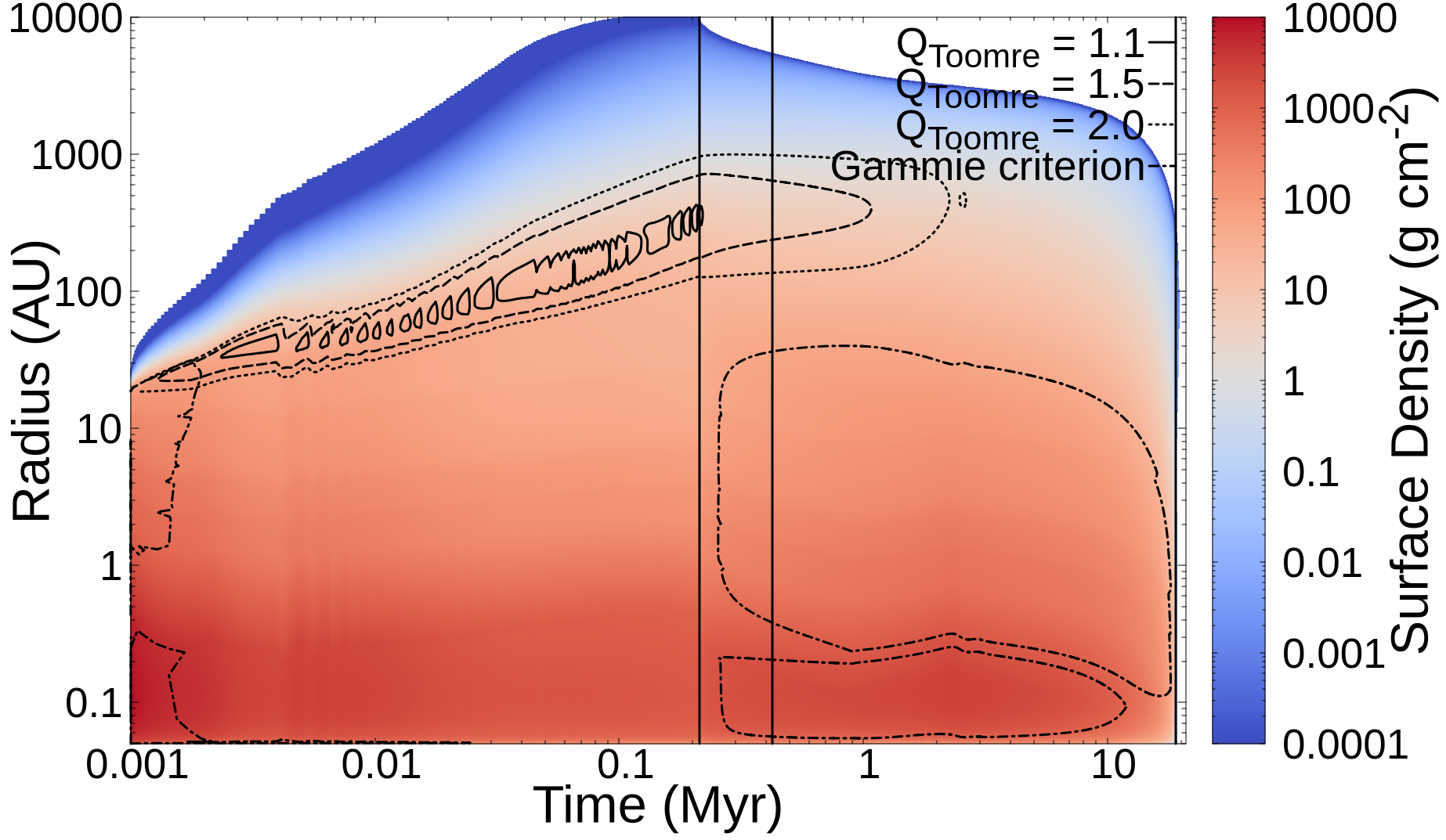}
  \end{subfigure}
  \caption{Time evolution of system 9992 from RUN\nobreakdash-5. Top left panel: stellar mass, disc mass and cumulative mass removed from the disc by fragmentation. Middle left panel: accretion and outflow rates. Bottom left panel: stellar luminosity. Top right panel: surface density at different times. Middle right panel: midplane temperature at different times. Bottom left panel: stellar luminosity (no contribution from accretion considered in this run). Bottom right panel: contour plot of the surface density with fragmentation criteria (see main text). The black vertical lines denote, in order of increasing time:
  $t_\mathrm{infall}\approx\SI{215}{kyr}$, $t_\mathrm{pms}\approx\SI{425}{kyr}$ and $t_\mathrm{NIR}\approx\SI{19.1}{Myr}$ (see Sect.~\ref{sec:example}).}
  \label{fig:example3}
\end{figure*}

Figure~\ref{fig:example3} displays the evolution of system 9992 from RUN\nobreakdash-5 (``MJ''), an example of a massive system. Initially it consists of a $\approx \SI{0.09}{\msun}$ star and a $\approx \SI{0.03}{\msun}$ disc. At the end of the simulation ($t_\mathrm{NIR} \approx \SI{19}{Myr}$) the stellar mass is $\SI{5}{\msun}$, the upper limit in our study (top left panel). 
The MCC is feeding the disc with material for $\SI{200}{kyr}$. 
The disc fragments $\num{72}$ times during the infall phase, but not afterwards. This is because  in RUN\nobreakdash-5 no accretion heating is considered, which facilitates fragmentation during and prevents a drop in temperature after the infall phase. Furthermore, $M_\mathrm{J,FR}$ is used as the initial fragment mass in this run. Therefore, fragmentation is not limited by the initial fragment mass. Approximately $\SI{2.1}{\msun}$ of mass is removed from the disc by fragmentation. This is substantial: almost one third of the total system mass. Like in the example from Sect.~\ref{sec:example}, the Gammie criterion never limits fragmentation (right bottom panel). The accretion rate of disc material onto the star (left middle panel) shows an interesting oscillating feature between $\num{4}$ and $\SI{10}{kyr}$. This is not caused by the accretion of clumps onto the star (not included in the figure) but by a decrease in accretion rate that starts when the conditions for fragmentation are satisfied somewhere in the disc. A substantial amount of mass ($\approx \SI{10}{\mj}$ or around $\SI{10}{\%}$ of $M_\mathrm{disc}$ at this time) is removed from the disc during a free-fall time (see Sect.~\ref{ssub:mfrag}). This reduces the accretion from disc to star. Surface densities and temperatures (right top and middle panels) are significantly higher in this system than in the first example (Sect.~\ref{sec:example}) due to the higher stellar mass and disc mass.

\section{Infall radii in RUN-2}\label{app:infallradii}

Table~\ref{table:infrad} gives the infall radii used in each stellar mass bin for RUN\nobreakdash-2 (see Sect.~\ref{sec:param}).

\begin{table}[ht]
\begin{tabular}{llll|llll}
\hline\hline
\bf{bin} & \bf{low} & \bf{up} & \bf{radius} & \bf{bin} & \bf{low} & \bf{up} & \bf{radius} \\
\hline
1   & 0.05  & 0.05  & 8.9    & 51  & 0.5   & 0.52  & 1.8    \\
2   & 0.05  & 0.05  & 8.3    & 52  & 0.52  & 0.55  & 1.8    \\
3   & 0.05  & 0.06  & 8.8    & 53  & 0.55  & 0.57  & 1.8    \\
4   & 0.06  & 0.06  & 6.7    & 54  & 0.57  & 0.6   & 1.8    \\
5   & 0.06  & 0.06  & 8.2    & 55  & 0.6   & 0.63  & 1.9    \\
6   & 0.06  & 0.07  & 8.7    & 56  & 0.63  & 0.66  & 1.9    \\
7   & 0.07  & 0.07  & 6.2    & 57  & 0.66  & 0.69  & 1.9    \\
8   & 0.07  & 0.07  & 6.6    & 58  & 0.69  & 0.72  & 2      \\
9   & 0.07  & 0.08  & 5.9    & 59  & 0.72  & 0.76  & 1.9    \\
10  & 0.08  & 0.08  & 6.1    & 60  & 0.76  & 0.79  & 2      \\
11  & 0.08  & 0.08  & 6.5    & 61  & 0.79  & 0.83  & 2      \\
12  & 0.08  & 0.09  & 5.5    & 62  & 0.83  & 0.87  & 2      \\
13  & 0.09  & 0.09  & 6.2    & 63  & 0.87  & 0.91  & 2      \\
14  & 0.09  & 0.1   & 5.6    & 64  & 0.91  & 0.95  & 2.1    \\
15  & 0.1   & 0.1   & 4.8    & 65  & 0.95  & 1     & 2.1    \\
16  & 0.1   & 0.1   & 4.2    & 66  & 1     & 1.04  & 2.1    \\
17  & 0.1   & 0.11  & 3.4    & 67  & 1.04  & 1.09  & 2.1    \\
18  & 0.11  & 0.11  & 3.1    & 68  & 1.09  & 1.15  & 2.1    \\
19  & 0.11  & 0.12  & 3      & 69  & 1.15  & 1.2   & 2.2    \\
20  & 0.12  & 0.13  & 2.8    & 70  & 1.2   & 1.26  & 2.2    \\
21  & 0.13  & 0.13  & 2.6    & 71  & 1.26  & 1.32  & 2.2    \\
22  & 0.13  & 0.14  & 2.6    & 72  & 1.32  & 1.38  & 2.2    \\
23  & 0.14  & 0.14  & 2.6    & 73  & 1.38  & 1.44  & 2.3    \\
24  & 0.14  & 0.15  & 2.3    & 74  & 1.44  & 1.51  & 2.2    \\
25  & 0.15  & 0.16  & 2.4    & 75  & 1.51  & 1.58  & 2.3    \\
26  & 0.16  & 0.17  & 2.4    & 76  & 1.58  & 1.66  & 2.3    \\
27  & 0.17  & 0.17  & 2.1    & 77  & 1.66  & 1.73  & 2.3    \\
28  & 0.17  & 0.18  & 2.3    & 78  & 1.73  & 1.82  & 2.3    \\
29  & 0.18  & 0.19  & 2.3    & 79  & 1.82  & 1.9   & 2.3    \\
30  & 0.19  & 0.2   & 2.2    & 80  & 1.9   & 1.99  & 2.3    \\
31  & 0.2   & 0.21  & 2.2    & 81  & 1.99  & 2.08  & 2.3    \\
32  & 0.21  & 0.22  & 2.2    & 82  & 2.08  & 2.18  & 2.3    \\
33  & 0.22  & 0.23  & 2.1    & 83  & 2.18  & 2.29  & 2.3    \\
34  & 0.23  & 0.24  & 2.1    & 84  & 2.29  & 2.39  & 2.3    \\
35  & 0.24  & 0.25  & 2      & 85  & 2.39  & 2.51  & 2.3    \\
36  & 0.25  & 0.26  & 2.1    & 86  & 2.51  & 2.62  & 2.3    \\
37  & 0.26  & 0.27  & 2      & 87  & 2.62  & 2.75  & 2.2    \\
38  & 0.27  & 0.29  & 2.1    & 88  & 2.75  & 2.88  & 2.2    \\
39  & 0.29  & 0.3   & 1.9    & 89  & 2.88  & 3.01  & 2.2    \\
40  & 0.3   & 0.32  & 1.9    & 90  & 3.01  & 3.15  & 2.1    \\
41  & 0.32  & 0.33  & 1.9    & 91  & 3.15  & 3.3   & 2.1    \\
42  & 0.33  & 0.35  & 1.9    & 92  & 3.3   & 3.46  & 2.1    \\
43  & 0.35  & 0.36  & 1.9    & 93  & 3.46  & 3.62  & 2.1    \\
44  & 0.36  & 0.38  & 1.9    & 94  & 3.62  & 3.79  & 2      \\
45  & 0.38  & 0.4   & 1.8    & 95  & 3.79  & 3.97  & 2      \\
46  & 0.4   & 0.42  & 1.9    & 96  & 3.97  & 4.16  & 2      \\
47  & 0.42  & 0.44  & 1.8    & 97  & 4.16  & 4.35  & 1.9    \\
48  & 0.44  & 0.46  & 1.8    & 98  & 4.35  & 4.56  & 1.9    \\
49  & 0.46  & 0.48  & 1.8    & 99  & 4.56  & 4.77  & 1.9    \\
50  & 0.48  & 0.5   & 1.8    & 100 & 4.77  & 5     & 2      \\
\hline
\end{tabular}
\caption{Bin index, lower and upper bin boundary (in $\msun$) and $r_i$ (in $\mathrm{au}$) for RUN\nobreakdash-2.}
\label{table:infrad}
\end{table}


\section{Initial conditions}\label{app:initdist}

The 35 systems from \citetalias{2018MNRAS.475.5618B} we used to obtain the probability distributions for $M_\mathrm{*,i}$, $M_\mathrm{disc,i}$ and $\dot{M}_\mathrm{in}$ are given in Table~\ref{table:35}. The 20 systems used for the distributions of $R_\mathrm{disc,i}$ and the expansion rate of $R_\mathrm{i}$ are given in Table~\ref{table:20}. The numbers given in these tables correspond to the protostar indices used in \citetalias{2018MNRAS.475.5618B}.

\begin{table}[ht]
\begin{tabular}{l|l|l|l|l}
\hline\hline
001 & 032 & 047 & 066 & 099 \\
002 & 033 & 049 & 069 & 101 \\
004 & 035 & 053 & 071 & 106 \\
006 & 036 & 057 & 078 & 108 \\
009 & 037 & 058 & 085 & 118 \\
010 & 040 & 061 & 086 & 119 \\
029 & 041 & 063 & 087 & 153 \\
\hline
\end{tabular}
\end{table}

\begin{table}[ht]
\caption{Systems from \citetalias{2018MNRAS.475.5618B} used for initial radii and expansion rates of infall radii.}
\label{table:20}
\begin{tabular}{l|l|l|l|l}
\hline\hline
001 & 029 & 041 & 069 & 106 \\
002 & 035 & 047 & 087 & 108 \\
004 & 037 & 061 & 099 & 118 \\
010 & 040 & 063 & 101 & 153 \\
\hline
\end{tabular}
\caption{Systems from \citetalias{2018MNRAS.475.5618B} used for initial masses and infall rates.}
\label{table:35}
\end{table}

\begin{figure*}[ht]
  \begin{subfigure}[t]{0.49\textwidth}
  \includegraphics[width=\linewidth]{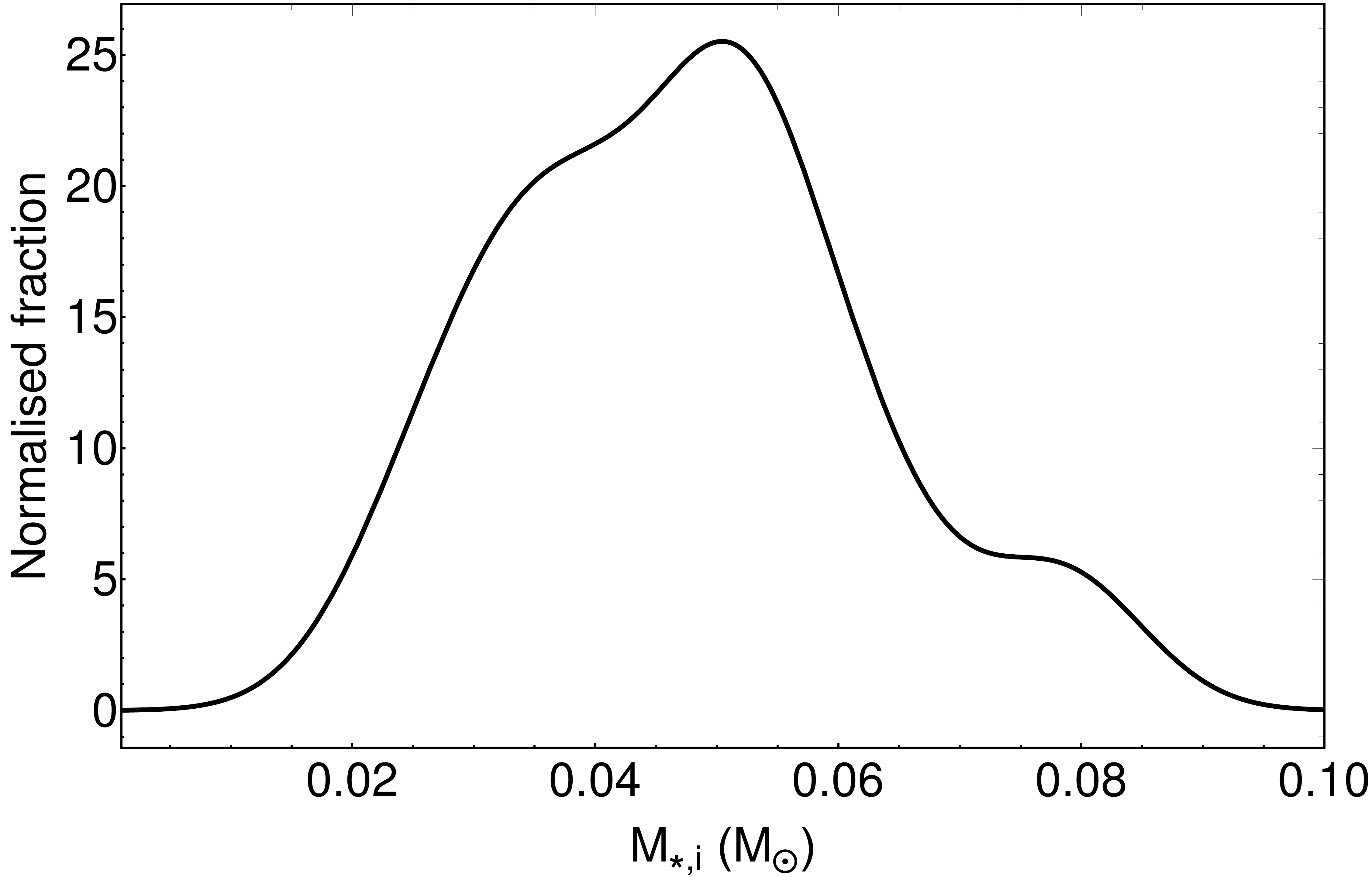}
  \end{subfigure}
  \begin{subfigure}[t]{0.49\textwidth}
  \includegraphics[width=\linewidth]{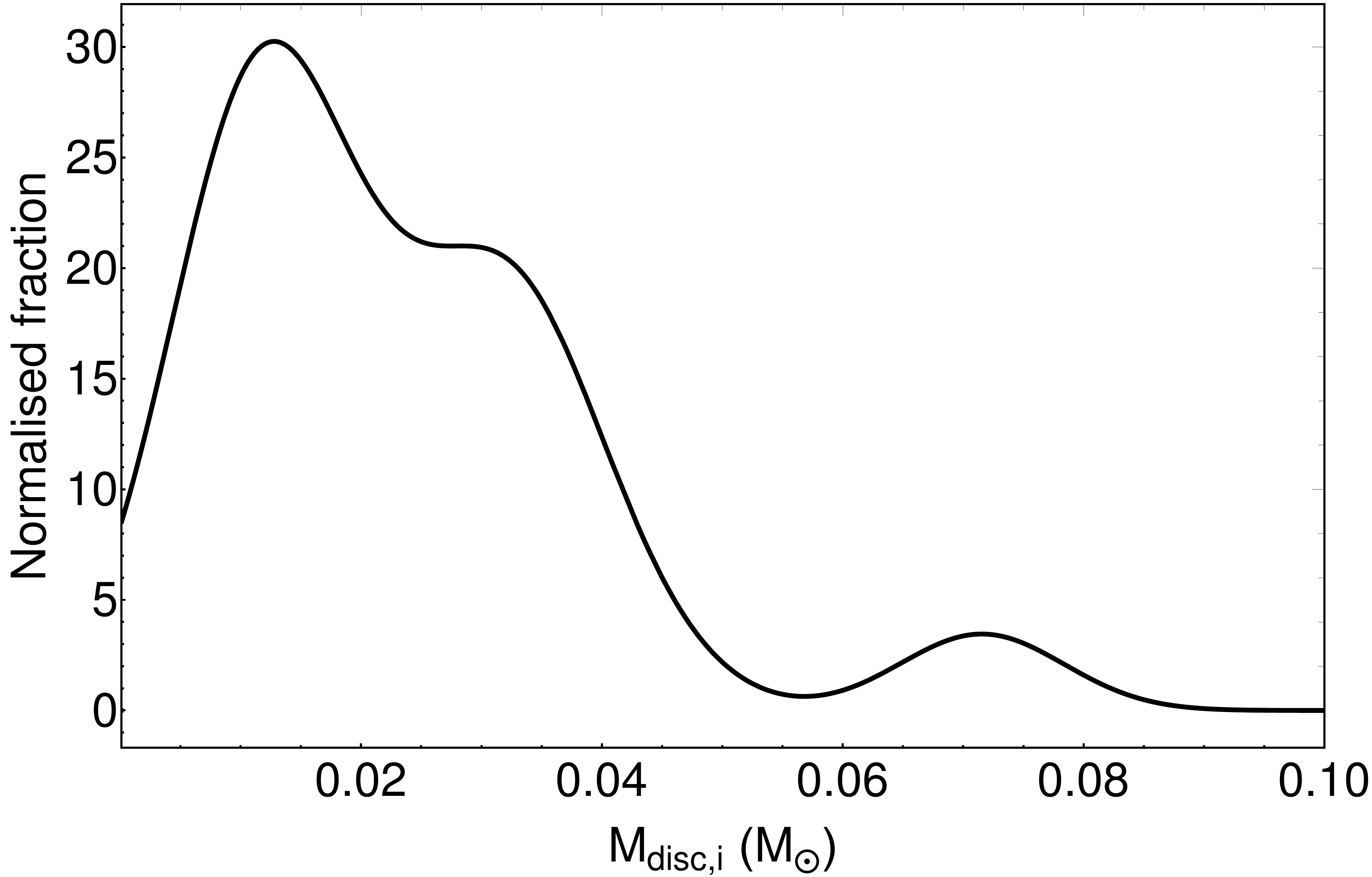}
  \end{subfigure}
  \begin{subfigure}[t]{0.49\textwidth}
  \includegraphics[width=\linewidth]{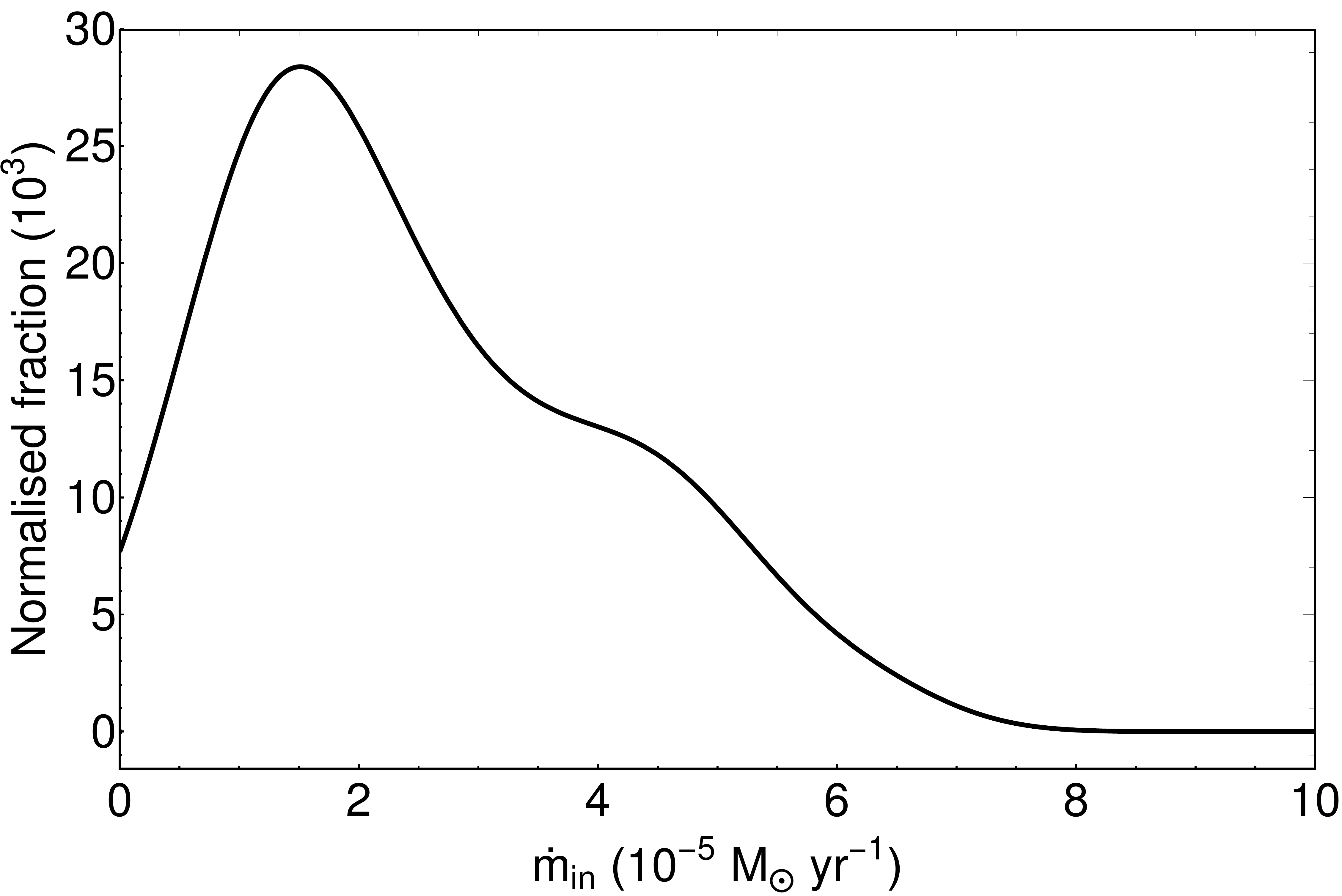}
  \end{subfigure}
  \begin{subfigure}[t]{0.49\textwidth}
  \includegraphics[width=\linewidth]{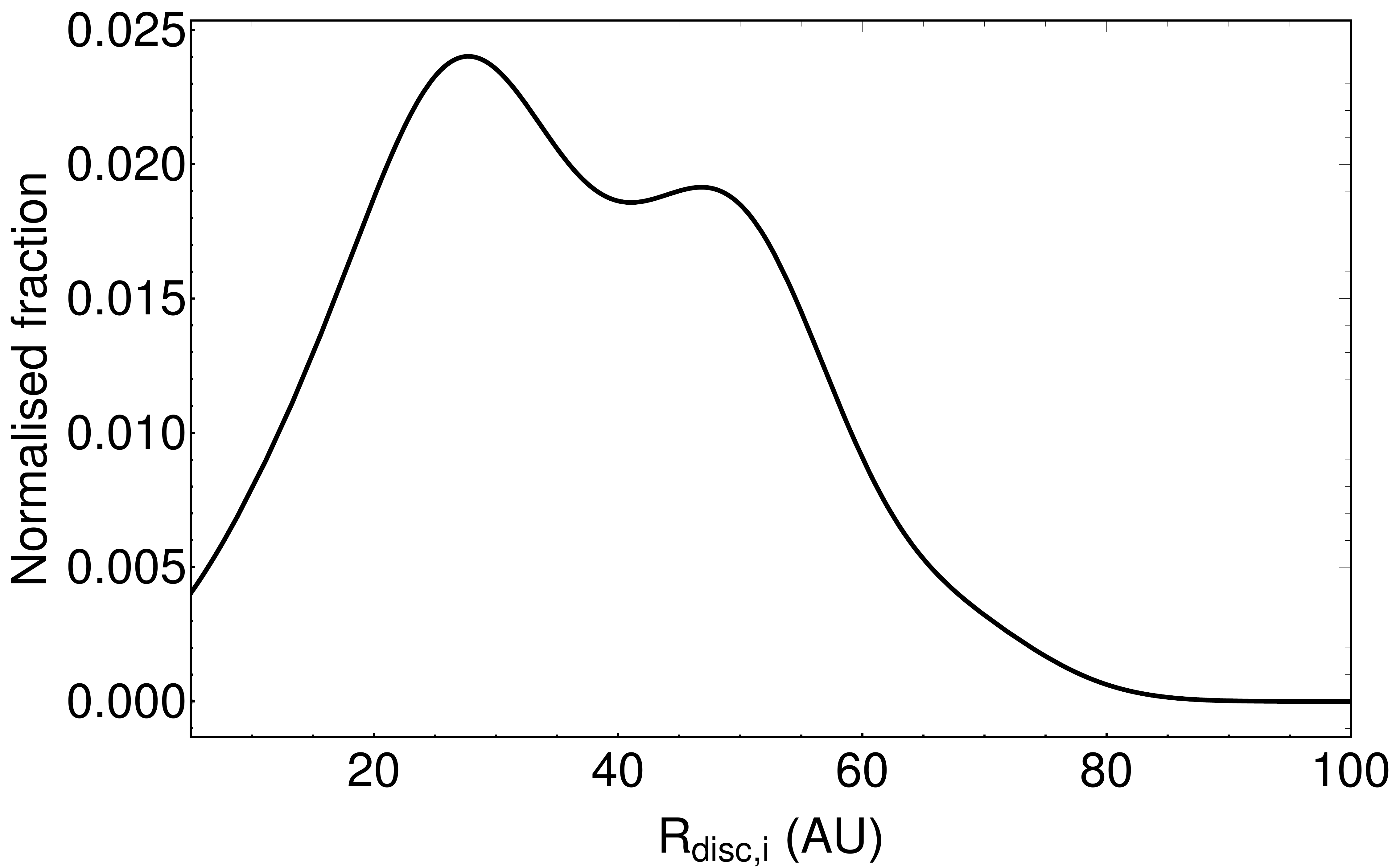}
  \end{subfigure}
  \begin{subfigure}[t]{0.49\textwidth}
  \includegraphics[width=\linewidth]{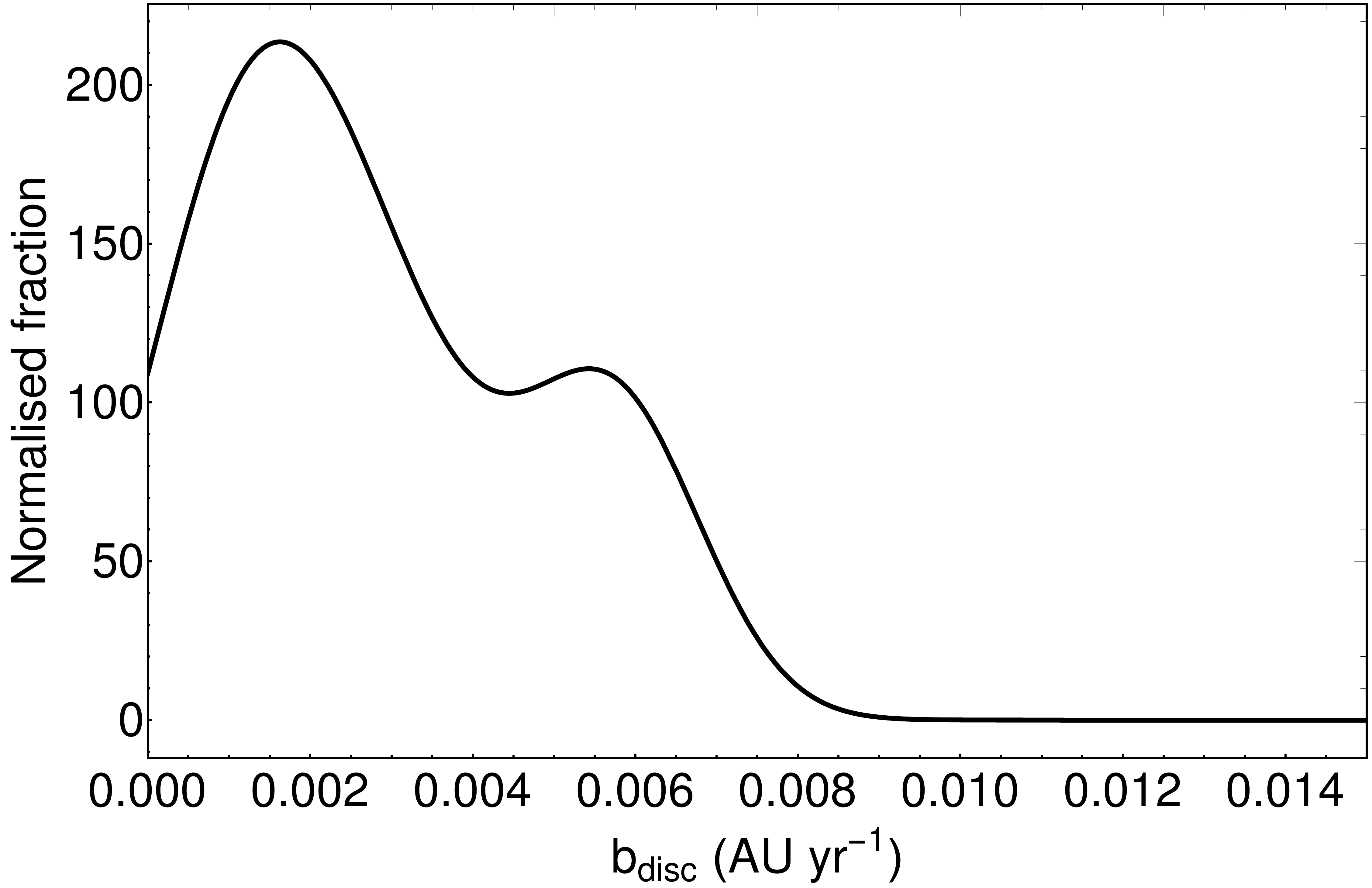}
  \end{subfigure}
  \begin{subfigure}[t]{0.49\textwidth}
  \includegraphics[width=\linewidth]{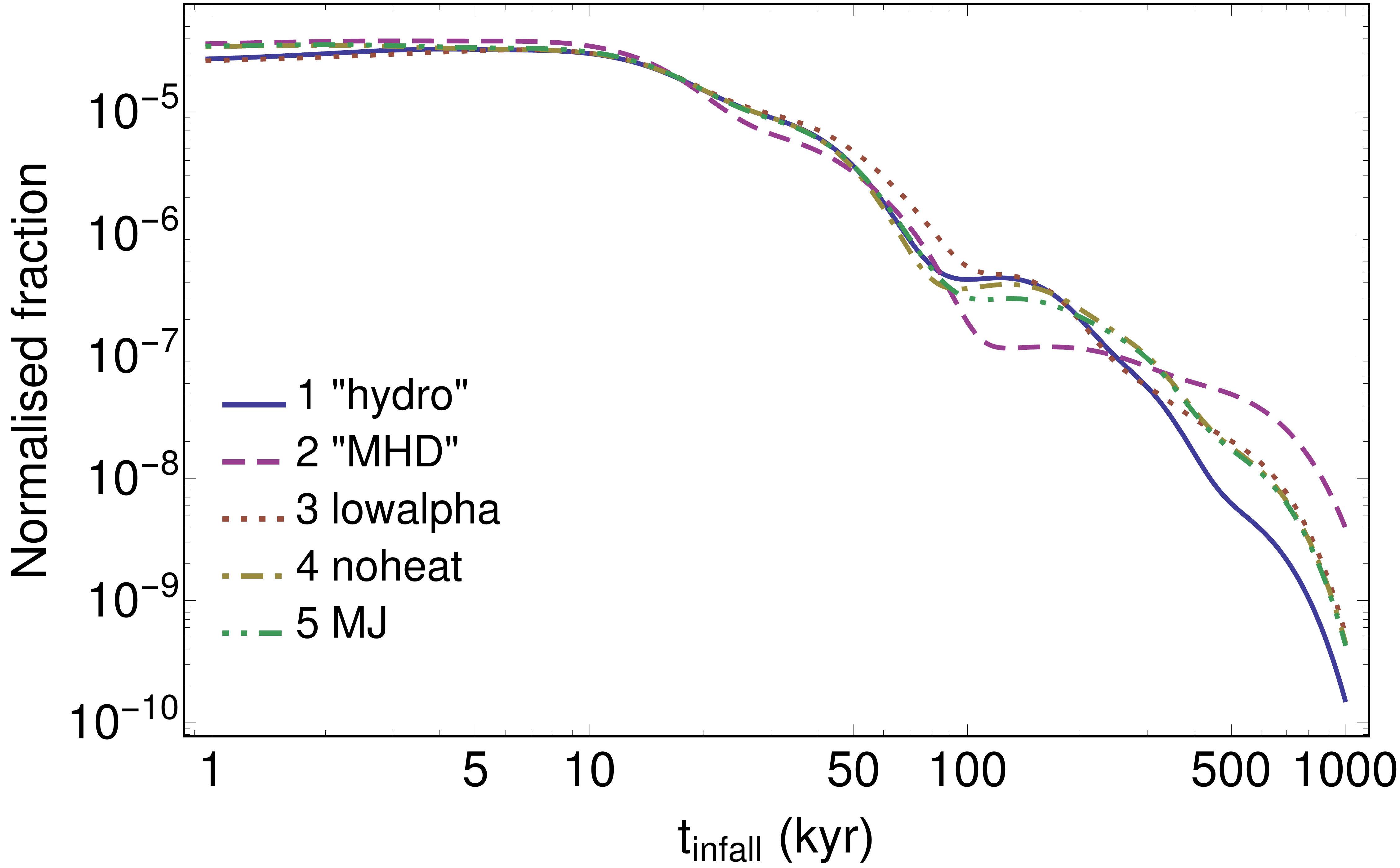}
  \end{subfigure}
  \caption{Kernel density estimate for the initial distributions (neglecting correlation). Top left: protostellar mass, top right: disc mass, middle left: infall rate, middle right: disc radius, bottom left: expansion rate of the infall radius, bottom right: $t_\mathrm{infall}$.}
  \label{fig:init}
\end{figure*}

Figure~\ref{fig:init} shows kernel density estimates for $M_\mathrm{*,i}$, $M_\mathrm{disc,i}$, $\dot{M}_\mathrm{in}$, $R_\mathrm{disc,i}$, $b_\mathrm{disc}$ and $t_\mathrm{infall}$, respectively. In order to create initial conditions for our runs, we do not use these estimates directly. Instead we use two multivariate distributions, one combining $M_\mathrm{*,i}$, $M_\mathrm{disc,i}$ and $\dot{M}_\mathrm{in}$, the other one combining $R_\mathrm{disc,i}$ and $b_\mathrm{disc}$.

\section{Reduced disc lifetimes}\label{app:lifetimes}

\begin{figure*}[ht]
  \begin{subfigure}[t]{0.49\textwidth}
  \includegraphics[width=\linewidth]{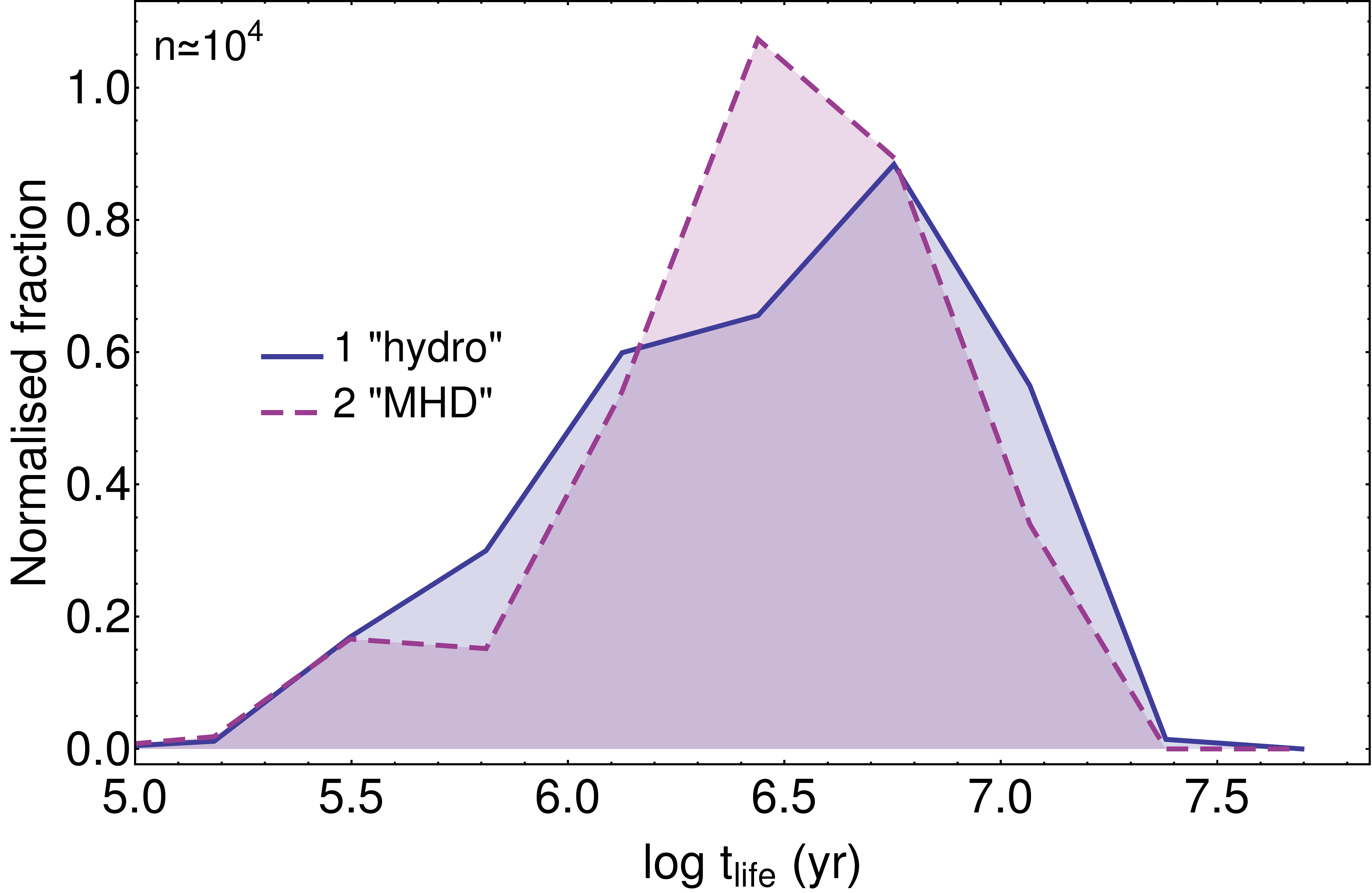}
  \end{subfigure}
  \begin{subfigure}[t]{0.49\textwidth}
  \includegraphics[width=\linewidth]{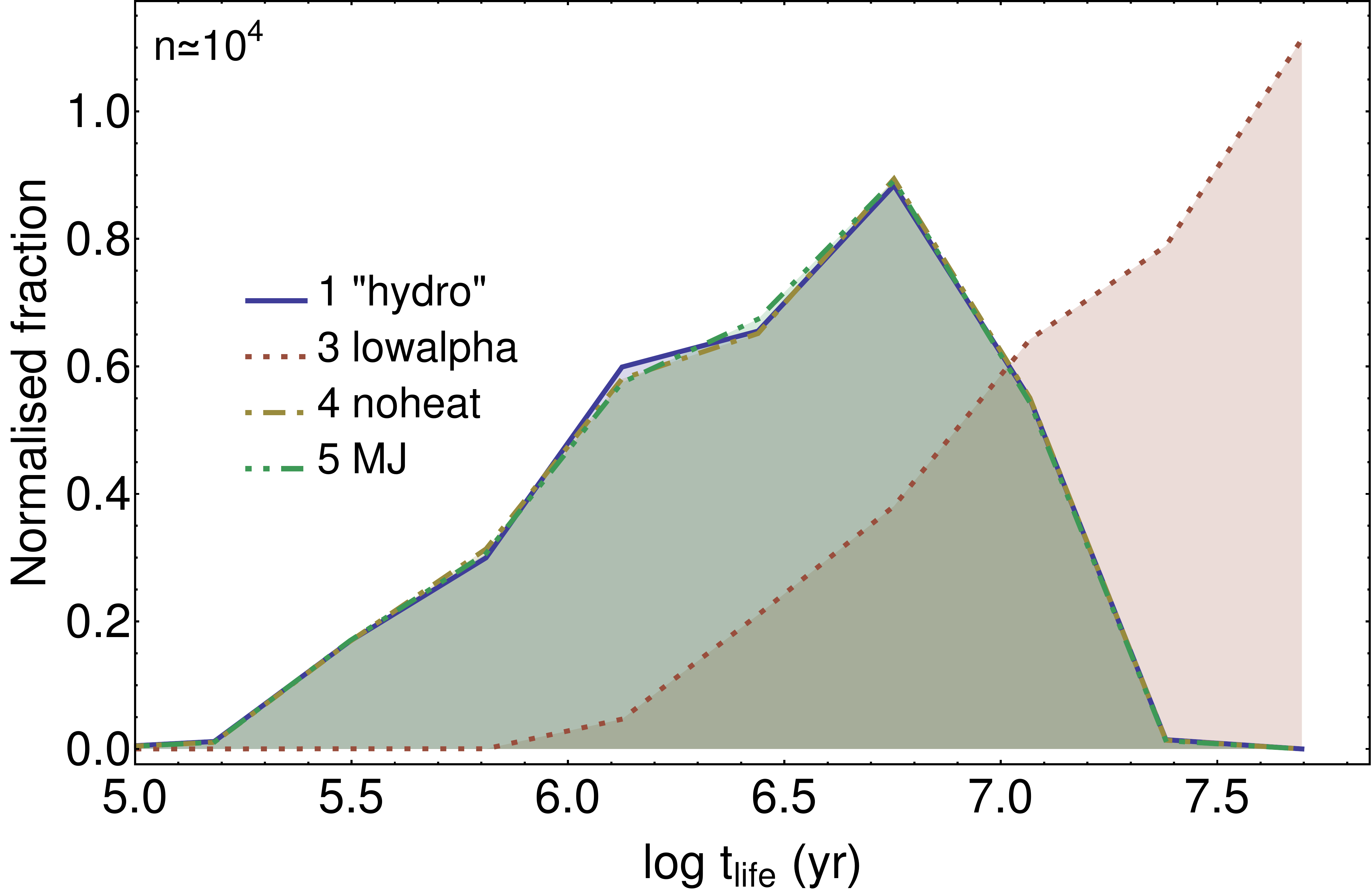}
  \end{subfigure}
  \begin{subfigure}[t]{0.49\textwidth}
  \includegraphics[width=\linewidth]{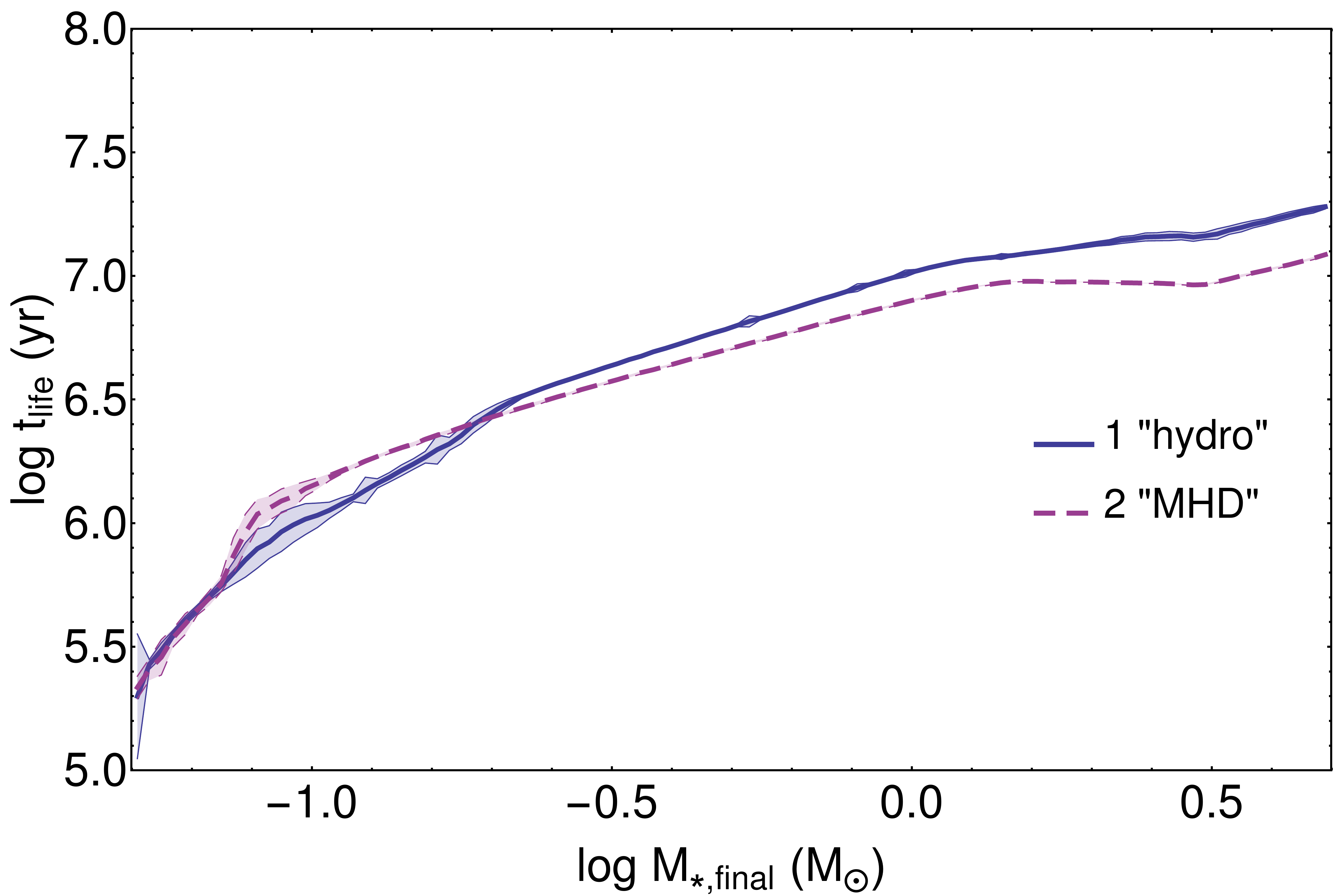}
  \end{subfigure}
  \begin{subfigure}[t]{0.49\textwidth}
  \includegraphics[width=\linewidth]{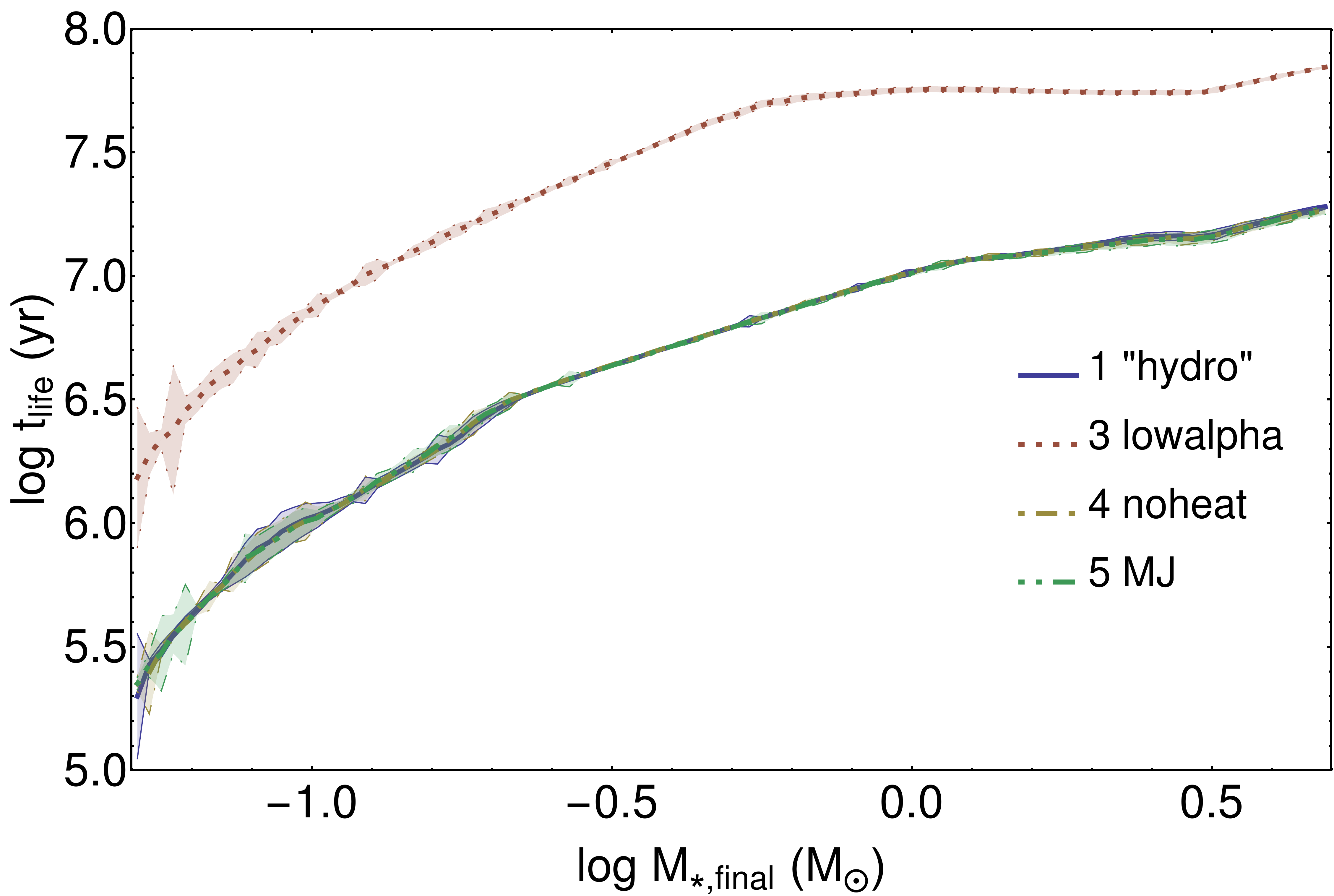}
  \end{subfigure}
  \caption{Top: distribution of disc lifetimes, reduced by $t_\mathrm{pms}$ (see Sect.~\ref{subs:disp}). RUN\nobreakdash-1 and RUN\nobreakdash-2 are shown in the left, RUN\nobreakdash-1, RUN\nobreakdash-3, RUN\nobreakdash-4 and RUN\nobreakdash-5 in the right panel. RUN\nobreakdash-1, RUN\nobreakdash-4 and RUN\nobreakdash-5 are almost indistinguishable. Bottom: Same runs, $t_\mathrm{life}$ as a function of final stellar mass.}
  \label{fig:tlife}
\end{figure*}

Here we show our results for the disc lifetimes when the reduction for the start of the PMS phase is applied (see Sect.~\ref{subs:disp} and Sect.~\ref{sect:Results}). Table~\ref{tab:tlife} gives the global mean lifetimes for all runs, Fig.~\ref{fig:tlife} shows the corresponding distributions.
The effect of the reduction is to decrease the disc lifetimes by $t_\mathrm{PMS}$, which depends on the accretion history of the system as explained in Sect.~\ref{subs:disp}. This leads to a stronger reduction for the runs based on hydrodynamic initial conditions (the PMS phase is reached later due to the higher disc masses and, hence, higher accretion rates). It is visible in the table as well as in the left panels of the figure. RUN\nobreakdash-1 and RUN\nobreakdash-2 are closer together although lifetimes in RUN\nobreakdash-1 are still higher. The effect on runs 1, 4 and 5 is very similar, so the lifetimes in these runs are still almost the same, as seen both in the table as well as in the right panels of the figure.

\begin{table}[ht]
\begin{tabular}{lllll}
\hline\hline
\multicolumn{2}{c}{Run} &
\begin{tabular}[c]{@{}l@{}} $t_\mathrm{life} (\mathrm{Myr})$ \end{tabular} \\
\hline
1 & ``hydro''    & $\num{4.6(1)}$ \\
2 & ``MHD''      & $\num{3.8(1)}$ \\
3 & lowalpha   & $\num{27.0(10)}$ \\
4 & noheat     & $\num{4.5(2)}$   \\
5 & MJ         &  $\num{4.5(2)}$  \\
\hline
\end{tabular}
\caption{Reduced disc lifetimes (global mean).}
\label{tab:tlife}
\end{table}


\section{Results in specific mass bins}\label{app:massbin}

Here we show the results from Sect.~\ref{subs:12} to \ref{ssec:dep}, the disc masses and disc radii at the end of the infall phase, as well as the disc lifetimes, in three specific mass bins. These are: $\num{0.48} - \SI{0.5}{\msun}$, $\num{1.00} - \SI{1.04}{\msun}$ and $\num{1.44} - \SI{1.51}{\msun}$.
We kept the same x-axis in all three bins for comparison. Also, some of the distributions are very narrow (see for example $t_\mathrm{NIR}$, compare to the bottom left panel of Fig.~\ref{fig:disc_2}). Sometimes all values fall into the same bin, hence the ``spiky'' nature of the figures shown here.

\begin{figure*}[ht]
  \begin{subfigure}[t]{0.49\textwidth}
  \includegraphics[width=\linewidth]{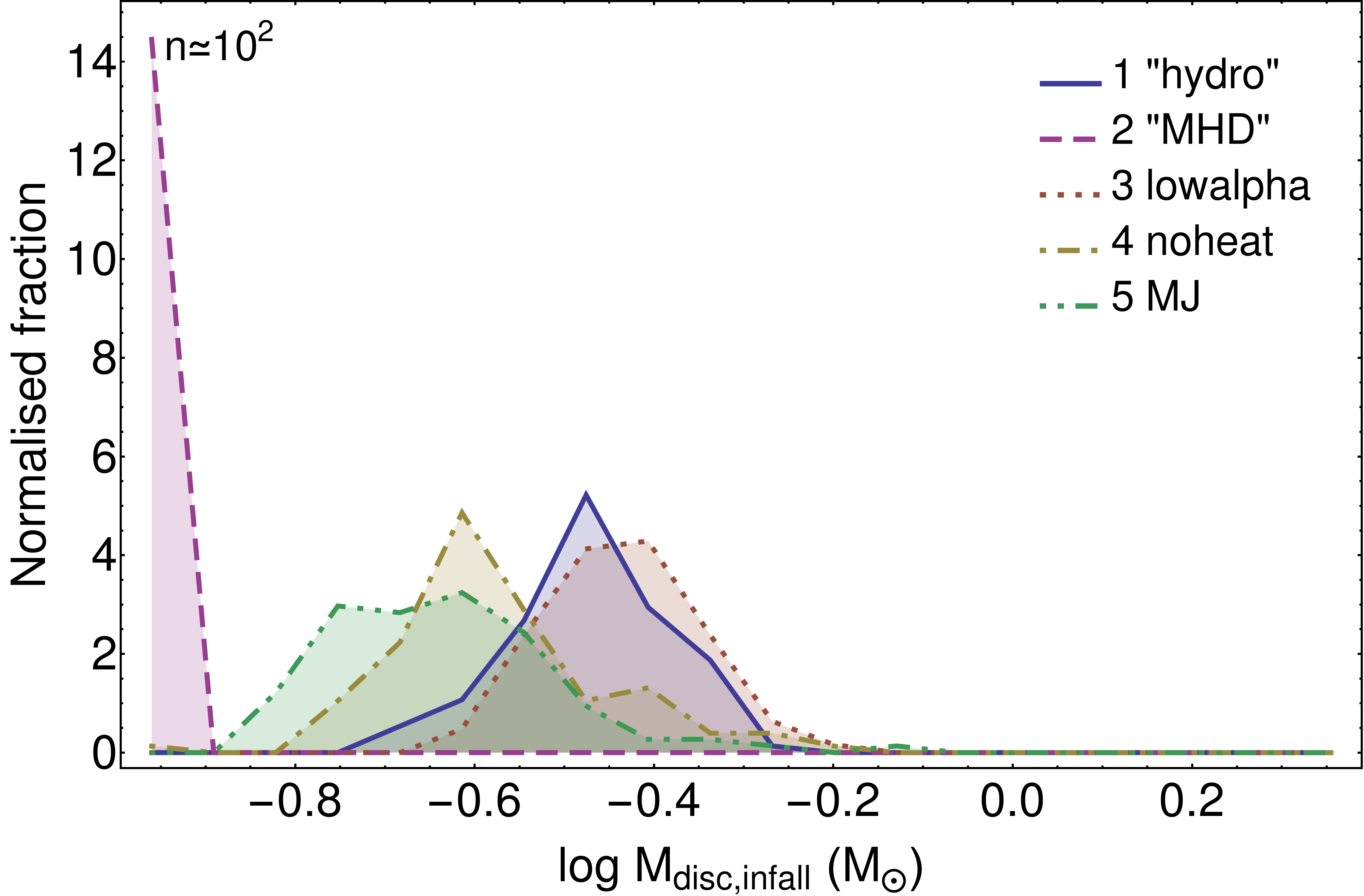}
  \end{subfigure}
  \begin{subfigure}[t]{0.49\textwidth}
  \includegraphics[width=\linewidth]{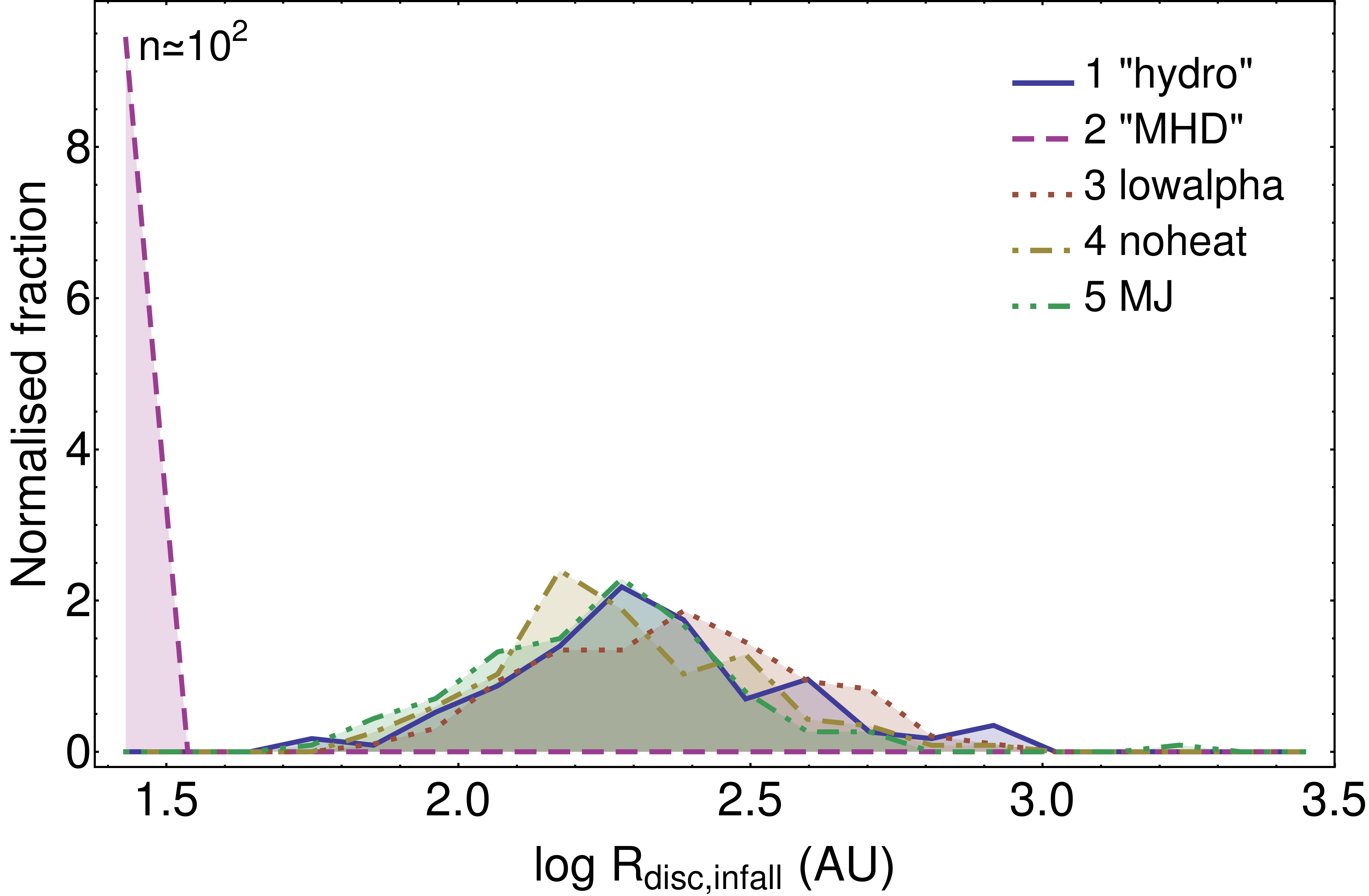}
  \end{subfigure}
  \begin{subfigure}[t]{0.49\textwidth}
  \includegraphics[width=\linewidth]{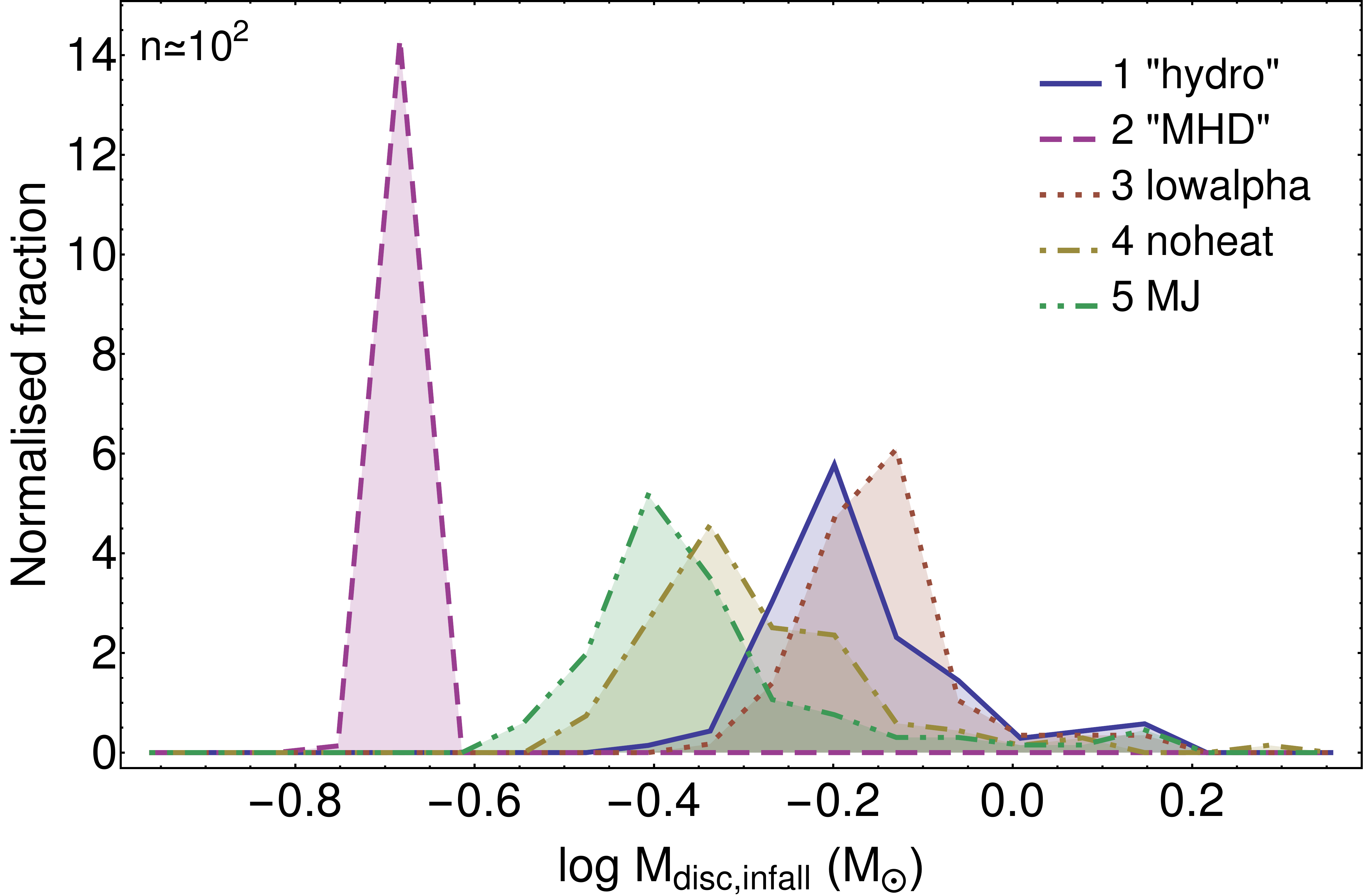}
  \end{subfigure}
  \begin{subfigure}[t]{0.49\textwidth}
  \includegraphics[width=\linewidth]{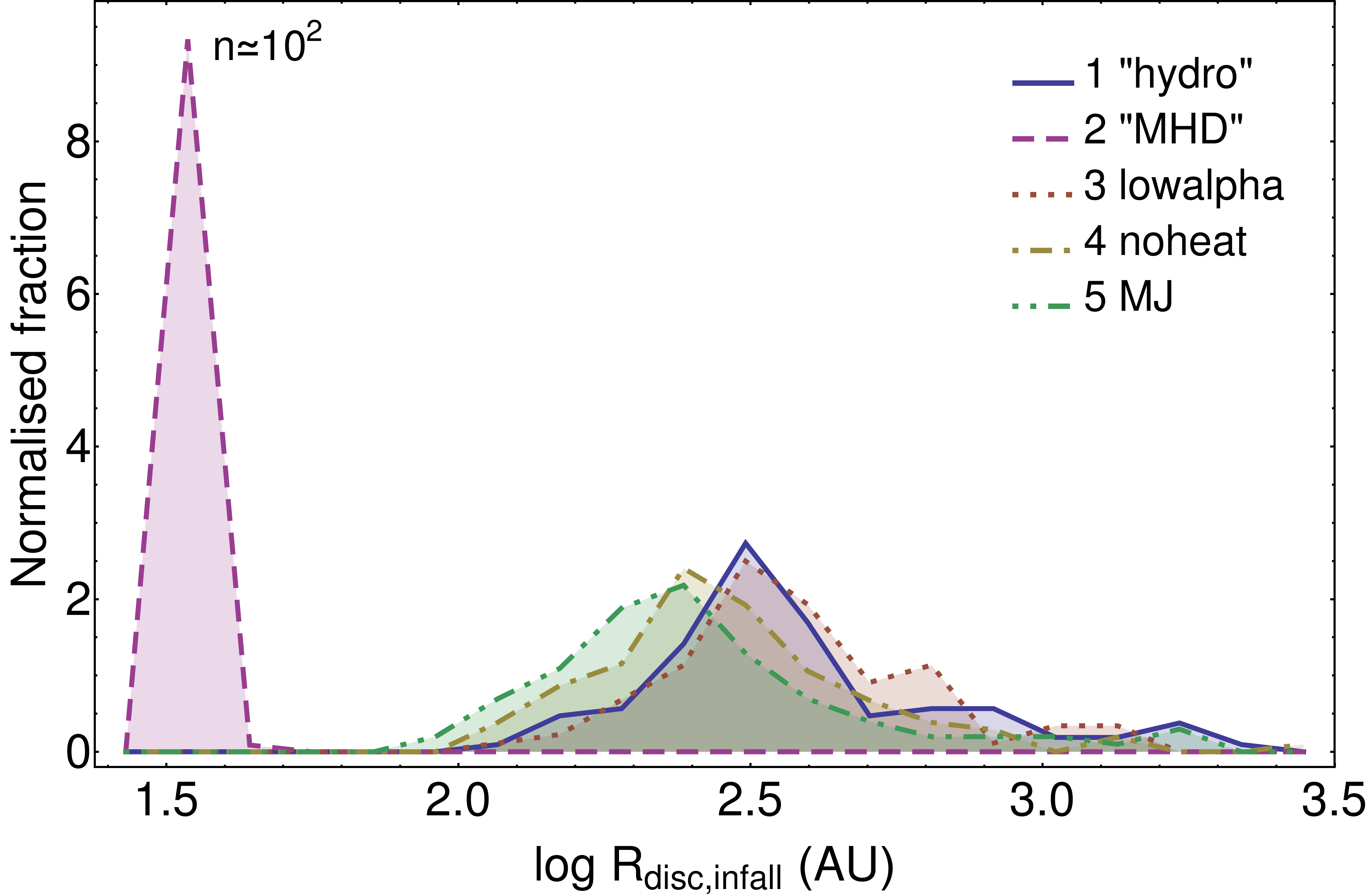}
  \end{subfigure}
  \begin{subfigure}[t]{0.49\textwidth}
  \includegraphics[width=\linewidth]{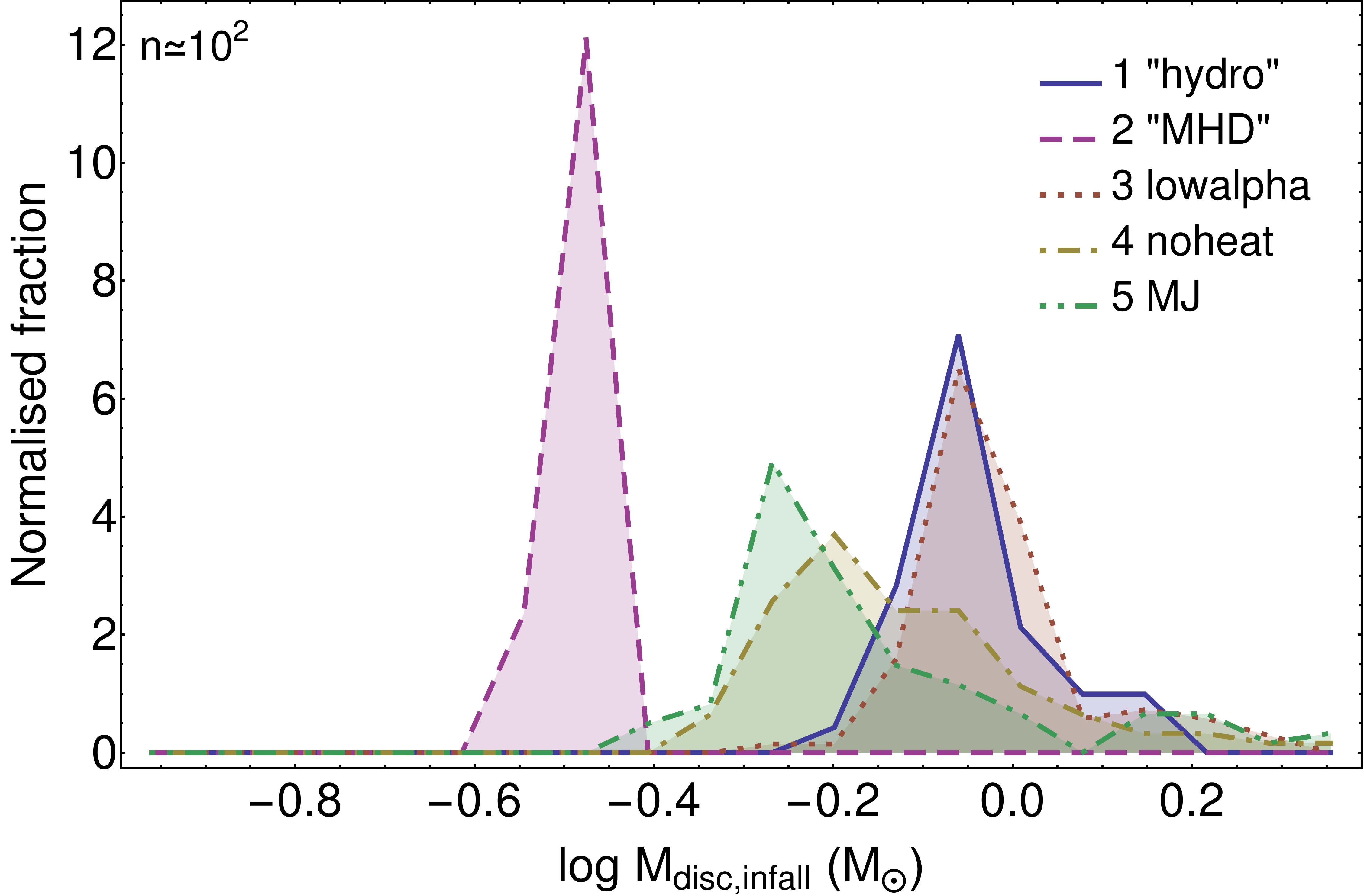}
  \end{subfigure}
  \begin{subfigure}[t]{0.49\textwidth}
  \includegraphics[width=\linewidth]{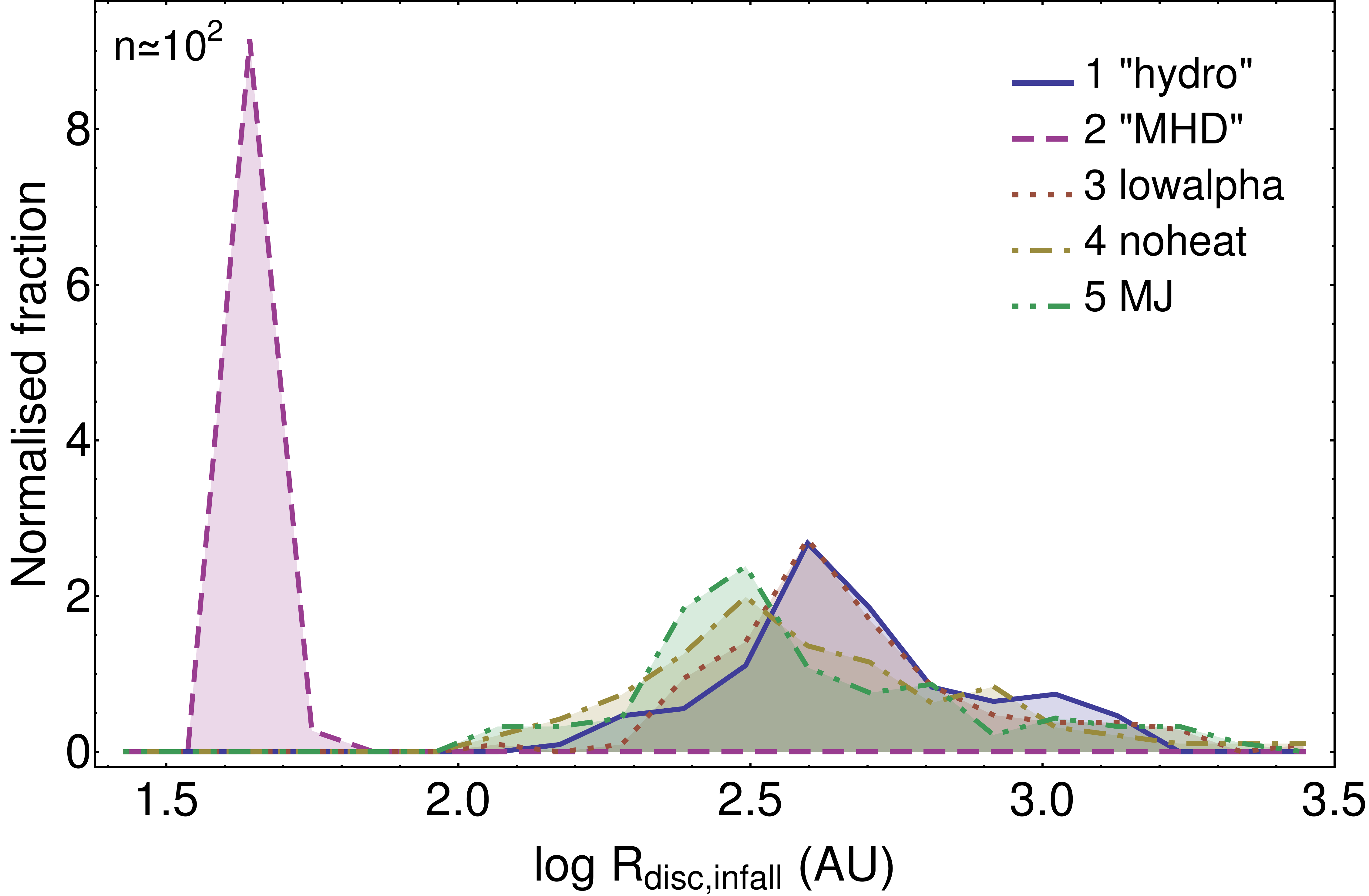}
  \end{subfigure}
  \caption{Distributions of disc masses (left column) and disc radii (right column) at the end of infall for specific mass bins. From top to bottom: \SI{0.5}{\msun}, \SI{1}{\msun} and \SI{1.5}{\msun}.}
  \label{fig:appmr}
\end{figure*}

\begin{figure*}[ht]
  \begin{subfigure}[t]{0.49\textwidth}
  \includegraphics[width=\linewidth]{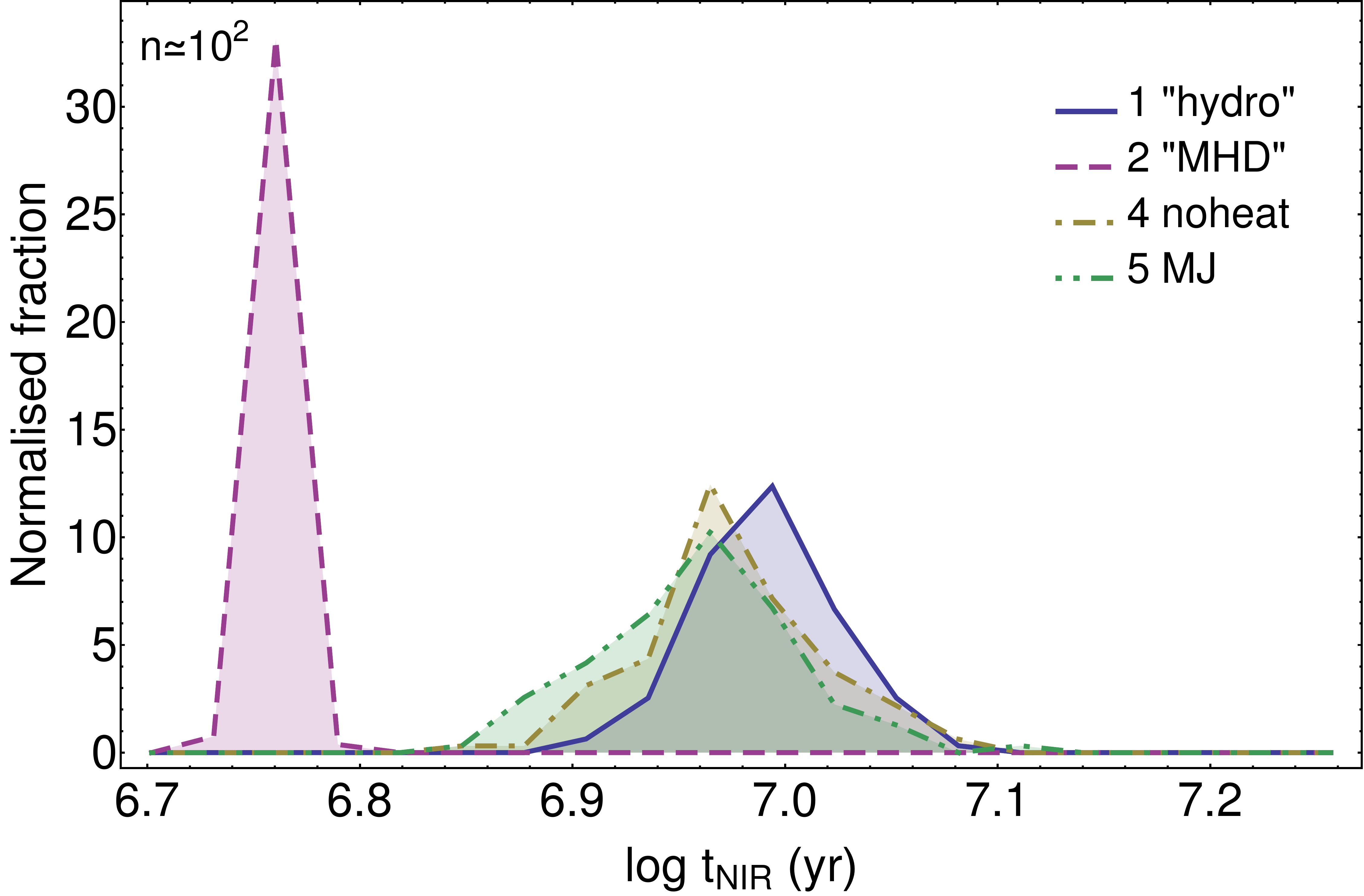}
  \end{subfigure}
  \begin{subfigure}[t]{0.49\textwidth}
  \includegraphics[width=\linewidth]{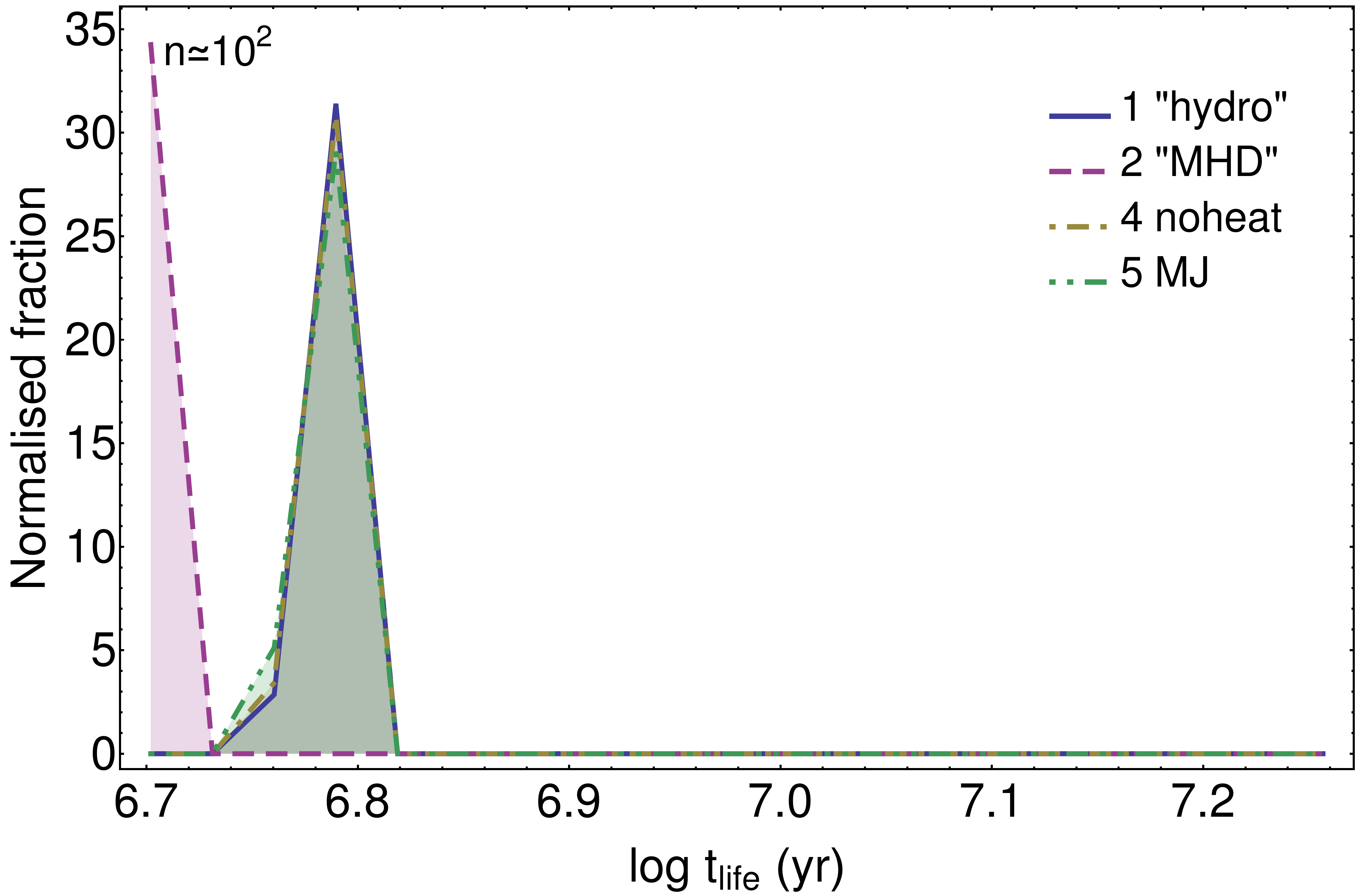}
  \end{subfigure}
  \begin{subfigure}[t]{0.49\textwidth}
  \includegraphics[width=\linewidth]{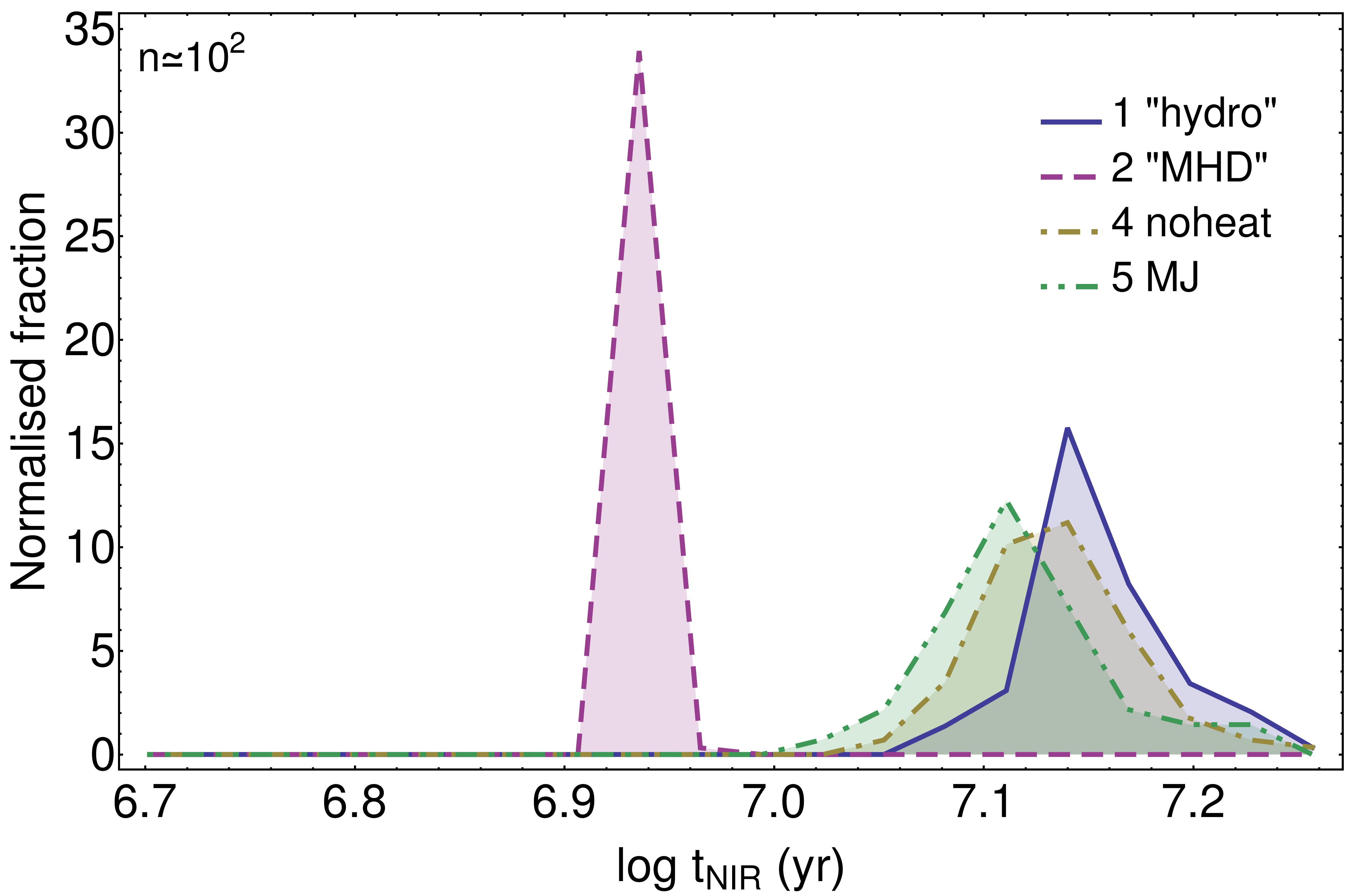}
  \end{subfigure}
  \begin{subfigure}[t]{0.49\textwidth}
  \includegraphics[width=\linewidth]{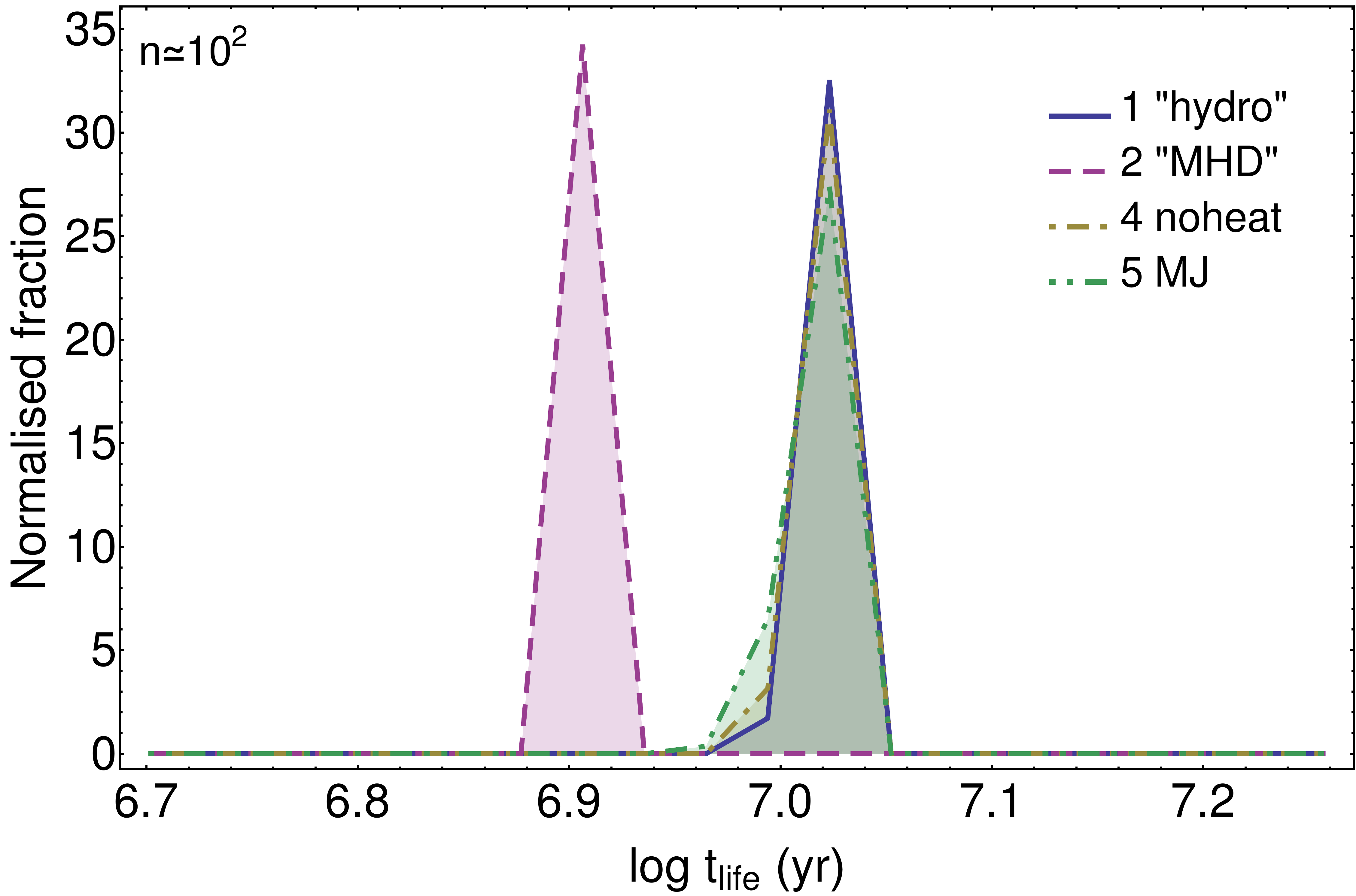}
  \end{subfigure}
  \begin{subfigure}[t]{0.49\textwidth}
  \includegraphics[width=\linewidth]{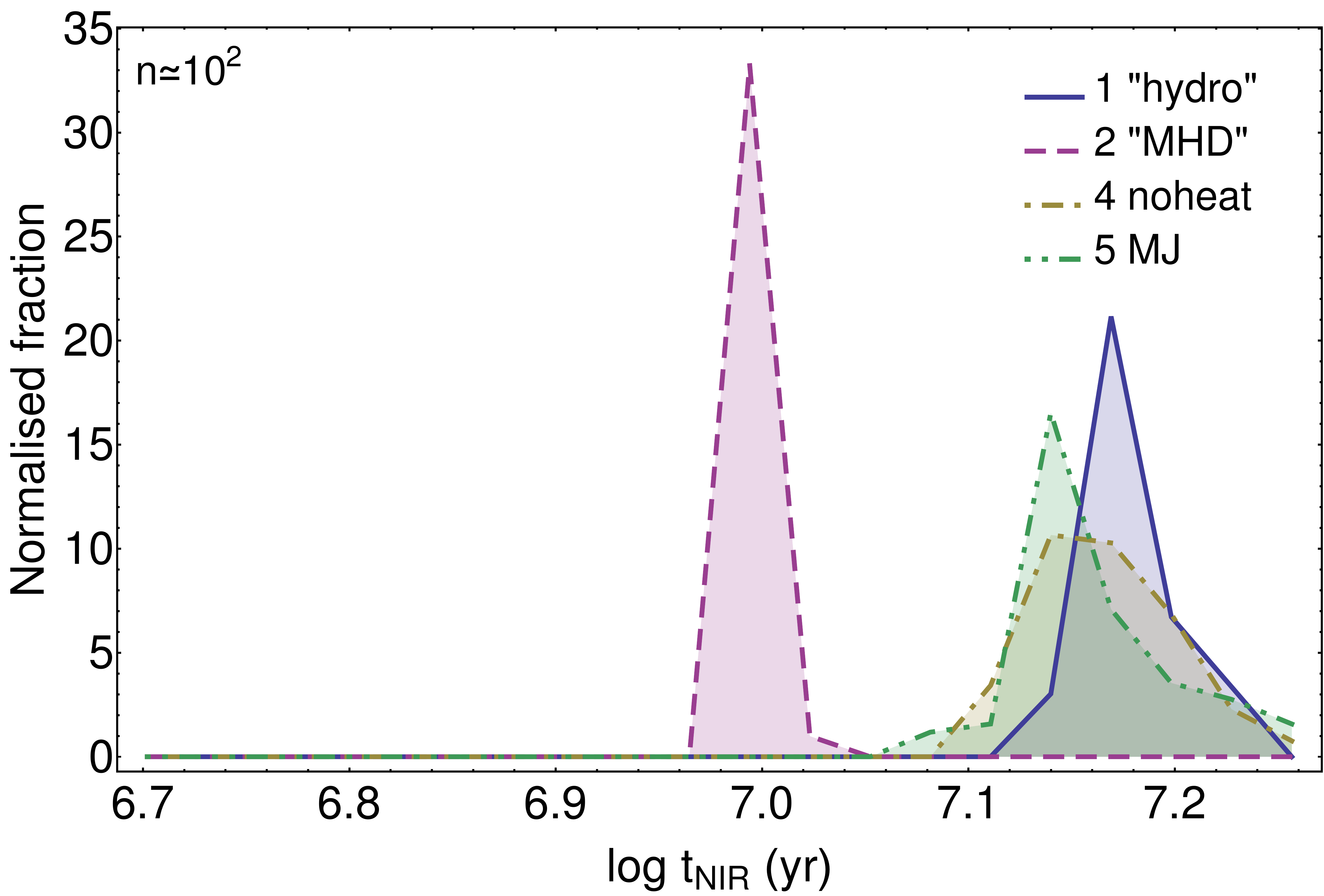}
  \end{subfigure}
  \begin{subfigure}[t]{0.49\textwidth}
  \includegraphics[width=\linewidth]{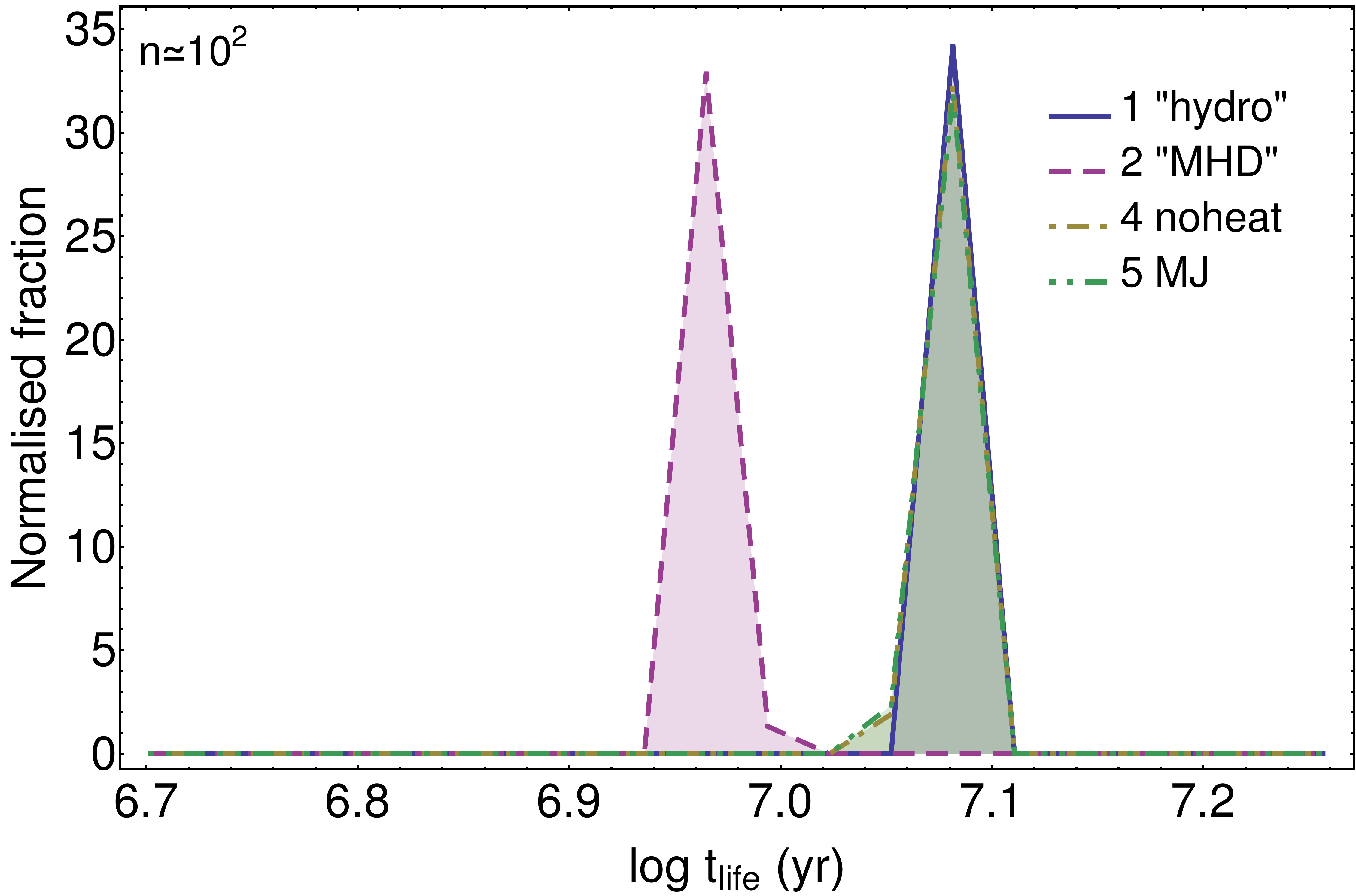}
  \end{subfigure}
  \caption{Distribution of $t_\mathrm{NIR}$ (left column) and $t_\mathrm{life}$ (right column) for specific mass bins. From top to bottom: \SI{0.5}{\msun}, \SI{1}{\msun} and \SI{1.5}{\msun}.}
  \label{fig:applt}
\end{figure*}

\end{appendix}

\end{document}